\documentclass[utf8]{FrontiersinHarvard}

\usepackage{url,hyperref,lineno,microtype,subcaption}
\usepackage{amsmath,amsthm,amssymb,amsfonts,latexsym,mathtools,bm,bbm}
\usepackage{braket}
\usepackage{xspace,xcolor}
\usepackage{ifthen,xifthen,delimset,bbold,mathbbol,multirow}
\usepackage{graphics,graphicx}
\usepackage[onehalfspacing]{setspace}

\def\keyFont{\fontsize{8}{11}\helveticabold }
\def\firstAuthorLast{Verriere {et~al.}} 
\def\Authors{
Marc Verriere$^{1}$, 
Nicolas Schunck$^{1,*}$, 
Irene Kim$^{2,3}$, 
Petar Marevi\'c$^{1,4,5}$, 
Kevin Quinlan$^{6}$, 
Michelle N. NGo$^{6,7}$, 
David Regnier$^{8,9}$, 
Raphael David Lasseri$^{4}$
}
\def\Address{%
  $^{1}$Nuclear and Data Theory group, Nuclear and Chemical Science Division, Lawrence Livermore National Laboratory, Livermore, CA-94550, USA%
  , \\
  $^{2}$Machine Intelligence group, Center for Applied Scientific Computing, Lawrence Livermore National Laboratory, Livermore, CA-94550, USA%
  , \\
  $^{3}$Sensing and Intelligent Systems group, Global Security Computing Applications Division, Lawrence Livermore National Laboratory, Livermore, CA-94550, USA%
  , \\
  $^{4}$Centre Borelli, ENS Paris-Saclay, Universit\'e Paris-Saclay, 91190 Gif-sur-Yvette, France%
  , \\
  $^{5}$Physics Department, Faculty of Science, University of Zagreb, 10000 Zagreb, Croatia%
  , \\
  $^{6}$Applied Statistics Group, Lawrence Livermore National Laboratory, Livermore, CA 94551, United States of America%
  , \\
  $^{7}$Center for Complex Biological Systems, University of California Irvine, Irvine, CA 92697, USA%
  , \\
  $^{8}$Universit\'e Paris-Saclay, CEA, Laboratoire Mati\`ere en Conditions Extr\^emes, 91680 Bruy\`eres-le-Ch\^atel, France%
  , \\
  $^{9}$CEA, DAM, DIF, 91297 Arpajon, France%
}
\def\corrAuthor{Corresponding Author}
\def\corrEmail{schunck1@llnl.gov}

\newcommand{\fvec}  [1]    {\boldsymbol{#1}}
\newcommand{\avec}         {\fvec{a}}
\newcommand{\bvec}         {\fvec{b}}

\newcommand{\qvec}         {\fvec{q}}

\newcommand{\rvec}         {\boldsymbol{r}}
\newcommand{\xvec}         {\mathbf{x}}
\newcommand{\diff}         {\operatorname{d}^{3}\!}

\newcommand{\op}    [1]    {\ensuremath{\hat{#1}}}

\newcommand{\jz}           {\ensuremath{\op{j}_{\rm z}}}

\newcommand{\idHO}         {\psi}
\newcommand{\idHFB} [1]    {\Phi({#1})}

\newcommand{\idPV}         {0}

\newcommand{\idNBodyOrig}  {\Phi}
\newcommand{\idNBodyRec}   {\Psi}
\newcommand{\idcan}        {\varphi}
\newcommand{\idcanR}       {\phi}
\newcommand{\orbCoefU}     {\ensuremath{\tilde{u}}}
\newcommand{\orbCoefV}     {\ensuremath{\tilde{v}}}
\newcommand{\olapCanRec}   {\ensuremath{\tilde{\tau}{}}}
\newcommand{\occCanU}      {\ensuremath{u}}
\newcommand{\occCanV}      {\ensuremath{v}}
\newcommand{\olapCan}      {\ensuremath{\tau}}

\newcommand{\ladder}[4]    {%
  \def\tempid{#1}%
  \def\tempac{#2}%
  \ifx\tempid\empty%
    {#3}_{#4}\ifx#2c^{\dagger}\fi%
  \else%
    {#3}^{(#1)\ifx#2c{\dagger}\fi}_{#4}%
  \fi%
}
\newcommand{\ladderc}[4]    {%
  \def\tempid{#1}%
  \def\tempac{#2}%
  \ifx\tempid\empty%
    {#3}\ifx#2c^{\dagger}\fi({#4})%
  \else%
    {#3}^{(#1)\ifx#2c{\dagger}\fi}({#4})%
  \fi%
}
\newcommand{\idP}           {c}
\newcommand{\aP}     [2][]  {\ladder{#1}{a}{\hat{\idP}}{#2}}
\newcommand{\cP}     [2][]  {\ladder{#1}{c}{\hat{\idP}}{#2}}

\newcommand{\idPc}          {a}
\newcommand{\aPc}    [2][]  {\ladder{#1}{a}{\hat{\idPc}}{#2}}
\newcommand{\cPc}    [2][]  {\ladder{#1}{c}{\hat{\idPc}}{#2}}

\newcommand{\idPyes}        {b}
\newcommand{\aPyes}  [2][]  {\ladder{#1}{a}{\hat{\idPyes}}{#2}}
\newcommand{\cPyes}  [2][]  {\ladder{#1}{c}{\hat{\idPyes}}{#2}}

\newcommand{\idPno}         {f}
\newcommand{\aPno}   [2][]  {\ladder{#1}{a}{\hat{\idPno}}{#2}}
\newcommand{\cPno}   [2][]  {\ladder{#1}{c}{\hat{\idPno}}{#2}}

\newcommand{\idQPno}        {\chi}
\newcommand{\aQPno}  [2][]  {\ladder{#1}{a}{\hat{\idQPno}}{#2}}
\newcommand{\cQPno}  [2][]  {\ladder{#1}{c}{\hat{\idQPno}}{#2}}

\newcommand{\idQPyes}       {\gamma}
\newcommand{\aQPyes} [2][]  {\ladder{#1}{a}{\hat{\idQPyes}}{#2}}
\newcommand{\cQPyes} [2][]  {\ladder{#1}{c}{\hat{\idQPyes}}{#2}}

\newcommand{\idQPc}         {\alpha}
\newcommand{\aQPc}   [2][]  {\ladder{#1}{a}{\hat{\idQPc}}{#2}}
\newcommand{\cQPc}   [2][]  {\ladder{#1}{c}{\hat{\idQPc}}{#2}}

\newcommand{\Nbasis}{N_{\rm basis}}
\newcommand{\Nqp}{N_{\rm qp}}
\newcommand{\Ncol}{N_{\rm col}}
\newcommand{\NPES}{N_{\rm p}}
\newcommand{\nsamp}{n_{\rm samples}}
\newcommand{\nfeat}{n_{\rm features}}
\newcommand{\Ntrain}{N_{\rm train}}
\newcommand{\Nval}{N_{\rm valid.}}
\newcommand{\Ngrid}{N_{\mathbf{r}}}
\newcommand{\Nz}{N_{z}}
\newcommand{\Nperp}{N_{\perp}}

\newcommand{\distGen}{\ensuremath{d}}
\newcommand{\distMSE}{\ensuremath{d_{\rm MSE}}}
\newcommand{\distInduced}{\ensuremath{d_{\rm I}}}
\newcommand{\distUnit}{\ensuremath{d_{\circ}}}
\newcommand{\distOrtho}{\ensuremath{d_\perp}}
\newcommand{\distProjUnit}{\ensuremath{D}}
\newcommand{\distProjOrtho}{\ensuremath{D_{\perp}}}

\newcommand{\dof}{d.o.f.}
\newcommand{\singp}{s.p.}
\newcommand{\quasp}{q.p.}
\newcommand{\Crr}[1]{C_{#1}^{\rho\rho}}
\newcommand{\Crt}[1]{C_{#1}^{\rho\tau}}
\newcommand{\CrDr}[1]{C_{#1}^{\rho\Delta\rho}}
\newcommand{\CrDJ}[1]{C_{#1}^{\rho\nabla J}}
\newcommand{\CJJ}[1]{C_{#1}^{JJ}}

\newcommand{\VZeroq}{V_{0}^{(\tau)}}

\newcommand{\ncanNv}{87}
\newcommand{\ncanPv}{60}
\newcommand{\ncanTv}{147}
\newcommand{\ncanN}{n_n}
\newcommand{\ncanP}{n_p}
\newcommand{\ncanT}{n_t}

\newcommand{\AEid}          {\Xi}
\newcommand{\AEtensor}  [1] {T^{({#1})}}
\newcommand{\AEcode}    [1] {\fvec{v}^{({#1})}}

\newcommand{\beqn}{\begin{eqnarray}}
\newcommand{\eeqn}{\end{eqnarray}}
\newcommand{\be}{\begin{equation}}
\newcommand{\ee}{\end{equation}}
\newcommand{\ba}{\begin{array}}
\newcommand{\ea}{\end{array}}
\newcommand{\tensor}[1]{\mathsf{#1}}
\newcommand{\gras}[1]{\boldsymbol{#1}}

\newcommand{\figref}  [1]{Fig.~\ref{#1}}
\newcommand{\secref}  [1]{Sec.~\ref{#1}}
\newcommand{\ssecref} [1]{Sec.~\ref{#1}}
\newcommand{\sssecref} [1]{Sec.~\ref{#1}}

\begin{document}
\onecolumn
\firstpage{1}

\title[Building surrogate models of nuclear DFT]{Building Surrogate Models of Nuclear Density Functional Theory with Gaussian Processes and Autoencoders} 

\author[\firstAuthorLast ]{\Authors} 
\address{} 
\correspondance{} 

\extraAuth{}

\maketitle

\begin{abstract} 
From the lightest Hydrogen isotopes up to the recently synthesized Oganesson 
($Z=118$), it is estimated that as many as about 3000 atomic nuclei could exist 
in nature. Most of these nuclei are too short-lived to be occurring on Earth, 
but they play an essential role in astrophysical events such as supernova 
explosions or neutron star mergers that are presumed to be at the origin of 
most heavy elements in the Universe. Understanding the structure, reactions, 
and decays of nuclei across the entire chart of nuclides is an enormous 
challenge because of the experimental difficulties in measuring properties of 
interest in such fleeting objects and the theoretical and computational issues 
of simulating strongly-interacting quantum many-body systems. Nuclear density 
functional theory (DFT) is a fully microscopic theoretical framework which has 
the potential of providing such a quantitatively accurate description of 
nuclear properties for every nucleus in the chart of nuclides. Thanks to 
high-performance computing facilities, it has already been successfully applied 
to predict nuclear masses, global patterns of radioactive decay like $\beta$ or 
$\gamma$ decay, and several aspects of the nuclear fission process such as, 
e.g., spontaneous fission half-lives. Yet, predictive simulations of nuclear 
spectroscopy -- the low-lying excited states and transitions between them -- or 
of nuclear fission, or the quantification of theoretical uncertainties and 
their propagation to basic or applied nuclear science applications, would 
require several orders of magnitude more calculations than currently possible. 
However, most of this computational effort would be spent into generating a 
suitable basis of DFT wavefunctions. Such a task could potentially be 
considerably accelerated by borrowing tools from the field of machine learning 
and artificial intelligence. In this paper, we review different approaches to 
applying supervised and unsupervised learning techniques to nuclear DFT.

\tiny
\keyFont{\section{Keywords:} 
Nuclear density functional theory, 
Gaussian process, 
Deep learning, 
Autoencoders, 
RESNET} 
\end{abstract}

\section{Introduction}
\label{sec:intro}

Predicting all the properties of every atomic nucleus in the nuclear chart, 
from Hydrogen all the way to superheavy elements, remains a formidable 
challenge. Density functional theory (DFT) offers a compelling framework to 
do so, since the computational cost is, in principle, nearly independent of the 
mass of the system \cite{eschrig1996fundamentals}. Because of our incomplete 
knowledge of nuclear forces and of the fact that the nucleus is a self-bound 
system, the implementation of DFT in nuclei is slightly different from other 
systems such as atoms or molecules and is often referred to as the energy 
density functional (EDF) formalism \cite{schunck2019energy}. 

Simple single-reference energy density functional (SR-EDF) calculations of 
atomic nuclei can often be done on a laptop. However, large-scale SR-EDF 
computations of nuclear properties or higher-fidelity simulations based on the 
multi-reference (MR-EDF) framework can quickly become very expensive 
computationally. Examples where such computational load is needed range from 
microscopic fission theory \cite{schunck2022theory,schunck2016microscopic} to 
parameter calibration and uncertainty propagation \cite{kejzlar2020statistical,
schunck2020calibration} to calculations at the scale of the entire chart of 
nuclides \cite{erler2012limits,ney2020global} relevant, e.g., for astrophysical 
simulations \cite{mumpower2016impact}. Many of these applications would benefit 
from a reliable emulator of EDF models.

It may be useful to distinguish two classes of quantities that such emulators 
should reproduce. What we may call ``integral'' quantities are 
quantum-mechanical observables such as, e.g., the energy, radius, or spin of 
the nucleus, or more complex data such as decay or capture rates. By contrast, 
we call ``differential'' quantities the basic degrees of freedom of the 
theoretical model. In this article, we focus on the Hartree-Fock-Bogoliubov 
(HFB) theory, which is both the cornerstone of the SR-EDF approach and provides 
the most common basis of generator states employed in MR-EDF calculations. In 
the HFB theory, all the degrees of freedom are encapsulated into three 
equivalent quantities: the quasiparticle spinors, as defined either on some 
spatial grid or configuration space; the full non-local density matrix 
$\rho(\rvec\sigma\tau,\rvec'\sigma'\tau')$ and pairing tensor 
$\kappa(\rvec\sigma\tau,\rvec'\sigma'\tau')$, where $\rvec$ refers to 
spatial coordinates, $\sigma=\pm 1/2$ to the spin projection and $\tau=\pm 1/2$ 
to the isopin projection \cite{perlinska2004local}; the full non-local HFB 
mean-field and pairing potentials, often denoted by $h(\rvec\sigma\tau,
\rvec'\sigma'\tau')$ and $\Delta(\rvec\sigma\tau,\rvec'\sigma'\tau')$.

Obviously, integral quantities have the clearest physical meaning and can be 
compared to data immediately. For this reason, they have been the focus of most 
of the recent efforts in applying techniques of machine learning and artificial 
intelligence (ML/AI) to low-energy nuclear theory, with applications ranging 
from mass tables \cite{utama2016nuclear,utama2017refining,utama2018validating,
niu2018nuclear,neufcourt2019neutron,lovell2022nuclear,mumpower2022physically}, 
$\beta$-decay rates \cite{niu2019predictions}, or fission product yields 
\cite{wang2019bayesian,lovell2020quantifying}. The main limitation of this 
approach is that it must be repeated for every observable of interest. In 
addition, incorporating correlations between such observables, for example the 
fact that $\beta$-decay rates are strongly dependent on $Q_{\beta}$-values 
which are themselves related to nuclear masses, is not easy. This is partly 
because the behavior of observables such as the total energy or the total spin 
is often driven by underlying shell effects that can lead to very rapid 
variations, e.g. at a single-particle crossing. Such effects could be very hard 
to incorporate accurately in a statistical model of integral quantities.

This problem can in principle be solved by emulating what we called earlier 
differential quantities. For example, single-particle crossings might be 
predicted reliably with a good statistical model for the single-particle 
spinors themselves. In addition, since differential quantities represent, by 
definition, all the degrees of freedom of the SR-EDF theory, any observable of 
interest can be computed from them, and the correlations between these 
observables would be automatically reproduced. In this sense, an emulator of 
differential quantities is truly an emulator for the entire SR-EDF approach. In 
the much simpler case of the Bohr collective Hamiltonian, such a strategy gave 
promising results \cite{lasseri2020taming}.

The goal of this paper is precisely to explore the feasibility of training 
statistical models to learn the degrees of freedom of the HFB theory. We have 
explored two approaches: a simple one based on independent, stationary Gaussian 
processes and a more advanced one relying on deep neural networks with 
autoencoders and convolutional layers. 

In Section \ref{sec:nDFT}, we briefly summarize the nuclear EDF formalism with 
Skyrme functionals with a focus on the HFB theory preserving axial symmetry. 
Section \ref{sec:GP} presents the results obtained with Gaussian processes. 
After recalling some general notions about Gaussian processes, we analyze the 
results of fitting HFB potential across a two-dimensional potential energy 
surface in $^{240}$Pu. Section \ref{sec:AE} is devoted to autoencoders. We 
discuss choices made both for the network architecture and for the training 
data set. We quantify the performance of autoencoders in reproducing canonical 
wavefunctions across a potential energy surface in $^{98}$Zr and analyze the 
structure of the latent space.

\section{Nuclear Density Functional Theory}
\label{sec:nDFT}

In very broad terms, the main assumption of density functional theory (DFT) for 
quantum many-body systems is that the energy of the system of interest can be 
expressed as a functional of the density of particles~\cite{parr1989density,
dreizler1990density,eschrig1996fundamentals}. Atomic nuclei are a somewhat 
special case of DFT, since the nuclear Hamiltonian is not known exactly and the 
nucleus is a self-bound system~\cite{engel2007intrinsicdensity,
barnea2007density}. As a result, the form of the energy density functional 
(EDF) is often driven by underlying models of nuclear forces, and the EDF is 
expressed as a function of non-local, symmetry-breaking, intrinsic 
densities~\cite{schunck2019energy}. In the single-reference EDF (SR-EDF) 
approach, the many-body nuclear state is approximated by a simple product state 
of independent particles or quasiparticles, possibly with some constraints 
reflecting the physics of the problem. We note $\ket{\idHFB{\qvec}}$ such as 
state, with $\qvec$ representing a set of constraints. The multi-reference EDF 
(MR-EDF) approach builds a better approximation of the exact many-body state by 
mixing together SR-EDF states.

\subsection{Energy Functional}
\label{subsec:EDF}

The two most basic densities needed to build accurate nuclear EDFs are the 
one-body density matrix $\rho$ and the pairing tensor $\kappa$ (and its complex 
conjugate $\kappa^{*}$). The total energy of the nucleus is often written as
\be
E[\rho,\kappa,\kappa^{*}] 
= 
E_{\rm nuc}[\rho] 
+ 
E_{\rm Cou}[\rho] 
+ 
E_{\rm pair}[\rho,\kappa,\kappa^{*}] \,,
\label{eq:etot}
\ee
where $E_{\rm nuc}[\rho]$ represents the particle-hole, or mean-field, 
contribution to the total energy from nuclear forces, $E_{\rm Cou}[\rho]$ the 
same contribution from the Coulomb force, and 
$E_{\rm pair}[\rho,\kappa,\kappa^{*}]$ the particle-particle contribution to 
the energy\footnote{The pairing contribution lumps together terms coming from 
nuclear forces, Coulomb forces and possibly rearrangement terms.}. In this 
work, we model the nuclear part of the EDF with a Skyrme-like term 
\be
E_{\rm nuc}[\rho] = \sum_{t=0,1} \int \diff{\rvec}\; \chi_t(\rvec) \,,
\ee
which includes the kinetic energy term and reads generically
\be
\chi_t(\rvec) 
= 
\Crr{t} \rho_t^2 
+ 
\Crt{t} \rho_t\tau_t 
+ 
\CJJ{t} \tensor{J}^{2}_t 
+ 
\CrDr{t}\rho_t\Delta\rho_t 
+ 
\CrDJ{t}\rho_t\gras{\nabla}\cdot\gras{J}_t \,.
\label{eq:edf}
\ee
In this expression, the index $t$ refers to the isoscalar ($t=0$) or isovector 
$(t=1$) channel and the terms $C_{t}^{uu'}$ are the coupling constants 
associated with the energy functional. The particle density $\rho_t(\rvec)$, 
kinetic energy density $\tau_t(\rvec)$, spin-current tensor 
$\tensor{J}_t(\rvec)$, and vector density $\gras{J}_t(\rvec)$ are all derived 
from the full one-body, non-local density 
$\rho(\rvec\sigma\tau,\rvec'\sigma'\tau')$ where $\rvec$ are spatial 
coordinates, $\sigma$ is the intrinsic spin projection, $\sigma = \pm 1/2$, and 
$\tau = \pm 1/2$ is the isospin projection; see \cite{engel1975timedependent,
dobaczewski1996timeodd,bender2003selfconsistent,perlinska2004local,
lesinski2007tensor} for their actual definition. Since we do not consider any 
proton-neutron mixing, all densities are diagonal in isospin space. The two 
remaining terms in \eqref{eq:etot} are treated in exactly the same way as in 
\cite{schunck2020bayesian}. In particular, the pairing energy is derived from a 
surface-volume density-dependent pairing force 
\be
V^{(\tau)}(\rvec,\rvec') = 
\VZeroq
\left[ 1 - \frac{1}{2}\frac{\rho(\rvec)}{\rho_{c}} \right ]
\delta(\rvec-\rvec') \,,
\label{eq:pairing}
\ee
where $\rho_{c} = 0.16 $ fm$^{-3}$ is the saturation density of nuclear matter.

\subsection{Hartree-Fock-Bogoliubov Theory}
\label{subsec:HFB}

The actual densities in~\eqref{eq:edf} are obtained by solving the 
Hartree-Fock-Bogoliubov (HFB) equation, which derives from applying a 
variational principle and imposing that the energy be minimal under variations 
of the densities \cite{schunck2019energy}. The HFB equation is most commonly 
solved in the form of a non-linear eigenvalue problem. The eigenfunctions 
define the quasiparticle (\quasp{}) spinors. Without proton-neutron mixing, we 
can treat neutrons and protons separately. Therefore, for any one type of 
particles, the HFB equation giving the $\mu$$^{\rm th}$ eigenstate reads in 
coordinate space \cite{dobaczewski1984hartreefockbogolyubov}
\begin{equation}
\addtolength{\arraycolsep}{-4pt}
\int \diff\rvec'
\sum_{\sigma'}\left(
\begin{array}{cc}
h(\rvec\sigma,\rvec'\sigma') - \lambda\delta_{\sigma\sigma'} & \tilde{h}(\rvec\sigma,\rvec'\sigma') 
\medskip \\
\tilde{h}^{*}(\rvec\sigma,\rvec'\sigma')  & -h(\rvec\sigma,\rvec'\sigma') + \lambda\delta_{\sigma\sigma'}
\end{array}
\right)
\left( \begin{array}{c}
U(E_{\mu},\rvec'\sigma') \medskip\\
V(E_{\mu},\rvec'\sigma')
\end{array}
\right) 
=
E_{\mu}
\left( \begin{array}{c}
U(E_{\mu},\rvec\sigma) \medskip\\
V(E_{\mu},\rvec\sigma)
\end{array}
\right) ,
\label{eq:HFB_eq}
\end{equation}
where $h(\rvec\sigma,\rvec'\sigma')$ is the mean field, 
$\tilde{h}(\rvec\sigma,\rvec'\sigma')$ the pairing field\footnote{
Following \cite{dobaczewski1984hartreefockbogolyubov,dobaczewski1996meanfield}, 
we employ the `russian' convention where the pairing field is defined from the 
pairing density $\tilde{\rho}(\rvec\sigma,\rvec'\sigma')$ rather than the 
pairing tensor. The quantity $\tilde{h}$ is related to the more traditional 
form of the pairing field $\Delta$ through:
$\tilde{h}(\rvec\sigma,\rvec'\sigma') =
-2\sigma'\Delta(\rvec\sigma,\rvec'-\sigma')$.} and $\lambda$ the Fermi energy. 
Such an eigenvalue problem must be solved for protons and for neutrons. 

For the case of Skyrme energy functionals and zero-range pairing functionals, 
both the mean field $h$ and pairing field $\tilde{h}$ become semi-local 
functions of $\rvec$ (semi-local refers to the fact that these potentials 
involve differential operators). We refer to \cite{vautherin1972hartreefock,
engel1975timedependent} for an outline of the derivations leading to the 
expressions of the mean field in the case of Skyrme functionals and to, e.g., 
\cite{dobaczewski1997solution,bender2009tensor,hellemans2012tensor,
ryssens2015solution} for the expression of the mean field in terms of coupling 
constants rather than the parameters of the Skyrme potential. In the following, 
we simply recall the essential formulas needed in the rest of the manuscript. 

Expression \ref{eq:HFB_eq} is written in coordinate space. In configuration 
space, i.e., when the {\quasp} spinors are expanded on a suitable basis of the 
single-particle ({\singp}) Hilbert space, the same equation becomes a
non-linear eigenvalue problem that can be written as
\be
\left(\begin{array}{cc}
h - \lambda & \tilde{h} \\
\tilde{h}^{*}  & -h^{*} + \lambda 
\end{array}\right)
\left(\begin{array}{cc}
U & V^{*} \\
V & U^{*}
\end{array}\right)
= 
\left(\begin{array}{cc}
U & V^{*} \\
V & U^{*}
\end{array}\right)
\left(\begin{array}{cc}
-E & 0 \\
 0 & E
\end{array}\right) ,
\ee
where $h$, $\tilde{h}$, $U$ and $V$ are now $\Nbasis\times\Nbasis$ matrices, 
with $\Nbasis$ the number of basis states. Eigenvalues are collected in the 
diagonal $\Nbasis\times\Nbasis$ matrix $E$. The set of all eigenvectors define 
the Bogoliubov matrix, 
\be
\mathcal{W} = \left(\begin{array}{cc}
U & V^{*} \\
V & U^{*}
\end{array}\right) ,
\label{eq:bogoW}
\ee
which is a unitary: $\mathcal{W}\mathcal{W}^{\dagger} = 
\mathcal{W}^{\dagger}\mathcal{W} = 1$. Details about the HFB theory can be 
found in the standard references \cite{valatin1961generalized,
mang1975selfconsistent,blaizot1985quantum,ring2004nuclear}.

\subsection{Mean-field and Pairing Potentials}
\label{subsec:pots}

The mean fields are obtained by functional differentiation of the 
scalar-isoscalar energy functional \eqref{eq:etot} with respect to all relevant 
isoscalar or isovector densities, $\rho_{0}$, $\rho_{1}$, $\tau_{0}$, etc. For 
the case of a standard Skyrme EDF when time-reversal symmetry is conserved, the 
corresponding mean-field potentials in the isoscalar-isovector representation
become semi-local \cite{dobaczewski1995timeodd,dobaczewski1997solution,
stoitsov2005axially,hellemans2012tensor}
\be
h_{t}(\rvec)
=
- \gras{\nabla} M_{t}^{*}(\rvec)\gras{\nabla}
+
U_{t}(\rvec)
+
\frac{1}{2i} \sum_{\mu\nu} \big( 
\nabla_{\mu}\sigma_{\nu} B_{t,\mu\nu}(\rvec) 
+ 
B_{t,\mu\nu}(\rvec)\nabla_{\mu}\sigma_{\nu} \big) ,
\label{eq:mean_field}
\ee
where, as before, $t=0,1$ refers to the isoscalar or isovector channel and the 
various contributions are
\begin{subequations}
\begin{align}
M_{t}(\rvec) & = \frac{\hbar^{2}}{2m} +  C_{t}^{\rho\tau}\rho_{t} ,
\label{eq:Meff}\\
U_{t}(\rvec) & =
2C_{t}^{\rho\rho}\rho_{t} 
+ 
C_{t}^{\rho\tau}\tau_{t} 
+ 
2C_{t}^{\rho\Delta\rho}\Delta\rho_{t} 
+ 
C_{t}^{\rho\nabla J}\gras{\nabla}\cdot\gras{J}_{t}
+
U_{t}^{(\rm rear)} ,
\label{eq:central}\\
B_{t,\mu\nu}(\rvec) & = 2C_{t}^{\rho J} J_{t,\mu\nu} - C_{t}^{\rho\Delta J} \nabla_{\mu}\rho_{t,\nu} .
\label{eq:SO}
\end{align}
\end{subequations}
In these expressions, $\mu, \nu$ label spatial coordinates and $\gras{\sigma}$ 
is the vector of Pauli matrices in the chosen coordinate system. For example, 
in Cartesian coordinates, $\mu, \nu \equiv x,y,z$ and $\gras{\sigma} = 
(\sigma_x, \sigma_y, \sigma_z)$. The term $U_{t}^{(\rm rear)}$ is the 
rearrangement potential originating from the density-dependent part of the 
energy. The resulting isoscalar and isovector mean-field and pairing potentials 
can then recombined to give the neutron and proton potentials,
\be
h^{(n)} = h_{0} + h_{1}, \qquad h^{(p)} = h_{0} - h_{1} .
\ee
Note that the full proton potential should also contain the contribution from 
the Coulomb potential.

The pairing field is obtained by functional differentiation of the same energy 
functional \eqref{eq:etot}, this time with respect to the pairing density. As a 
result, one can show that it is simply given by
\be
\tilde{h}^{(\tau)}(\rvec) = 
\VZeroq \left[ 1 - \frac{1}{2}\frac{\rho_{0}(\rvec)}{\rho_{c}} \right ]
\tilde{\rho}^{(\tau)}(\rvec) .
\label{eq:pairing_field}
\ee

\subsection{Collective Space}
\label{subsec:collective}

Nuclear fission or nuclear shape coexistence are two prominent examples of 
large-amplitude collective motion of nuclei \cite{schunck2022theory,
heyde2011shape}. Such phenomena can be accurately described within nuclear DFT 
by introducing a small-dimensional collective manifold, e.g., associated with 
the nuclear shape, where we assume the nuclear dynamics is confined 
\cite{nakatsukasa2016timedependent,schunck2019energy}. The generator coordinate 
method (GCM) and its time-dependent extension (TDGCM) provide quantum-mechanical 
equations of motion for such collective dynamics \cite{griffin1957collective,
wawong1975generatorcoordinate,reinhard1987generator,bender2003selfconsistent,
verriere2020timedependent}. In the GCM, the HFB solutions are generator states, 
i.e., they serve as a basis in which the nuclear many-body state is expanded. 
The choice of the collective manifold, that is, of the collective variables, 
depends on the problem at hand. For shape coexistence or fission, these 
variables typically correspond to the expectation value of multipole moment 
operators on the HFB state. A pre-calculated set of HFB states with different 
values for the collective variables defines a potential energy surface (PES).

In practice, PES are obtained by adding constraints to the solutions of the HFB 
equation. This is achieved by introducing a set of constraining operators 
$\hat{Q}_{a}$ capturing the physics of the problem at hand. The set of all such 
constraints $\qvec \equiv (q_1,\dots,q_N)$ defines a point in the PES. In this 
work, our goal is to design emulators capable of reproducing the HFB solutions 
at any given point $\qvec$ of a PES. Throughout this article, we consider 
exclusively two-dimensional collective spaces made spanned by the expectation 
values of the axial quadrupole $\hat{Q}_{20}$ and axial octupole $\hat{Q}_{30}$ 
moment operators. In the presence of constraints, the mean-field potential in 
the HFB equation is modified as follows
\be
h(\rvec\sigma,\rvec'\sigma') - \lambda\delta_{\sigma\sigma'}
\quad
\rightarrow
\quad
h(\rvec\sigma,\rvec'\sigma') - \Big( \lambda + \sum_{a}\lambda_{a}Q_{a}(\rvec)\Big)\delta_{\sigma\sigma'} .\
\ee
As well known, the Fermi energies play in fact the role of the Lagrange 
parameters $\lambda_a$ for the constraints on particle number. When performing 
calculations with constraints on the octupole moment, it is also important to 
fix the position of the center of mass. This is typically done by adding a 
constraint on the dipole moment $\hat{Q}_{10}$. In the following, we note 
$q_{\lambda\mu}$ the expectation value of the operator $\hat{Q}_{\lambda\mu}$ 
on the quasiparticle vacuum, $q_{\lambda\mu} = 
\langle{\idHFB{\qvec}} | \hat{Q}_{\lambda\mu} | \idHFB{\qvec}\rangle$.

Potential energy surfaces are a very important ingredient in a very popular 
approximation to the GCM called the Gaussian overlap approximation (GOA) 
\cite{brink1968generator,onishi1975local,une1976collective}. By assuming, among 
other things, that the overlap between two HFB states with different 
collective variables $\qvec$ and $\qvec'$ is approximately Gaussian, the GOA 
allows turning the integro-differential Hill-Wheeler-Griffin equation of the 
GCM into a much more tractable Schr\"odinger-like equation. The time-dependent 
version of this equation reads as \cite{verriere2020timedependent}
\be
i \hbar\frac{\partial}{\partial t} g(\qvec,t) 
= 
\left[ 
- \frac{\hbar^2}{2} \sum_{\alpha\beta} \frac{\partial }{\partial q_{\alpha}} B_{\alpha\beta}(\qvec) \frac{\partial}{\partial q_{\beta}} 
+
V(\qvec)
\right] g(\qvec,t),
\label{eq:GCM_GOA}
\ee
where $g(\qvec,t)$ is the probability to be at point $\qvec$ of the collective 
space at time $t$, $V(\qvec)$ is the actual PES, typically the HFB energy as a 
function of the collective variables $\qvec$ (sometimes supplemented by some 
zero-point energy correction) and $B_{\alpha\beta}(\qvec)$ the collective 
inertia tensor. In \eqref{eq:GCM_GOA}, indices $\alpha$ and $\beta$ run 
from 1 to the number $\Ncol$ of collective variables. While the HFB energy 
often varies smoothly with respect to the collective variables, the collective 
inertia tensor can exhibit very rapid variations near level crossings.

\subsection{Canonical Basis}
\label{subsec:canonical}

The Bloch-Messiah-Zumino theorem states that the Bogoliubov matrix 
$\mathcal{W}$ of \eqref{eq:bogoW} can be decomposed into a product of three 
matrices \cite{ring2004nuclear,bloch1962canonical,zumino1962normal}
\be
\mathcal{W} 
= \mathcal{D}\bar{\mathcal{W}}\mathcal{C}
= 
\left(\begin{array}{cc}
D & 0 \\
0 & D^{*}
\end{array}\right)
\left(\begin{array}{cc}
\bar{U} & \bar{V} \\
\bar{V} & \bar{U}
\end{array}\right)
\left(\begin{array}{cc}
C & 0 \\
0 & C^{*}
\end{array}\right)\, ,
\label{eq:W}
\ee
where $D$ and $C$ are unitary matrices. The matrices $\bar{U}$ and $\bar{V}$ 
take the very simple canonical form
\be
\bar{U} = \left(\begin{array}{cccccc}
0 & & & & \\
& \ddots & & & \\
& & u_{k} & 0 & & \\  
& & 0 & u_{\bar{k}} & & \\  
& & & \ddots & \\
& & & & 0
\end{array}\right), 
\qquad
\bar{V} = \left(\begin{array}{cccccc}
0 & & & & \\
& \ddots & & & \\
& & 0 & v_{k} & & \\  
& & v_{\bar{k}} & 0 & & \\  
& & & \ddots & \\
& & & & 0
\end{array}\right)\, . 
\ee
Starting from an arbitrary {\singp} basis $(\aP{},\cP{})$ of the Hilbert space, 
the transformation characterized by the matrix $\mathcal{D}$ leads to a new 
basis $(\aPc{},\cPc{})$ that diagonalizes the density matrix $\rho$ and puts 
the pairing tensor $\kappa$ into the canonical form similar to that of 
$\bar{V}$. This new basis is called the canonical basis of the HFB theory. 
Properties of the canonical basis are discussed in details in the literature; 
see, e.g., \cite{ring2004nuclear,schunck2019energy}. In the HFB theory, 
quasiparticles are superpositions of particle operators $\cPc{}$ and hole 
operators $\aPc{}$. Thus, the canonical basis is transformed according to the 
matrix $\bar{\mathcal{W}}$ to obtain a set of quasiparticle operators 
$(\aQPc{},\cQPc{})$. There is another transformation of these operators 
associated with the matrix $\mathcal{C}$. However, the most important property 
for the purpose of this paper is that physical observables associated with HFB 
solutions do not depend on that last transformation.

In addition to simplifying the calculation of many-body observables, the 
canonical basis is also computationally less expensive than the full Bogoliubov 
basis\footnote{This statement is obviously not true when solving the HFB 
equation directly in coordinate space. In the case of the local density 
discussed here, the expression $\rho(\rvec)=\sum_{\sigma}\sum_{\mu} 
V_{\mu}(\rvec,\sigma) V^{*}_{\mu}(\rvec,\sigma)$ is just as computationally 
expensive as the canonical basis expression $\rho(\rvec) = \sum_{\sigma} 
\sum_{\mu} v_{\mu}^{2} |\idcan_{\mu}(\rvec,\sigma)|^2$.}. As an illustration, 
let us take the example of the local density $\rho(\rvec)$. Assuming the 
{\singp} basis $(\aP{},\cP{})$ is represented by the basis functions 
$\{ \idHO_{n}(\rvec,\sigma) \}_{n\in\mathbb{N}}$, the local density (for 
isospin $\tau$) is obtained from the matrix of the Bogoliubov transformation by
\be
\rho(\rvec) = 
\sum_{\sigma} \sum_{\mu} \sum_{mn} V_{m\mu}^{*} V_{n\mu}\,
\idHO_{m}(\rvec,\sigma) \idHO^{*}_{n}(\rvec,\sigma) .
\ee
Notwithstanding the constraints imposed by the orthonormality of the {\quasp} 
spinors, the number of independent parameters in this expression approximately 
scales like $2\times\Nbasis^2 \times\Nqp\times\Ngrid$, where $\Nbasis$ is the 
size of the {\singp} basis, $\Nqp$ the number of {\quasp} states $\mu$ and 
$\Ngrid$ the total number of points in the spatial grid $\rvec$ (which depends 
on the symmetries imposed). In the canonical basis, and assuming that the state 
$\cPc{\mu}\ket{\idPV}$ is associated with the wavefunction 
$\idcan_{\mu}(\rvec,\sigma)$, the same object is represented by 
\be
\rho(\rvec) = 
\sum_{\sigma} \sum_{\mu} v_{\mu}^{2} 
|\idcan_{\mu}(\rvec,\sigma)|^2 .
\ee
The number of data points now scales like $2\times\Nqp\times\Ngrid + \Nqp$, or 
about $\Nbasis^2$ smaller than before. For calculations with $\Nbasis\approx 
1000$ the compression enabled by the canonical basis is of the order of 
$10^{6}$.

\subsection{Harmonic Oscillator Basis}
\label{subsec:ho}

All calculations in this article were performed with the HFBTHO code 
\cite{marevic2022axiallydeformed}. Recall that HFBTHO works by expanding the 
solutions on the axially-deformed harmonic oscillator basis 
\cite{stoitsov2005axially}. Specifically, the HO basis functions are written
\be
\idHO_{\gras{n}}(\rvec,\sigma) = 
\idHO_{n_{r}}^{\Lambda}(r)
\idHO_{n_{z}}(z) 
\frac{e^{i\Lambda \theta}}{\sqrt{2\pi}}
\chi_{\Sigma}(\sigma),
\label{eq:HOfunctions}
\ee
where $\gras{n} \equiv (n_{r}, n_z, \Lambda, \Omega=\Lambda\pm\Sigma)$ are the quantum numbers 
labeling basis states and
\begin{subequations}
\begin{align}
\idHO_{n_{r}}^{\Lambda}(r) & = N_{n_{r}} \beta_{\perp} \sqrt{2} \eta^{|\Lambda|/2}e^{-\eta/2} L_{n_{r}}^{|\Lambda|}(\eta), \\
\idHO_{n_{z}}(z) & = N_{n_z} \sqrt{\beta_{z}} e^{-\xi^2/2} H_{n_{z}}(\xi),
\end{align}
\end{subequations}
with $\eta = \beta_{\perp}^2 r^2$ and $\xi = \beta_z z$ dimensionless variables, 
$L_{n_{r}}^{|\Lambda|}$ the associated Laguerre polynomials of order $n_{r}$ 
and $H_{n_{z}}$ the Hermite polynomial of order $n_{z}$. The oscillator scaling 
factors $\beta_{\perp}$ and $\beta_{z}$ are the inverse of the oscillator 
lengths, i.e., $\beta_z = 1/b_z$.

All integrations are performed by Gauss quadrature, namely Gauss-Hermite for 
integrations along the $\xi$-axis of the intrinsic reference frame and 
Gauss-Laguerre for integrations along the perpendicular direction characterized 
by the variable $\eta$. In the following, we note $\Nz$ the number of 
Gauss-Hermite nodes and $\Nperp$ the number of Gauss-Laguerre nodes.

\section{Supervised Learning with Gaussian Processes}
\label{sec:GP}

Gaussian processes (GPs) are a simple yet versatile tool for regression that 
has found many applications in low-energy nuclear theory over the past few 
years, from determining the nuclear equation of state \cite{drischler2020how} 
to nuclear cross sections calculations \cite{kravvaris2020quantifying,
acharya2022gaussian} to modeling of neutron stars \cite{pastore2017new}. In the 
context of nuclear DFT, they were applied to build emulators of $\chi_2$ 
objective functions in the UNEDF project \cite{kortelainen2010nuclear,
kortelainen2012nuclear,kortelainen2014nuclear,higdon2015bayesian,
mcdonnell2015uncertainty,schunck2020calibration}, of nuclear mass models 
\cite{neufcourt2018bayesian,neufcourt2019neutron,neufcourt2020proton,
neufcourt2020quantified} or of potential energy surfaces in actinides 
\cite{schunck2020bayesian}. In this section, we test the ability of GPs to 
learn directly the HFB potentials across a large, two-dimensional collective 
space.

\subsection{Gaussian Processes}
\label{sec:GP:gp}

Gaussian processes are commonly thought of as the generalization of 
normally-distributed random variables (Gaussian distribution) to functions. 
There exists a considerable field of applications for GPs and we refer to the 
reference textbook by Rasmussen \& Williams for a comprehensive review of the 
formalism and applications of GPs \cite{rasmussen2006gaussian}. For the purpose 
of this work, we are only interested in the ability of GPs to be used as a 
regression analysis tool and we very briefly outline below some of the basic 
assumptions and formulas.

We assume that we have a dataset of observations $\{ \mathbf{y} = 
y_{i} \}_{i=1,\dots,n}$ and that these data represent $n$ realizations of 
\be
y = f(\xvec) + \epsilon, 
\ee
where $f: \xvec\mapsto f(\xvec)$ is the unknown function we are seeking to 
learn from the data. Saying that a function $f$ is a Gaussian process means 
that every finite collection of function values $\mathbf{f} = (f(\xvec_1),
\dots,f(\xvec_p))$ follows a $p$-dimensional multivariate normal distribution. 
In other words, we assume that the unknown function $f$ follows a normal 
distribution in `function space'. This is denoted by
\be
f(\xvec) \sim \mathcal{GP}\big( m(\xvec),k(\xvec,\xvec') \big),
\ee
where $m: \xvec\mapsto m(\xvec)$ is the mean function and $k: (\xvec,\xvec') 
\mapsto k(\xvec,\xvec')$ the covariance function, which is nothing but the 
generalization to functions of the standard deviation,
\be
k(\xvec,\xvec') 
= 
\mathbb{E}\left[ \big(f(\xvec)-m(\xvec)\big) \big(f(\xvec')-m(\xvec')\big)\right].
\ee
Thanks to the properties of Gaussian functions, the mean and covariance 
functions have analytical expressions as a function of the test data 
$\mathbf{y}$ and covariance $k$; see Eqs.(2.25)-(2.26) in 
\cite{rasmussen2006gaussian}.

The covariance function is the central object in GP regression. It is typically 
parametrized both with a functional form and with a set of free parameters 
called hyperparameters. The hyperparameters are determined from the observed 
data by maximizing the likelihood function. In our tests, the covariance matrix 
is described by a standard Mat\'ern 5/2 kernel, 
\be
k(\xvec,\xvec') = 
\left(1 + \frac{\sqrt{5}}{\ell} || \xvec - \xvec'||
+ 
\frac{5}{3 \ell^2} || \xvec - \xvec'||^2 \right) 
\exp \left( - \frac{ \sqrt{5}}{ \ell } || \xvec - \xvec'||  \right) \,,
\ee
where $\ell$ is the length-scale that characterizes correlations between values 
of the data at different locations. The length-scale is a hyper-parameter that 
is optimized in the training phase of the Gaussian process. In this work, we 
only considered stationary GPs: the correlation between data points $\xvec$ and 
$\xvec'$ only depends on the distance $|| \xvec - \xvec'||$ between these 
points, not on their actual value. This approximation may be too restrictive.

\subsection{Study Case}
\label{sec:GP:results}

\subsubsection{HFB Potentials}
\label{subsec:dof:pots}

Section \ref{subsec:HFB} showed that the HFB mean-field potential involves 
several differential operators. When the HFB matrix is constructed by computing 
expectation values of the HFB potential on basis functions, differentiation is 
carried over to the basis functions and computed analytically -- one of the 
many advantages of working with the HO basis. In practice, this means that the 
elements of the HFB matrix are computed by multiplying spatial kernels with 
different objects representing either the original HO functions or their 
derivatives. This means that we cannot consider a single emulator for the 
entire HFB potential. Instead, we have to build several different ones for each 
of its components: the central potential $U$ (derivative of the EDF with 
respect to $\rho$), the $r$- and $z$-derivatives of the effective mass $M^{*}$ 
(derivative with respect to the kinetic density $\tau$), the $r$- and 
$z$-derivatives of the spin-orbit potential $W$, and the pairing field 
$\tilde{h}$. There are six such functions for neutrons and another six for 
protons. We denote this set of twelve functions as $\{ f_i \}_{i=1,\dots,12}$. 

At any given point $\qvec$ of the {\em collective} space, these functions are 
all local, scalar functions of $\eta$ and $\xi$, $f_{i}(\qvec)\equiv f_{i}: 
(\eta,\xi) \mapsto f_{i}(\eta,\xi; \qvec)$ where $(\eta,\xi)$ are the nodes of 
the Gauss-Laguerre and Gauss-Hermite quadrature grid. We note generically 
$f_{ik}(\qvec)$ the value at point $k$ of the quadrature grid (linearized) of 
the sample at point $\qvec$ of the function $f_{i}$. When fitting Gaussian 
process to reproduce mean-field and pairing potentials, we consider a 
quadrature grid of $\Nz\times\Nperp = 3200$ points. Our goal is thus to build 
3200 different emulators, one for each point $k$ of that grid, for each of the 
12 local functions characterizing the mean-field and pairing potentials. This 
gives a grand total of 38 400 emulators to build. While this number is large, 
it is still easily manageable on standard computers. It is also several orders 
of magnitude smaller than emulating the full set of quasiparticle spinors, as 
we will see in the next section.

In addition, the value of all the Lagrange parameters used to set the 
constraints must also be included in the list of data points. In our case, we 
have $5$ of them: the two Fermi energies $\lambda_{n}$ and $\lambda_{p}$ 
and the three constraints on the value of the dipole, quadrupole and octupole 
moments, $\lambda_1$, $\lambda_2$ and $\lambda_3$, respectively. Finally, we 
also fit the expectation value of the three constraints on $\hat{Q}_{10}$, 
$\hat{Q}_{20}$ and $\hat{Q}_{30}$. We thus have a grand 
total of 38 408 functions of $\qvec$ to emulate.

\begin{figure}[!htb]
\begin{center}
\includegraphics[width=0.7\textwidth]{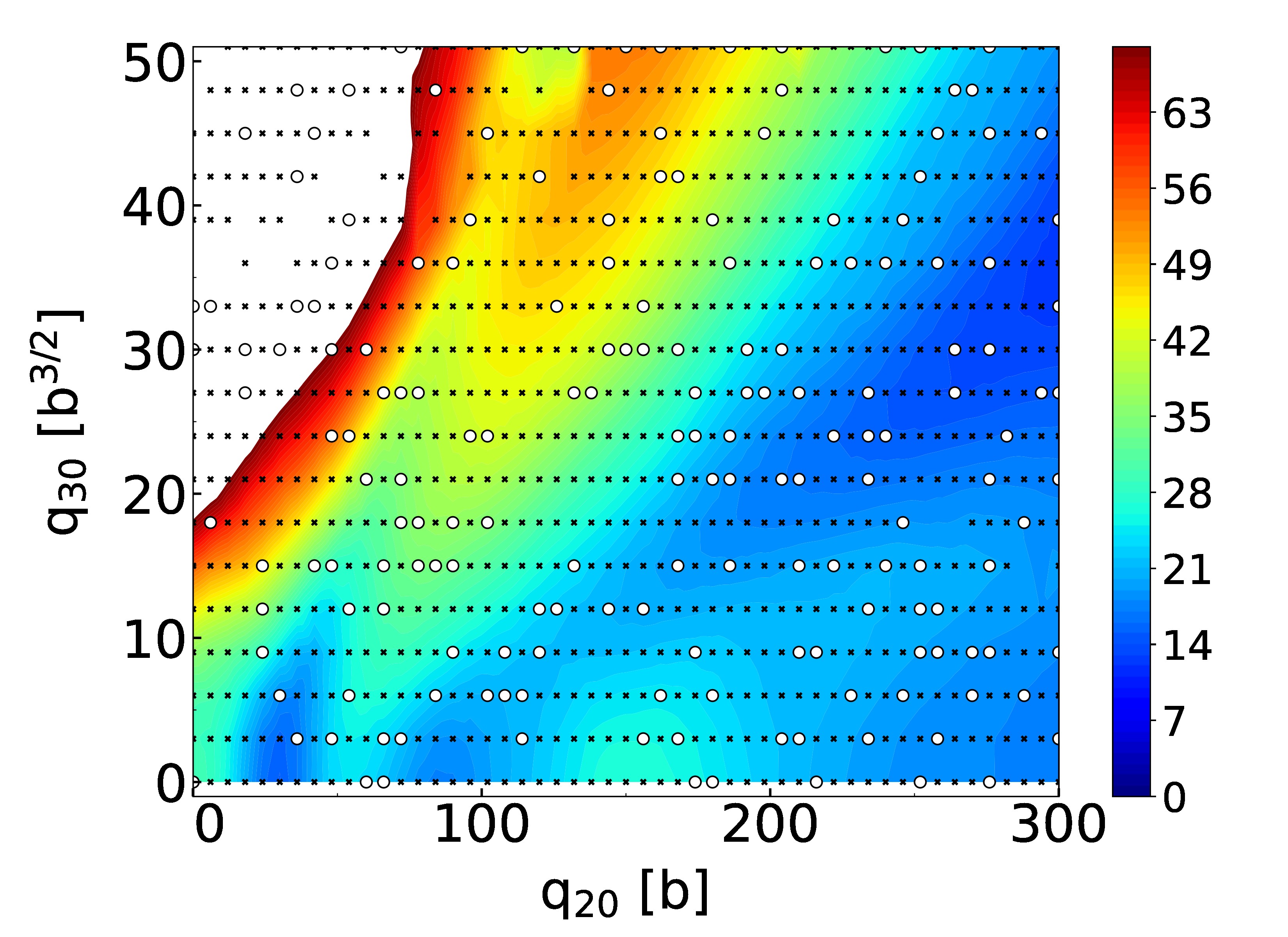}
\caption[]{Potential energy surface of $^{240}$Pu with the SkM* EDF for the 
grid $(q_{20},q_{30}) \in [ 0\, {\rm b}, 300\, {\rm b}] \times 
[0\, {\rm b}^{3/2}, 51\, {\rm b}^{3/2}]$ with steps $\delta q_{20}=6\, {\rm b}$ 
and $\delta q_{30}= 3\, {\rm b}^{3/2}$. The black crosses are the training 
points, the white circles the validation points. Energies indicated by the 
color bar are in MeV relatively to -1820 MeV.}
\label{fig:pes_converged}
\end{center}
\end{figure}

\subsubsection{Training Data and Fitting Procedure}
\label{sec:GP:results:training}

We show in Fig.~\ref{fig:pes_converged} the potential energy surface that 
we are trying to reconstruct. This PES is for the $^{240}$Pu nucleus and was 
generated with the SkM* parameterization of the Skyrme energy functional 
\cite{bartel1982better}. The pairing channel is described with the zero-range, 
density-dependent pairing force of Eq.~\eqref{eq:pairing} that has exactly the 
same characteristics as in \cite{schunck2014description}. 

We imposed constraints on the axial quadrupole and octupole moments such that: 
$0\, {\rm b} \leq q_{20}\leq 300\, {\rm b}$ and $0\, {\rm b}^{3/2} \leq q_{30}
\leq 51\, {\rm b}^{3/2}$ with steps of $\delta q_{20} = 6\, {\rm b}$ and 
$\delta q_{30} = 3\, {\rm b}^{3/2}$, respectively. The full PES should thus 
contain 918 collective points. In practice, we obtained $\NPES = 887$ fully 
converged solutions. Calculations were performed with the HFBTHO solver by 
expanding the solutions on the harmonic oscillator basis with $N_{\rm max}= 28$ 
deformed shells and a truncation in the number of states of $\Nbasis = 1000$. 
At each point of the PES, the frequency $\omega_{0}$ and deformation
$\beta_{2}$ of the HO basis are set according to the empirical formulas given 
in \cite{schunck2014description}. Following standard practice, we divided the 
full $\NPES = 887$ dataset of points into a training (80\% of the points) and 
validation (20\% of the points) set. The selection was done randomly and 
resulted in $\Ntrain = 709$ training points and $\Nval = 178$ validation 
points. The training points are marked as small black crosses in 
Fig.~\ref{fig:pes_converged} while the validation points are marked as larger 
white circles.

Based on the discussion in Section \ref{subsec:dof:pots}, we fit a Gaussian 
process to each of the 38 408 variables needed to characterize completely the 
HFB matrix. Since we work in a two-dimensional collective space, we have two 
features and the training data is represented by a two-dimensional array $X$ of 
dimension $(\nsamp, \nfeat)$ with $\nsamp = \NPES$ and $\nfeat = 2$. The target 
values $Y$ (= the value at point $k$ on the quadrature grid of any of the 
functions $f_i$) are contained in a one-dimensional array of size $\NPES$. 
Prior to the fit, the data is normalized between 0 and 1. The GP is based on a 
standard Mat\'ern kernel with $\nu = 2.5$ and length-scale $\ell$. In practice, 
we use different length-scales for the $q_{20}$ and $q_{30}$ directions so 
that $\ell = \gras{\ell}$ is a vector. We initialized these values at the 
spacing of the grid $\gras{\ell}=(\delta q_{20}, \delta q_{30})$. We added a 
small amount of white noise to the Mat\'ern kernel to account for the global 
noise level of the data.

\subsubsection{Performance}
\label{sec:GP:results:perf}

\begin{figure}[!htb]
\begin{center}
\includegraphics[width=\textwidth]{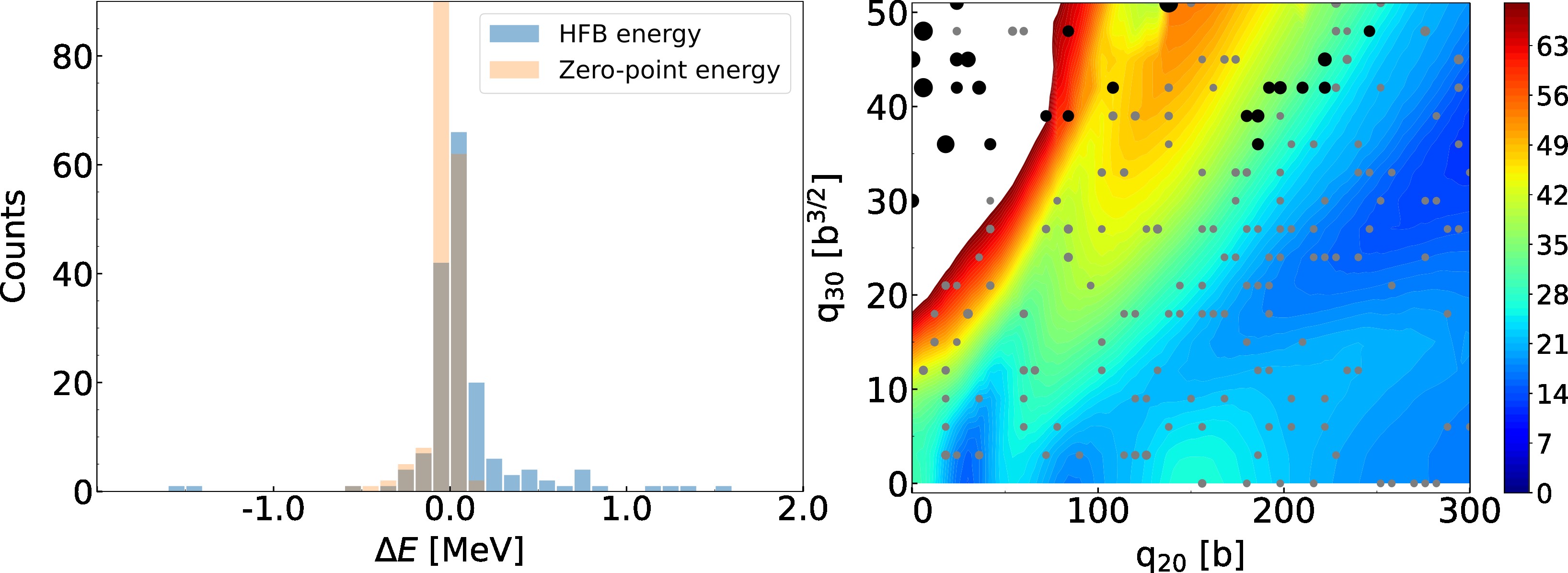}
\caption[]{Left: Histogram of the error on the GP-predicted total HFB energy 
and zero-point energy correction across the validation points. Bin size is 100 
keV. Right: Size of the error on the GP-predicted total HFB energy across the 
validation set. Gray circles have an error lower than 500 keV and the size of 
the markers correspond to energy bins of 100 keV. Black circles have an error 
greater than 500 keV and are binned by 400 keV units. Energies indicated by the 
color bar are in MeV relatively to -1820 MeV.}
\label{fig:error_energies}
\end{center}
\end{figure}

Once the GP has been fitted on the training data, we can estimate its 
performance on the validation data. For each of the $\Nval = 178$ validation 
points, we used the GP-fitted HFB potentials to perform a single iteration of 
the HFB self-consistent loop and extract various observables from this single 
iteration. Figure~\ref{fig:error_energies} focuses on the total HFB energy and 
the zero-point energy correction $\varepsilon_{0}$. Together, these two 
quantities define the collective potential energy in the collective 
Hamiltonian~\eqref{eq:GCM_GOA} of the GCM. The left panel of the figure shows 
the histogram of the error $\Delta E = E_{\rm HFB}^{\rm (true)} - 
E_{\rm HFB}^{\rm (GP)}$, where $E_{\rm HFB}^{\rm (true)}$ is the result from 
the fully converged HFB solution and $E_{\rm HFB}^{\rm (GP)}$ is the value 
predicted by the Gaussian process. The bin size is 100 keV. Overall, we find 
that the large majority of the error is within $\pm 200$ keV. This is a rather 
good result considering the span of the PES and the fact that basis truncation 
errors can easily amount to a few MeV \cite{schunck2013density}.

To gain additional insight, we draw in the right panel of 
Fig.~\ref{fig:error_energies} each of the validation points with a marker, the 
size of which is proportional to the error of the prediction. To further 
distinguish between most points and the few outliers, we show in gray the 
points for which the absolute value of the error is less than 500 keV and in 
black the points for which it is greater than 500 keV. For the gray points, we 
use 5 different marker sizes corresponding to energy bins of 100 keV: the 
smaller grey symbol corresponds to an error smaller than 100 keV, the larger 
one between 400 and 500 keV. Similarly, the larger black circles have all an 
error greater than 500 keV and are ordered by bins of 400 keV (there are only 
two points for which the error is larger than 4 MeV). Interestingly, most of 
the larger errors are concentrated in the region of small elongation 
$q_{20} < 80\, {\rm b}$ and high asymmetry $q_{30} > 30\, {\rm b}^{3/2}$. This 
region of the collective space is very high in energy (more than 100 MeV above 
the ground state) and plays no role in the collective dynamics.

\begin{figure}[!htb]
\begin{center}
\includegraphics[width=0.8\textwidth]{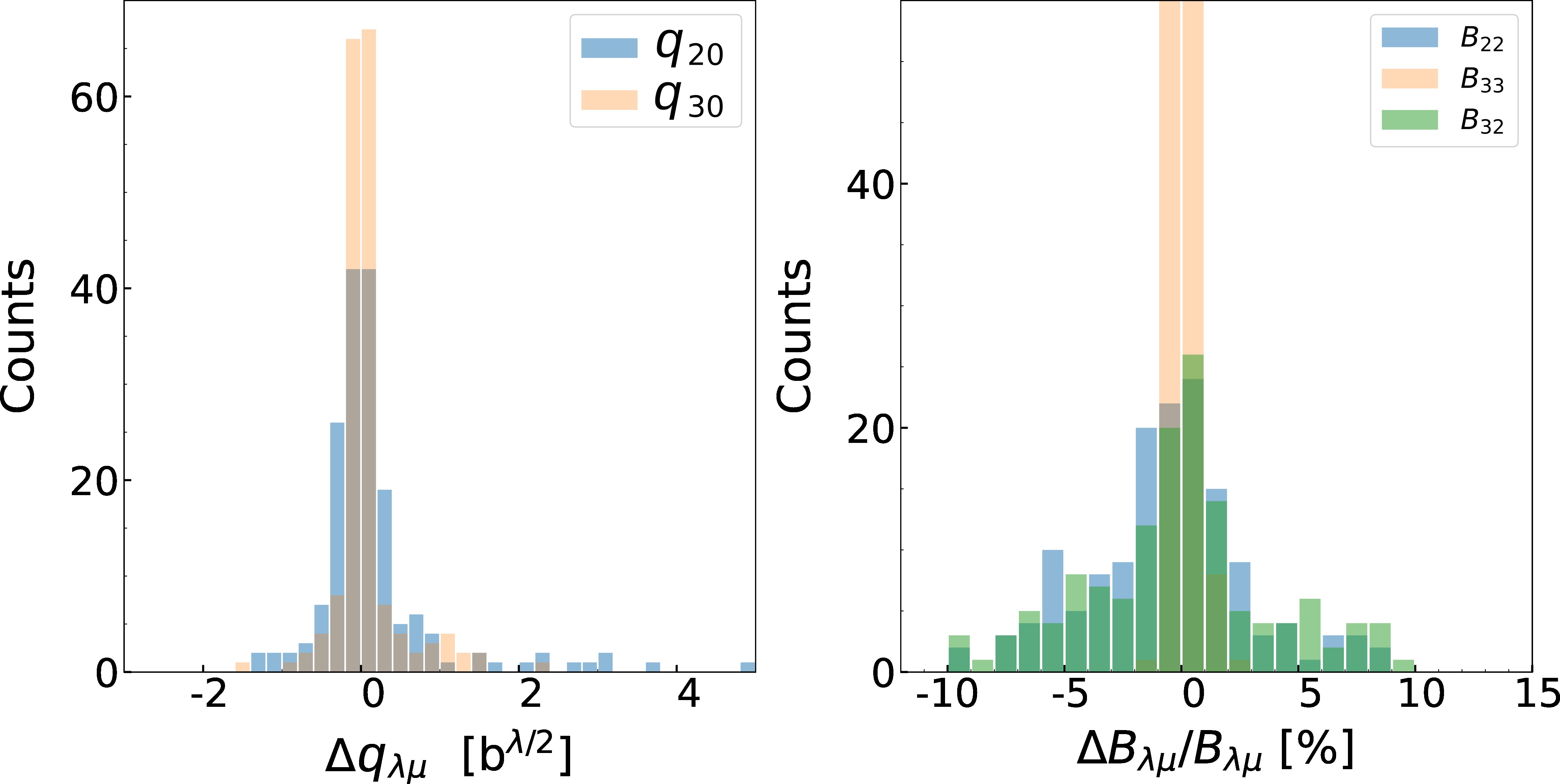}
\caption[]{Left: Histogram of the error on the GP-predicted values of the 
multipole moments. The bin size is 0.2 $b^{\lambda/2}$ with $\lambda=2$ 
(quadrupole moment) or $\lambda=3$ (octupole moment). Right: Histogram of the 
relative error, in percents, on the GP-predicted values of the components of 
the collective inertia tensor. The bin size is 1, corresponding to 1\% relative 
errors.}
\label{fig:histo_others}
\end{center}
\end{figure}

Note that the expectation values of the multipole moments themselves are not 
reproduced exactly by the GP: strictly speaking, the contour plot in the right 
panel of Fig.~\ref{fig:error_energies} is drawn based on the requested values 
of the constraints, not their actual values as obtained by solving the HFB 
equation once with the reconstructed potentials. The histogram in the left 
panel of Fig.~\ref{fig:histo_others} quantifies this discrepancy. It shows the 
absolute error $\Delta q_{\lambda\mu} = q_{\lambda\mu}^{(\rm true)} - 
q_{\lambda\mu}^{(\rm GP)}$, where $q_{\lambda\mu}^{(\rm true)}$ is the result 
from the fully converged HFB solution and $q_{\lambda\mu}^{(\rm true)}$ is the 
value predicted by the Gaussian process. On average, the error remains within 
$\pm 0.5\,\mathrm{b}$ for $q_{20}$ and $\pm 0.5\,\mathrm{b}^{3/2}$ for 
$q_{30}$, which is significantly smaller than the mesh size.

The collective potential energy is only one of the two ingredients used to 
simulate fission dynamics. As mentioned in Sec.~\ref{subsec:collective}, see 
Eq.~\eqref{eq:GCM_GOA}, the collective inertia tensor is another essential 
quantity \cite{schunck2016microscopic,schunck2022theory}. In this work, we 
computed the collective inertia at the perturbative cranking approximation 
\cite{schunck2016microscopic}. Since we work in two-dimensional collective 
spaces, the collective inertia tensor $\tensor{B}$ has three independent 
components, hereafter labeled $B_{22}$, $B_{33}$ and $B_{32} = B_{23}$. Figure 
\ref{fig:histo_others} shows the relative error on these quantities, defined as 
$\epsilon = (B_{\lambda\lambda'}^{(\rm true)} - B_{\lambda\lambda'}^{(\rm GP)}) 
/ B_{\lambda\lambda'}^{(\rm true)}$. Overall, the error is more spread than for 
the energy but rarely exceeds five percents\footnote{Note that $B_{32}$ 
vanishes for axially-symmetric shapes. As a result, the relative error can be 
artificially large for values of $q_{30} \approx 0\,\mathrm{b}^{3/2}$.}.

\begin{figure}[!htb]
\begin{center}
\includegraphics[width=1\textwidth]{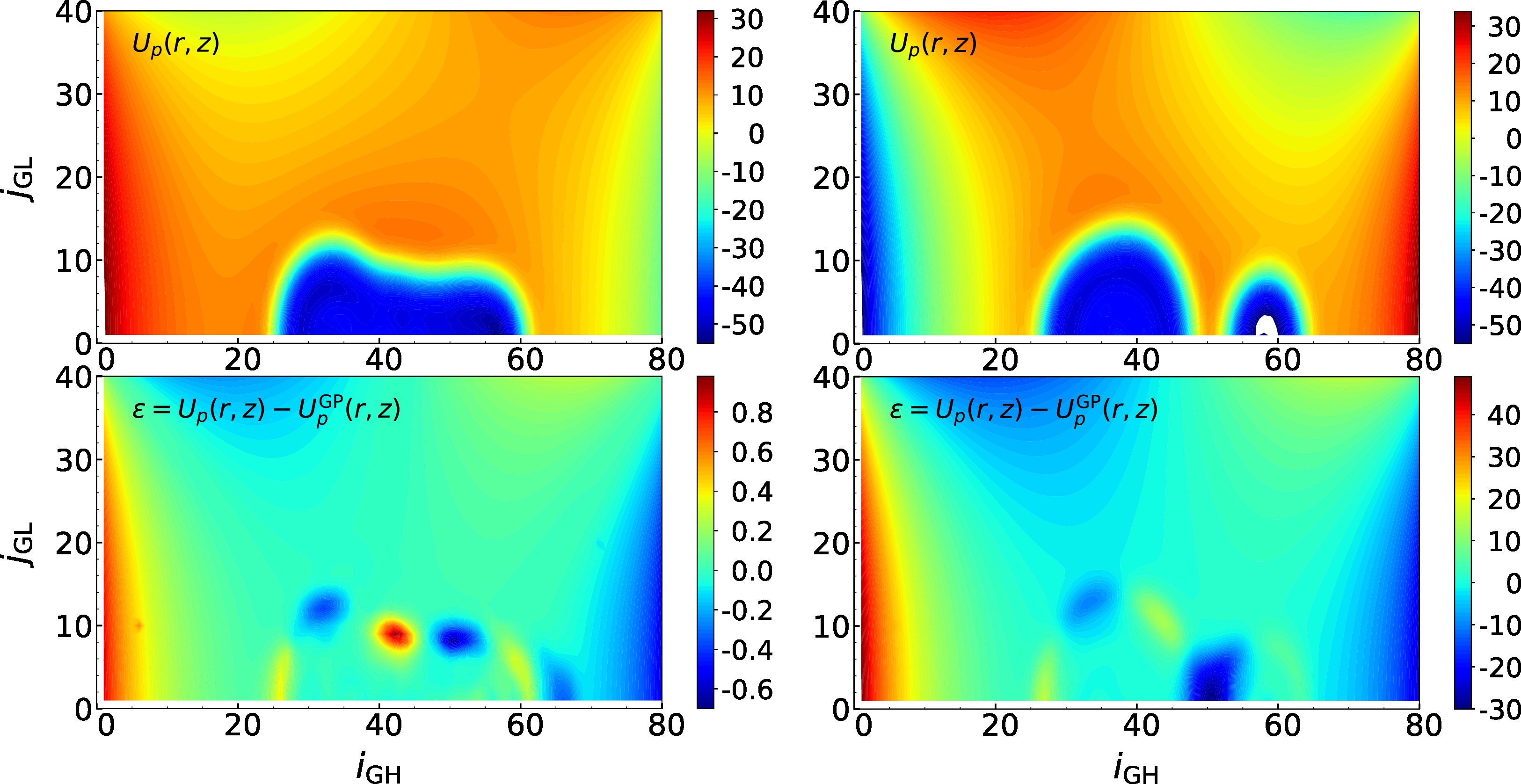}
\caption[]{Upper left: Central part of the mean-field potential for protons, 
$U_p(r,z)$ for the configuration $(q_{20},q_{30}) = (198\, \mathrm{b}, 30\, 
\mathrm{b}^{3/2})$; bottom left: Error in the GP fit for that same 
configuration. Upper right: Central part of the mean-field potential for 
protons, $U_p(r,z)$ for the configuration $(q_{20},q_{30}) = (138\, \mathrm{b}, 
51\, \mathrm{b}^{3/2})$; bottom right: Error in the GP fit for that same 
configuration. For all figures, $i_{\rm GH}$ and $j_{\rm GL}$ refer to the 
index $i$ and $j$ on the Gauss-Hermite and Gauss-Laguerre quadrature grid, and 
the energy given by the error bar is in MeV.}
\label{fig:Up_error}
\end{center}
\end{figure}

Both the total energy and the collective mass tensor are computed from the HFB 
solutions. However, since the GP fit is performed directly on the mean-field 
and pairing potentials, one can analyze the error on these quantities directly. 
In Fig.~\ref{fig:Up_error}, we consider two different configurations. The 
configuration $\mathcal{C}_1 = (q_{20},q_{30}) = (198\, \mathrm{b}, 30\,
\mathrm{b}^{3/2})$ is very well reproduced by the GP with an error in the HFB 
energy of 4.4 keV and a relative error on $B_{22}$ of -0.43 \% and $B_{22}$ of 
-0.84 \% only. In contrast, the configuration $\mathcal{C}_2 = (q_{20},q_{30})
= (138\, \mathrm{b},51\, \mathrm{b}^{3/2})$ is one of the worst possible cases, 
with a total error on the HFB energy of 9.0 MeV and relative errors on $B_{22}$ 
of -71.0 \% and $B_{22}$ of -13.7 \%. For each of these two configurations, we 
look at the central part of the mean-field potential for protons, the term 
$U_{p} = U_{0} - U_{1}$ of \eqref{eq:central}. The left side of 
Fig.~\ref{fig:Up_error} shows, respectively, the actual value of $U_{p}(r,z)$ 
across the quadrature grid (top panel) and the difference between the true 
value and the GP fit (bottom panel) for the configuration $\mathcal{C}_1$. The 
right side of the figure shows the same quantity for the configuration 
$\mathcal{C}_2$. In all four plots, the energy scale is in MeV.

We see that for the `good' configuration $\mathcal{C}_1$, the error is between 
-0.6 MeV and 1.0 MeV but is mostly occurring at the surface of the nucleus and 
at the edges of the domain. Conversely, the `bad' configuration $\mathcal{C}_2$ 
actually corresponds to a scissioned configuration: the mean-field potential 
(upper right panel) shows two different regions corresponding to fully 
separated fragments\footnote{This particular scission configuration corresponds 
to what is called cluster radioactivity \cite{warda2011microscopic,
warda2018cluster,matheson2019cluster}. The heavy fragment is much larger than 
the light one. Here, $\langle A_{\rm H}\rangle = 205.6$, 
$\langle A_{\rm L}\rangle = 34.4$}. Such a geometric configuration is very 
different from the rest of the potential energy surface shown in 
Fig.~\ref{fig:pes_converged}, which contains mostly non-scissioned 
configurations. As a result, the error in the GP fit is very large in the 
region between the two fragments since it predicts this configuration to be 
non-scissioned. Note that in HFBTHO, the representation of the potentials on 
the quadrature points does not contain the exponential factor 
$\exp(-\beta_z \xi^2)\exp(-\beta^2_{\perp}\xi^2)$ which is factored in in the 
quadrature weights. Therefore, the large errors at the edges of the domain, for 
$i_{\rm GH} \approx 0$, $i_{\rm GH} \approx 80$ or $j_{\rm GL} \approx 40$ are 
not significant since they are entirely absorbed by this exponential factor.

Overall, Gaussian processes seem to provide an efficient way to predict HFB 
solutions across potential energy surfaces. Their primary advantage is that 
they are very simple to implement, with several popular programming 
environments offering ready-to-use, full GP packages, and very fast to train (a 
few minutes at most for a few hundreds of samples). As our examples suggest, 
GPs are very good at interpolating across a domain where solutions behave 
smoothly. In the case of PES, this implies that the training data must not 
contain, e.g., scissioned and non-scissioned configurations. More generally, it 
should not feature too many discontinuities \cite{dubray2012numerical}. When 
these conditions are met, GPs can be used to quickly and precisely densify a 
PES, e.g., to obtain more precise fission paths in spontaneous fission 
half-live calculations \cite{sadhukhan2020microscopic}.

However, Gaussian processes are intrinsically limited. In our example, we 
treated the value of each potential at each point of the quadrature mesh as an 
independent GP. Yet, such data are in reality heavily correlated. To 
incorporate such correlations requires generalizing from scalar GPs to vector, 
or multi-output GPs \cite{bruinsma2020scalable}. In our example of nuclear 
potentials, the output space would be $\mathbb{R}^{D}$ with $D \approx 32 008$. 
An additional difficulty is related to choosing the kernel that is appropriate 
to describe the correlated data and identifying what the prior distribution 
should be \cite{alvarez2012kernels}. Yet another deficiency of standard 
Gaussian processes, especially in contrast to the deep-learning techniques 
discussed below, is that they are not capable to learn a latent representation 
of the data. For these reasons, we consider such techniques helpful mostly to 
densify existing potential energy surfaces.

\section{Deep Learning with Autoencoders}
\label{sec:AE}

Even though self-consistent potential energy surfaces are key ingredients in 
the microscopic theory of nuclear fission \cite{bender2020future}, we must 
overcome two significant obstacles to generate reliable and complete PES. 
First, the computational cost of nuclear DFT limits the actual number of 
single-particle {\dof}. When solving the HFB equation with basis-expansion 
methods, for example, the basis must be truncated (up to a maximum of about a 
few thousand states, typically), making the results strongly 
basis-dependent~\cite{schunck2013microscopic}; even in mesh-based methods, the 
size of the box and lattice spacing also induce truncation effects 
\cite{ryssens2015numerical,jin2017coordinatespace}. Most importantly, the 
number of collective variables that can be included in the PES is also limited: 
in spontaneous fission calculations, which do not require a description of the 
PES up to scission, up to $\Ncol = 5$ collective variables have been 
incorporated \cite{sadhukhan2020microscopic}; when simulating the PES up to 
scission, only 2 collective variables are included with only rare attempts to 
go beyond\cite{regnier2017microscopic,zhao2021microscopic}. As a consequence, 
the combination of heavily-truncated collective spaces and the adiabatic 
hypothesis inherent to such approaches leads to missing regions in the PES and 
spurious connections between distinct channels with unknown effects on physics 
predictions \cite{dubray2012numerical,lau2022smoothing}. The field of deep 
learning may offer an appealing solution to this problem by allowing the 
construction of low-dimensional and continuous surrogate representations of 
potential energy surfaces. In the following, we test the ability of 
autoencoders -- a particular class of deep neural networks -- to generate 
accurate low-dimensional representations of HFB solutions.

\subsection{Network Architecture}
\label{sec:AE:archi}

The term `deep learning' encompasses many different types of mathematical and 
computational techniques that are almost always tailored to specific 
applications. In this section, we discuss some of the specific features of the 
data we seek to encode in a low-dimensional representation, which in turn help 
constrain the network architecture. The definition of a proper loss function 
adapted to quantum-mechanical datasets is especially important.


\subsubsection{Canonical States}
\label{subsec:dof:canonicals}

We aim at building a surrogate model for determining canonical wavefunctions as 
a function of a set of continuous constraints. Canonical states are denoted 
generically $\idcan^{(\tau)}_{\mu}(\rvec,\sigma)$ with $\rvec \equiv 
(r,z,\theta)$ the cylindrical coordinates and $\sigma=\pm 1/2$ the spin. Fully 
characterizing an HFB state requires the set of canonical wavefunctions for 
both neutrons and protons, which are distinguished by their isospin quantum 
number $\tau = +1/2$ (neutrons) and $\tau=-1/2$ (protons). As mentioned 
in~\secref{sec:nDFT}, an HFB solution ${\ket{\idHFB{\fvec{q}}}}$ is entirely 
determined up to a global phase by the set of all canonical states 
${\{ \idcan^{(\tau)}_{\mu}(\fvec{r}, \sigma) \}_{\mu}}$ and their associated 
occupation amplitudes ${\{v^{(\tau)}_\mu\}_{\mu}}$.

In this work, we restrict ourselves to axially-symmetric configurations. In 
that case, the canonical wavefunctions are eigenstates of the projection of the 
total angular momentum on the symmetry axis $\jz$ with eigenvalue $\Omega$ and 
acquire the same separable structure \eqref{eq:HOfunctions} as the HO basis 
functions,
\begin{equation}
\idcan^{(\tau)}_{\mu}(\fvec{r}, \sigma)
= 
\idcan^{(\tau)}_{\mu}(r, z, \sigma) \frac{e^{i\Lambda\theta}}{\sqrt{2\pi}},
\end{equation}
where $\idcan^{(\tau)}_{\mu}(r, z, \sigma)$ is the canonical wavefunction at 
$\theta=0$. In this initial work, we only consider even-even nuclear systems 
and time-reversal symmetric nuclear Hamiltonians. Therefore, Kramer's 
degeneracy ensures that paired particles in the canonical basis are 
time-reversal partners of each other: $\idcan^{(\tau)}_{\bar{\mu}}(\fvec{r}, 
\sigma) = 2\sigma \idcan^{(\tau)*}_{\mu}(\fvec{r}, -\sigma)$. This guarantees 
that the canonical wavefunction at $\theta=0$ can be chosen purely real. 
Incidentally, it also means that we only need to describe one wavefunction per 
pair of particles. Using these properties, we can completely describe a 
canonical wavefunction in our model by only predicting a single pair of 
real-valued functions (one for each spin projection $\sigma$). 

As shown by \eqref{eq:mean_field}, \eqref{eq:pairing_field} and 
\eqref{eq:Meff}-\eqref{eq:SO}, all mean-field and pairing potentials are 
functions of the Skyrme densities. The kinetic energy density $\tau(r,z)$, 
spin-current tensor $\tensor{J}(r,z)$, and vector density $\gras{J}(r,z)$ 
involve derivatives of the quasiparticle spinors or, in the canonical basis, of 
the canonical wavefunctions on the quadrature grid \cite{stoitsov2005axially}. 
We compute these derivatives by first extracting the coefficients 
$\alpha^{(\tau)}_{\gras{n}\mu}$ of the expansion of the canonical wavefunctions 
$\idcan^{(\tau)}_{\mu}(\rvec,\sigma)$ in the HO basis
\be
\idcan^{(\tau)}_{\mu}(\rvec,\sigma) 
= 
\sum_{\gras{n}} \alpha^{(\tau)}_{\gras{n}\mu} \idHO_{\gras{n}}(\rvec,\sigma)
\Rightarrow
\alpha^{(\tau)}_{\gras{n}\mu} 
= 
\int \diff\rvec\, \idHO^{*}_{\gras{n}}(\rvec,\sigma)\idcan^{(\tau)}_{\mu}(\rvec,\sigma), 
\label{eq:HOexpansion}
\ee
using Gauss-Laguerre and Gauss-Hermite quadrature. Since all the derivatives of 
the HO functions can be computed analytically, the expansion 
\eqref{eq:HOexpansion} makes it very easy to compute partial derivatives with 
respect to $r$ or $z$, for example,
\be
\frac{\partial\idcan^{(\tau)}_{\mu}}{\partial z}(\rvec,\sigma) 
= 
\sum_{\gras{n}} \alpha^{(\tau)}_{\gras{n}\mu} \frac{\partial\idHO_{\gras{n}}}{\partial z}(\rvec,\sigma) .
\ee

\subsubsection{Structure of the Predicted Quantity}
\label{sec:AE:archi:struct}

In the ideal case, the canonical wavefunctions evolve smoothly with the 
collective variables. The resulting continuity of the many-body state with 
respect to collective variables is a prerequisite for a rigorous description of 
the time evolution of fissioning systems, yet it is rarely satisfied in 
practical calculations. We discuss below the three possible sources of 
discontinuity of the canonical wavefunctions in potential energy surfaces.

First, the canonical wavefunctions are invariant through a global phase. Since
the quantity we want to predict is real, the orbitals can be independently 
multiplied by an arbitrary sign. Even though this type of discontinuity does 
not impact the evolution of global observables as a function of deformation, it 
affects the learning of the model: since we want to obtain continuous 
functions, a flipping of the sign would be seen by the neural network as a 
discontinuity in the input data. We address this point through the choice of 
the loss, as discussed in \sssecref{sec:AE:archi:loss}, and through the 
determination of the training set, as detailed in \ssecref{sec:AE:training}.

\begin{figure}[!htb]
\centering
\includegraphics[width=0.7\linewidth]{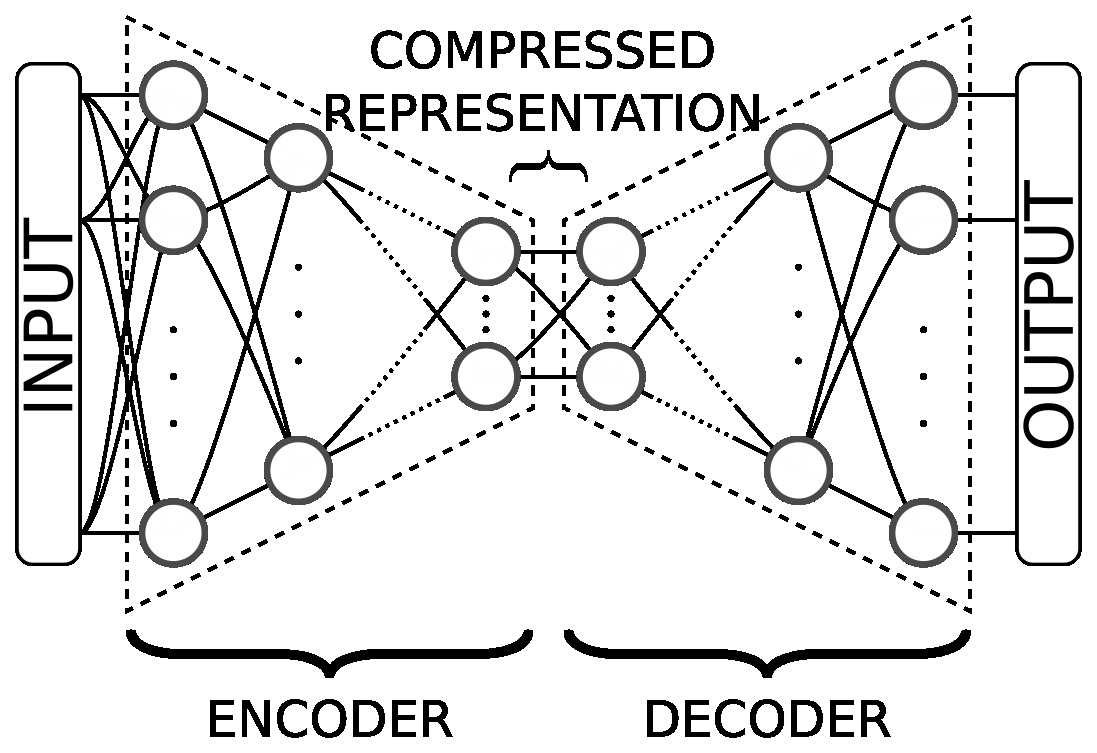}
\caption{An autoencoder is the association of two blocks. The first one, on the 
left, compresses the input data into a lower-dimensional representation, or 
code, in the latent space. The second one, on the right, decompresses the code 
back into the original input.}
\label{fig:AE:archi:struct:AE}
\end{figure}

Second, we work within the adiabatic approximation, which consists in building 
PES by selecting the {\quasp} vacuum that minimizes the energy at each point. 
When the number $\Ncol$ of collective variables of the PES is small, this 
approximation may lead to discontinuities~\cite{dubray2012numerical}. These 
discontinuities correspond to missing regions of the collective space and are 
related to the inadequate choice of collective variables. Since we want to 
obtain a continuous description of the fission path, we must give our neural 
network the ability to choose the relevant degrees of freedom. This could be 
achieved with autoencoders. Autoencoders are a type of neural networks 
analogous to the zip/unzip programs. They are widely used and greatly 
successful for representation learning -- the field of Machine Learning that 
attempts to find a more meaningful representation of complex data 
\cite{baldi2012autoencoders,burda2015importance,chen2012marginalized,
gong2019memorizing,bengio2013representation,zhang2014saliency,yu2017multitask} 
and can be viewed as a non-linear generalization of principal component 
analysis (PCA). As illustrated in \figref{fig:AE:archi:struct:AE}, an 
autoencoder $\AEid$ typically consists of two components. The encoder 
$E(\AEtensor{\idcan})$ encodes complex and/or high-dimensional data 
$\AEtensor{\idcan}$ to a typically lower-dimensional representation 
$\AEcode{\idcan}$. The \textit{latent space} is the set of all possible such 
representations. The decoder $D(\AEcode{\idcan})$ takes the low-dimensional 
representation of the encoder and uncompresses it into a tensor 
$\AEtensor{\idcanR}$ as close as possible to $\AEtensor{\idcan}$. Such 
architectures are trained using a {\em loss function} that quantifies the 
discrepancy between the initial input and the reconstructed output,
\begin{equation}
  \label{eq:AE:archi:struct:loss}
  \mathcal{L}_{\rm rec.}(\AEtensor{\idcanR})
  = d(\AEtensor{\idcan}, \AEtensor{\idcanR}),
\end{equation}
where $\distGen(., .)$ defines the metric in the space of input data. We 
discuss the choice of a proper loss in more details in 
\sssecref{sec:AE:archi:loss}.

Third, the evolution of the \quasp{} wavefunctions as a function of the
collective variables $\qvec$ may lead to specific values $\qvec_{i}$ where the
\quasp{} solutions are degenerate. These degeneracies form a sub-manifold of
dimension at most $D-2$, where $D$ is the number of collective {\dof}s. As a
consequence, they cannot appear in one-dimensional PES: \quasp{} solutions with
the same symmetry ``cannot cross'' (the famous no-crossing rule 
\cite{vonneuman1929uber}). In multi-dimensional spaces, this rule does not hold
anymore: when following a closed-loop trajectory around such a degeneracy, the
sign of the \quasp{} wavefunctions is flipped, in a similar manner that we flip
side when winding around a Moebius strip \cite{teller1937crossing,
longuet-higgins1958studies,longuet-higgins1975intersection}. In the field of
quantum chemistry, such degeneracies are referred to as diabolical points or
conical intersections \cite{domcke2011conical,larson2020intersections}. The
practical consequence of conical intersections for deep learning is that the
manifold of all the \quasp{} wavefunctions cannot be embedded in a
$D$-dimensional latent space. Such singularities can be treated in two ways:
(i) by using a latent space of higher dimension than needed or (ii) by
implementing specific neural network layers capable of handling such cases. For
now, we do not consider these situations.

\subsubsection{Loss Functions and Metrics}
\label{sec:AE:archi:loss}
As already discussed in~\sssecref{sec:AE:archi:struct}, autoencoders are 
trained through the minimization of a loss function that contains a 
reconstruction term of the form~\eqref{eq:AE:archi:struct:loss}. As suggested
by its name, this term ensures that the autoencoder can correctly reconstruct
the input tensor $\AEtensor{\idcan}$ from its compressed representation. It
depends on a definition for the metric $\distGen(.,.)$ used to compare the
different elements of the input space. Since our canonical wavefunctions
$\idcan_{\mu}$ are expanded on the axial harmonic oscillator basis of 
Sec.~\ref{subsec:ho}, they are discretized on the Gauss quadrature mesh without 
any loss of information. Therefore, both the input and output tensors of our 
surrogate model are a rank-3 tensor 
$\AEtensor{\idcanR} \equiv \AEtensor{\idcan} \equiv T_{ijk}$ of dimensions
$\Nperp\times\Nz\times 2$, where $i$ is the index of the Gauss-Laguerre node 
along the $r$-axis, $j$ the index of the Gauss-Hermite node along the
$z$-axis, and $k$ the index of the spin component.

A standard loss used with autoencoders is the mean-square-error (MSE). Because
of the structure of our input data, see \sssecref{subsec:dof:canonicals}, the
MSE loss reads in our case
\begin{equation}
  \label{eq:AE:archi:MSE}
  \distMSE(\AEtensor{\idcan}, \AEtensor{\idcanR}) =
    \frac{1}{\Nperp\times\Nz\times 2}
    \sum_{i=0}^{\Nperp-1}
    \sum_{j=0}^{\Nz-1}
    \sum_{i=0}^{1}
      \big( \AEtensor{\idcanR}_{ijk} -\AEtensor{\idcan}_{ijk} \big)^2.
\end{equation}
The MSE is very general and can be thought of, quite simply, as the mean
squared ``distance'' between the initial and reconstructed data. However, this
generality implies that it does not contain any information about the
properties of the data one tries to reconstruct.

Indeed, we can define a metric that is better suited to the physics we aim to
describe. Let us recall that our goal is to compute potential energy surfaces
that can be used, e.g., for fission simulations. These PES are nothing but
generator states for the (TD)GCM mentioned in \secref{subsec:collective}. The
GCM relies on the norm kernel
$\mathcal{N}(\fvec{q}, \fvec{q}')$ and the Hamiltonian kernel
$\mathcal{H}(\fvec{q}, \fvec{q}')$, which are defined as
\begin{align}
  \mathcal{N}(\fvec{q}, \fvec{q}') &=
    \braket{\idHFB{\fvec{q}} | \idHFB{\fvec{q}'}}, \\
  \mathcal{H}(\fvec{q}, \fvec{q}') &=
    \braket{\idHFB{\fvec{q}} | \op{H} | \idHFB{\fvec{q}'}}.
\end{align}
Since the norm kernel involves the standard inner product in the many-body
space, it represents the topology of that space. Therefore, it should be
advantageous to use for the loss a metric induced by the same inner product
that defines the norm kernel. 

In our case, we want to build an AE where the encoder $\AEcode{\idcan} =
E(\AEtensor{\idcan})$ compresses the {\it single-particle}, canonical orbitals 
$\{\idcan_{\mu}\}_{\mu}$ associated with $\ket{\idNBodyOrig}$ into a 
low-dimensional vector $\AEcode{\idcan}$ and where the decoder 
$\AEtensor{\idcanR} = D(\AEcode{\idcan})$ is used to compute the set of 
reconstructed canonical orbitals $\{\idcanR_{\mu}\}_{\mu}$. Most importantly, 
this reconstruction should be such that the reconstructed {\em many-body} state 
$\ket{\idNBodyRec}$ is as close as possible to the original state 
$\ket{\idNBodyOrig}$. In other words, we need to use a loss that depends on the 
norm overlap (between many-body states) but since we work with single-particle 
wavefunctions, we must have a way to relate the norm overlap to these {\singp} 
wavefunctions. This can be achieved with Equations~(5.4) and (5.6) of 
\cite{haider1992microscopic}, which relate the inner product 
$\braket{\idNBodyOrig | \idNBodyRec}$ in the many-body space with the inner 
product (overlap) $\braket{\idcan_{\mu} | \idcanR_{\nu}}$ between the related 
canonical orbitals $\idcan_{\mu}$ and $\idcanR_{\nu}$,
\begin{equation}
\braket{\idcan_{\mu} | \idcanR_{\nu}}
\equiv
  \tau^{(\idcan\idcanR)}_{\mu\nu}
  = \{\cPc[\idcan]{\mu}, \aPc[\idcanR]{\nu}\}
  = \sum_\sigma
    \int\diff{\fvec{r}}\,
      \idcan_{\mu}^*(\fvec{r}, \sigma)\idcanR_{\nu}(\fvec{r}, \sigma)
\end{equation}
and with the occupation amplitudes. However, it assumes that the canonical 
wavefunctions of each many-body state are orthogonal. This property is not 
guaranteed for our {\em reconstructed} canonical wavefunctions. In fact, 
because of this lack of orthogonality, the reconstructed wavefunctions cannot 
be interpreted as representing the canonical basis of the Bloch-Messiah-Zumino 
decomposition of the quasiparticle vacuum and the Haider \& Gogny formula 
cannot be applied `as is'. However, we show in Appendix~\ref{apx:nonortho} that 
it is possible to find a set of transformations of the reconstructed 
wavefunctions that allows us to define such as genuine canonical basis. 

We want the loss function to depend only on the error associated with the 
reconstructed orbital $\idcanR_{\mu}$. Therefore, we should in principle 
consider the many-body state $\ket{\tilde{\idNBodyOrig}_{\mu}}$ where only the 
orbital $\idcan_{\mu}$ is substituted by its reconstruction $\idcanR_{\mu}$. We 
can then compute the inner product between $\ket{\idNBodyOrig}$ and 
$\ket{\tilde{\idNBodyOrig}_{\mu}}$ using Appendix~\ref{apx:nonortho} and deduce 
any induced metric $f$
\begin{equation}
  \label{eq:AE:archi:loss:exactdist}
  d_{\rm exact}^f(T^{(\idcan)}, T^{(\idcanR)})
  = f%
    \left(
      \frac{\braket{\idNBodyOrig | \tilde{\idNBodyOrig}_{\mu}}}%
           {\sqrt{\braket{\tilde{\idNBodyOrig}_{\mu} | \tilde{\idNBodyOrig}_{\mu}}}}
    \right).
\end{equation}
However, computing this metric is too computationally involved to be carried 
out explicitly for each training data at each epoch. Instead, we keep this 
metric for comparing \textit{a posteriori} the performance of our model.

Instead of explicitly determining $d_{\rm exact}^f(T^{(\idcan)}, 
T^{(\idcanR)})$, we focus on reproducing canonical orbitals using the metrics 
of the one-body Hilbert space. In practice we considered the distance noted 
$\distUnit^{(0)}$ that is induced by the inner product between normalized 
functions in the one-body Hilbert space, that is,
\be
\distUnit^{(0)}(\idcan, \idcanR)
\equiv \left(
\frac{\bra{\idcan}}{\sqrt{\braket{\idcan|\idcan}}} 
- 
\frac{\bra{\idcanR}}{\sqrt{\braket{\idcanR|\idcanR}}} 
\right)%
\left(
\frac{\ket{\idcan}}{\sqrt{\braket{\idcan|\idcan}}} 
- 
\frac{\ket{\idcanR}}{\sqrt{\braket{\idcanR|\idcanR}}}
\right)\,, 
\ee
which is nothing but
\be
\distUnit^{(0)}(\idcan, \idcanR) = \sum_\sigma\int\diff{\fvec{r}}
\big|
\idcan(\fvec{r}, \sigma) - \idcanR(\fvec{r}, \sigma)
\big|^2 \, ,
\ee
where the $\idcan(\fvec{r},\sigma)$ and $\idcanR(\fvec{r},\sigma)$ have been 
normalized. Since all wavefunctions are discretized on the Gauss quadrature 
mesh, this distance reads
\be
\distUnit^{(0)}(\idcan, \idcanR)
= \sum_{n_{\perp}n_{z}n_{\sigma}} W_{n_{\perp}n_{z}}
\big|
T^{(\idcan)}_{n_{\perp}n_{z}n_{\sigma}} - T^{(\idcanR)}_{n_{\perp}n_{z}n_{\sigma}}
\big|^2,
\label{eq:AE:archi:fullGaussQdist}
\ee
where the weights $W$ are given by
\begin{equation}
    W_{n_{\perp}n_{z}}
    = \frac{w^{\rm GL}_{n_{\perp}}}{2b_\perp^2}
      \times 2\pi
      \times\frac{w^{\rm GH}_{n_{z}}}{b_z} .
\end{equation}
These weights, which depend on the indices $n_{\perp}$ and $n_{z}$ in the
summation, are the only difference between the squared distance loss
\eqref{eq:AE:archi:fullGaussQdist} and the MSE loss \eqref{eq:AE:archi:MSE}. 
Although the distance \eqref{eq:AE:archi:fullGaussQdist} is norm-invariant, it 
still depends on the global phase of each orbital. We have explored other 
possible options for the loss based on norm- and phase-invariant distances; see 
Appendix~\ref{apx:metrics} for a list. However, we found in our tests that the 
distance $\distUnit^{(0)}$ systematically outperformed the other ones and, for 
this reason, only show results obtained with this one.

\subsubsection{Physics-Informed Autoencoder}
\label{sec:AE:archi:resnet}

From a mathematical point of view, deep neural networks can be thought of as a
series of compositions of functions. Each composition operation defines a new
layer in the network. Networks are most often built with alternating linear and
nonlinear layers. The linear part is a simple matrix multiplication. Typical
examples of nonlinear layers include sigmoid, tanh, Rectified Linear Unit
(ReLU) functions. In addition to these linear and nonlinear layers, there could
be miscellaneous manipulations of the model for more specific purposes, such as
adding batch normalization layers~\cite{ioffe2015batch}, applying
dropout~\cite{srivastava2014dropout} to some linear layers, or skip
connection~\cite{he2016deep} between layers.

\begin{figure}[!htb]
\begin{center}
\includegraphics[width=0.7\textwidth]{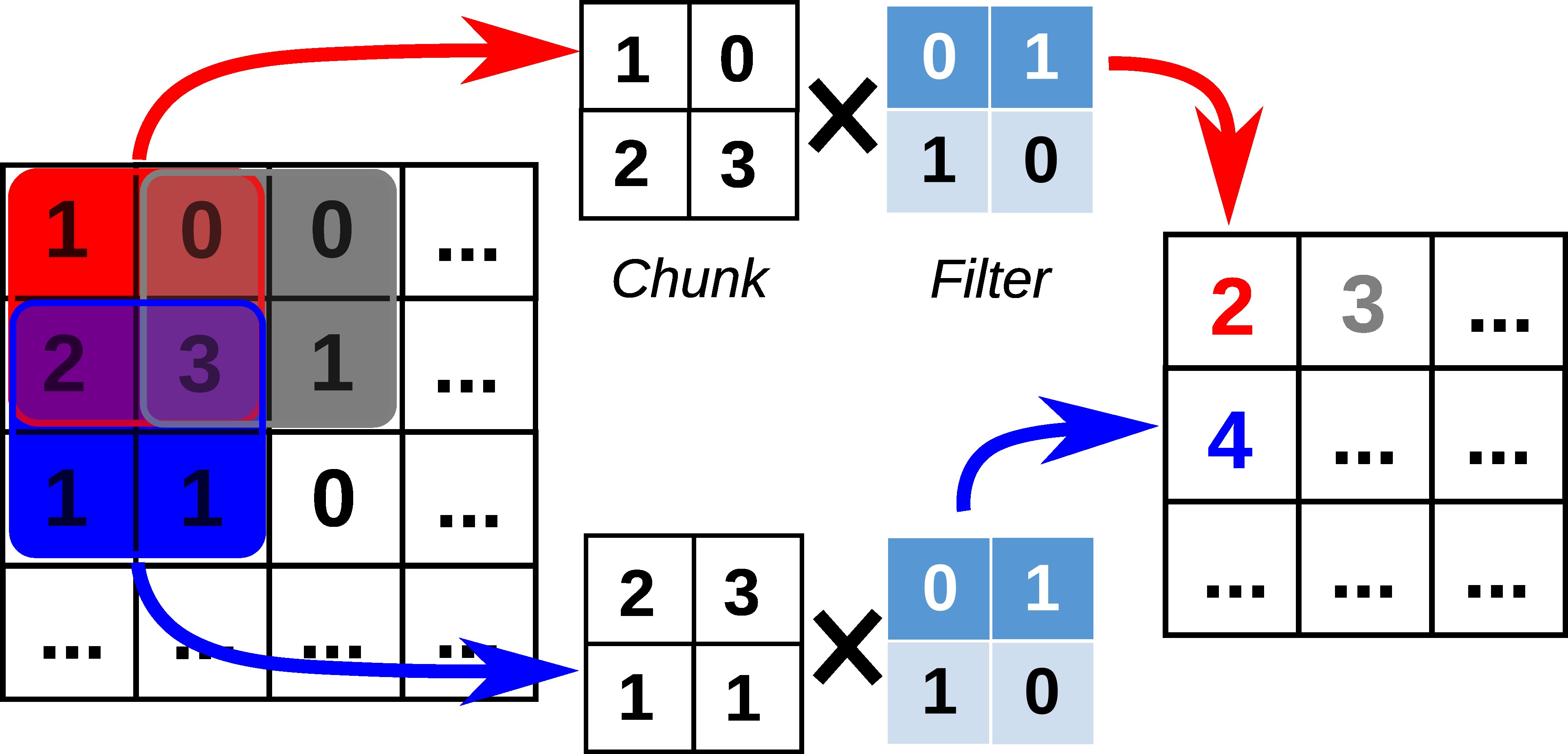}
\caption{Schematic example of a convolutional layer. For any $2\times 2$ chunk
$C$ of the input image on the left, this convolutional layer performs the
point-wise multiplication of $C$ with the filter $F$ followed by the addition
of all elements. This compresses the initial chunk of the image into a single
integer.}
\label{fig:conv}
\end{center}
\end{figure}

Our data is a smooth function defined over a $\Nperp\times\Nz = 60\times 40$ 
grid and is analogous to a small picture. For this reason, we chose a 2D 
convolutional network architecture. Convolutional layers are popular for image 
analysis, because they incorporate the two-dimensional pixel arrangement in the 
construction of the weights of the network. These two-dimensional weights, or 
filters, capture local shapes and can model the dependent structure in nearby 
pixels of image data. Given a 2D $m\times m$ input array, a 2D filter $F$ is a 
$n\times n$ matrix, usually with $n\ll m$. If we note $I^n$ the space of 
$n\times n$ integer-valued matrices, then the convolutional layer $\mathcal{C}$ 
is an operation of the $\mathcal{C}: (I^n, I^n) \rightarrow \mathbb{N}$ that is 
applied to all pairs $(F,C)$ where $C$ is any $n\times n$ chunk of the input 
image; see Fig.~\ref{fig:conv} for an example. This way, the resulting output 
summarizes the strength and location of that particular filter shape within the 
image. As the model gets trained, the filter parameters are fitted to a shape 
that is learned to be important in the training data. Convolutional neural 
network are very effective for image analysis and are currently widely 
used~\cite{krizhevsky2012imagenet,zeiler2014visualizing,sermanet2013overfeat,
szegedy2017inception}

In this work, we used the Resnet 18 model as our encoder and constructed the
decoder from a transposed convolution architecture of the Resnet 18. The Resnet
18 model was first introduced by He {\it et al.} as a convolutional neural
network for image analysis \cite{he2016deep}. It was proposed as a solution to
the degradation of performance as the network depth increases. Resnet branches 
an identity-function addition layer to sub-blocks (some sequential layers of 
composition) of a given network. While a typical neural network sub-block input 
and output could be represented by $x$ and $f(x)$, respectively, a Resnet 
sub-block would output $f(x) + x$ for the same input $x$, as in Figure 2 of 
\cite{he2016deep}. This architecture is called `skip connection' and was shown 
to be helpful for tackling multiple challenges in training deep neural network 
such as vanishing gradient problem and complex loss function 
\cite{li2018visualizing,he2016deep}. Since then, the Resnet architecture has 
been widely successful, often being used as a baseline for exploring new 
architectures \cite{zhang2022resnet,radosavovic2020designing} or as the central 
model for many analyses \cite{cubuk2020randaugment,yun2019cutmix, 
zhang2017mixup}. In a few cases, it was also combined with autoencoders for 
feature learning from high-dimensional data \cite{wickramasinghe2021resnet}.

\begin{figure}[!htb]
\begin{center}
\includegraphics[width=0.8\textwidth]{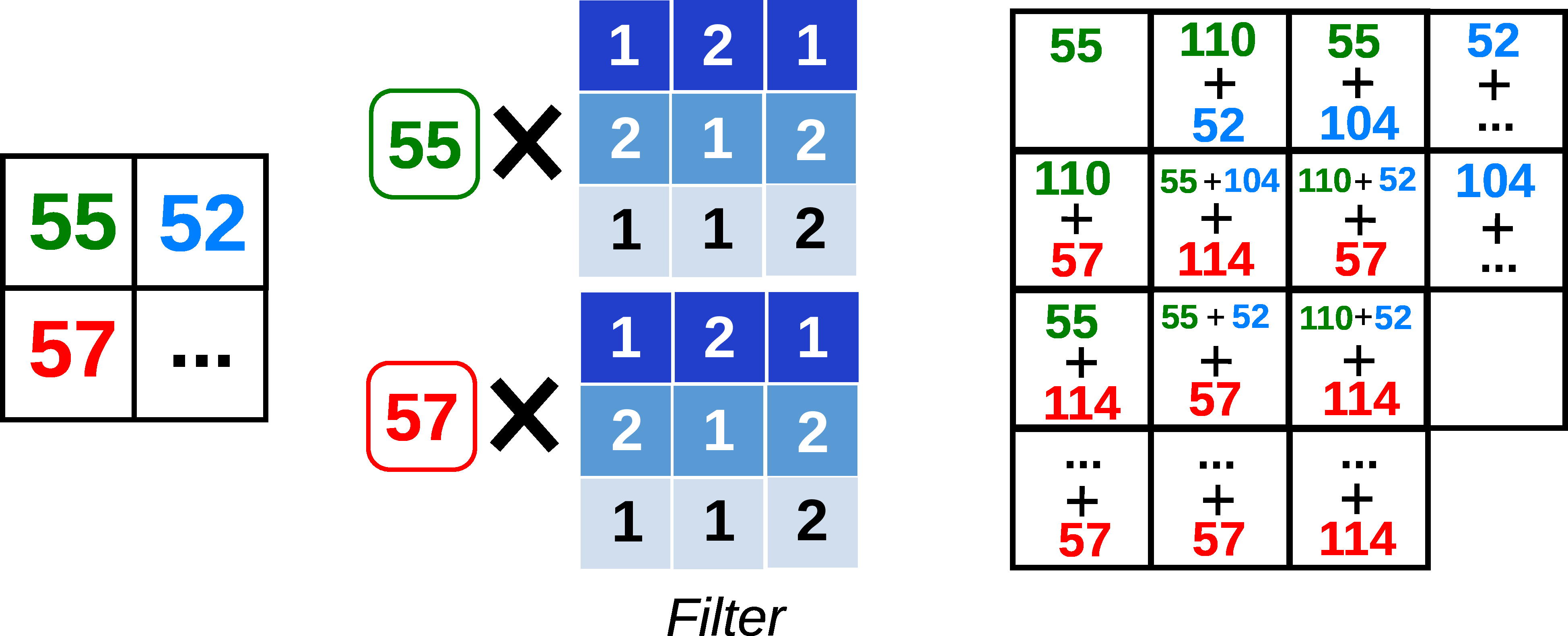}
\caption{Schematic illustration of the 2D-transposed convolution. Each input
value, e.g., 55, 57, etc., is multiplied by the entire kernel resulting in a
$3\times 3$ matrix. These matrices are then added to one another in a sliding
and overlapping way.}
\label{fig:deconv}
\end{center}
\end{figure}

\begin{figure}[!htb]
\begin{center}
\includegraphics[width=0.564\textwidth]{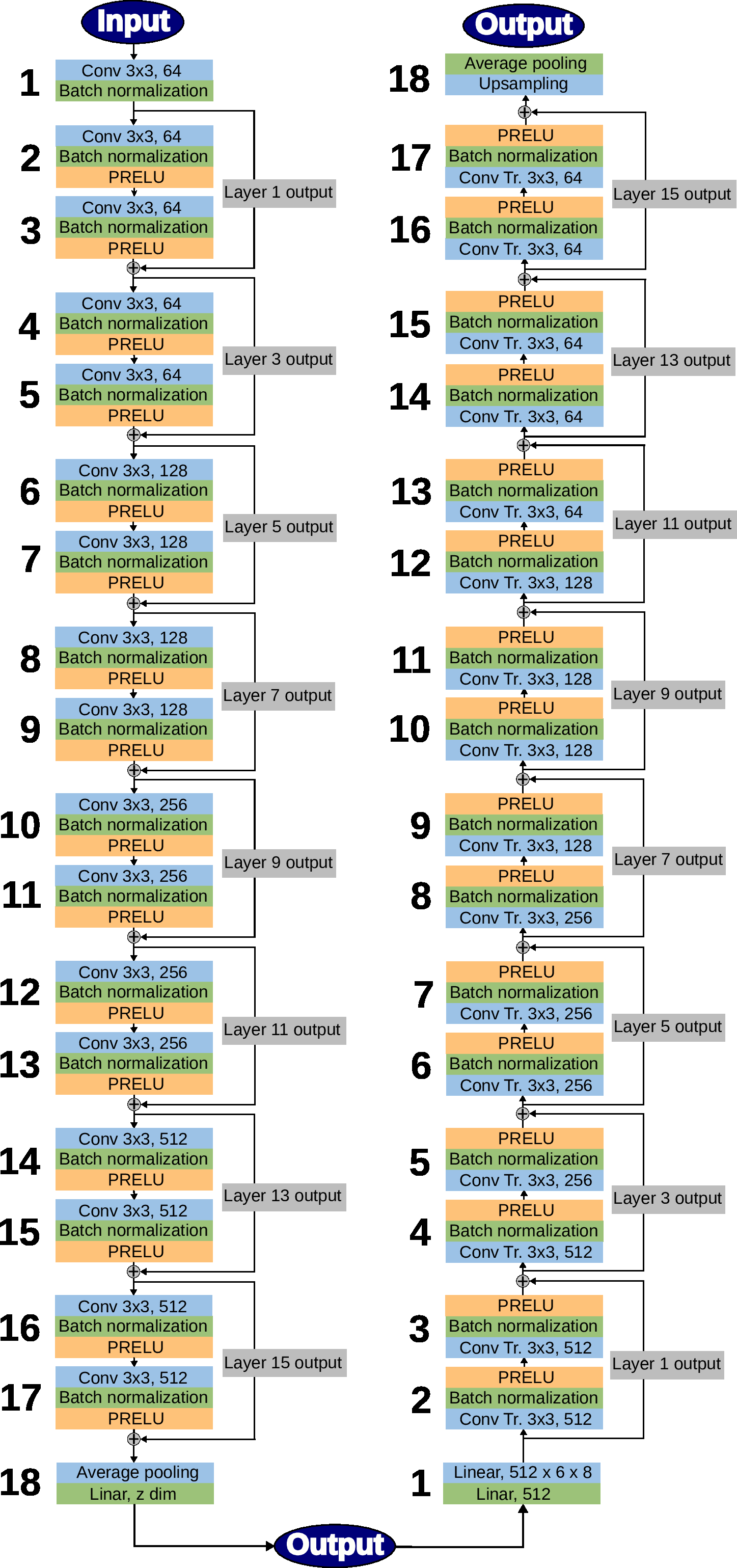}
\caption{Representation of our modified Resnet 18 architecture for the encoder 
(left) and the decoder (right). Large numbers on the left of each side label 
the different layers. Numbers such as 64, 128, etc. refer to the size of the 
filer; see text for a discussion of some of the main layers.}
\label{fig:resnet}
\end{center}
\end{figure}

For the decoder part, we designed a near-mirror image of the encoder using 
transposed convolution. Transposed convolution is essentially the opposite 
operation to convolution in terms of input and output dimensions. Here the 
meaning of transpose refers to the form of the filter matrix when the 
convolution layer is represented by a 1D vector input obtained from linearizing 
the 2D input. Note that the mirror-located filters in the decoder are 
independent parameters and not the  actual transposed filter matrix of the 
encoder. Such a construction ensures symmetrical encoder and decoder models, 
making the decoder model close to the inverse shape of the encoder model. 
Figure \ref{fig:deconv} illustrates the operation: one input value is 
multiplied by the entire kernel (filter) and is added to the output matrix at 
its corresponding location. The corresponding output location for each colored 
input number are color-coded and show how the addition is done.

The first and the last layer of the Resnet architecture are mostly for resizing 
and were minimally modified from the original Resnet 18 model since the size of 
our input data is significantly smaller than typical image sizes used for 
Resnet image classification analyses. We also modified the number of input 
channels of the first layer of the encoder to be 2 (for each of the spin 
components of the nuclear wavefunction) instead of the usual number 3 (for the 
RGB colors of colored images) or 1 (for black-and-white images). The spin 
components are closely related to each other with covariance structure, similar 
to how colors interact within an image. Therefore, we treat a pair of spin 
components as a single sample and treat each component as an input channel. The 
same applies to the output channel of the decoder.

The full network is represented schematically in Fig.~\ref{fig:resnet}. 
Parametric Rectified Linear Unit, or PRELU, layers were added to impose 
nonlinearity in the model \cite{he2015delving}. PRELU layers are controlled by 
a single hyperparameter that is trained with the data. Batch normalization is a 
standardizing layer that is applied to each batch by computing its mean and 
standard deviation. It is known to accelerate training by helping with 
optimization steps~\cite{ioffe2015batch}. The average pooling layer (bottom 
left) averages each local batch of the input and produces a downsized output. 
The upsampling layer (top right) upsamples the input using a bilinear 
interpolation.

\subsection{Training}
\label{sec:AE:training}

As mentioned in Section~\ref{sec:AE:archi:resnet}, the loss is the discrepancy 
between the input of the encoder and the output of the decoder. The 
minimization of the loss with respect to all the model parameters $w$, such as 
the filter parameters, is the training process. We used the standard 
back-propagation algorithm to efficiently compute the gradient of the loss 
function with respect to the model parameters. The gradient computation is done 
with the chain rule, iterating from the last layer in the backward direction. 
We combined this with the mini-batch gradient descent algorithm: ideally, one 
would need the entire dataset to estimate the gradient at the current model 
parameter value. However, with large datasets, this becomes computationally 
inefficient. Instead we use a random subset of the entire data, called 
mini-batch, to approximate the gradient, and expedite the convergence of the 
optimization. For each mini-batch, we update each parameter $w$ by taking small 
steps of gradient descent, $w_{t+1} = w_{t} - \alpha \frac{\displaystyle 
\partial\mathcal{L}}{\displaystyle \partial w_{t}}$. At step $t$, or at 
$t^{\mathrm{th}}$ mini-batch, the average loss $\mathcal{L}$ and the gradient 
with respect to current model parameter $w_{t}$ are computed. Then 
$\alpha$-sized gradient descent step is taken to update the model parameters. 
Instead of using the current gradient for the update, one can use a weighted 
average of past gradients. We employed the well-known Adam algorithm, which 
uses the exponential moving average of current and past gradients 
\cite{kingma2014adam}. 

Iterating over the entire dataset once, using multiple mini-batches, is called 
an epoch. Typically a deep neural network needs hundreds to thousands of epochs 
for the algorithm to converge. Parameters such as the batch size or learning 
rate, the parameters of the optimizer itself (Adam's or other), and the number 
of epochs are hyper-parameters that must be tuned for model fitting. For our 
training, we used the default initialization method in PyTorch for the model 
parameters. The linear layers were initialized with a random uniform 
distribution over $[-1/k, 1/k]$, where $k$ is the size of the weight. For 
example, if there are 2 input channels and 3$\times$3 convolution filters are 
used, $k = 2\times 3\times 3$. PRELU layers were initialized with their default 
PyTorch value of $0.25$. We proceeded with mini-batches of size 32 with the 
default $\beta_1 = 0.9, \beta_2 = 0.999$ and $\epsilon = 10^{-8}$: all these 
numbers refer to the PyTorch implementation of the Adam's optimizer. For 
$\alpha$, we used $0.001$ as starting value and used a learning rate scheduler, 
which reduces the $\alpha$ value by a factor of 0.5 when there is no 
improvement in the loss for 15 epochs. After careful observation of the loss 
curves, we have estimated that at least 1000 epochs are needed to achieve 
convergence.

To mitigate the problem of the global phase invariance of the canonical 
wavefunctions discussed in \sssecref{sec:AE:archi:struct}, we doubled the size 
of the dataset: at each point $\qvec$ of the collective space (=the sample), we 
added to each canonical wavefunction $\idcan_{\mu}(\rvec,\sigma)$ the same 
function with the opposite sign $-\idcan_{\mu}(\rvec,\sigma)$. The resulting 
dataset was then first split into three components, training, validation and 
test datasets, which represent 70\%, 15\%, and 15\% of the entire data 
respectively. Training data is used for minimizing the loss with respect to the 
model parameters as explained above. Then we choose the model at the epoch that 
performs the best with the validation dataset as our final model. Finally, the 
model performance is evaluated using the test data.

\subsection{Results}
\label{sec:AE:results}

In this section, we summarize some of the preliminary results we have obtained 
after training several variants of the AE. In \sssecref{sec:AE:results:perf}, 
we give some details about the training data  and the quality of the 
reconstructed wavefunctions. We discuss some possible tools to analyze the 
structure of the latent space in \sssecref{sec:AE:results:latent}. In these two 
sections, we only present results obtained for latent spaces of dimension 
$D=20$. In \sssecref{sec:AE:results:validation}, we use the reconstructed 
wavefunctions to recalculate HFB observables with the code HFBTHO. We show the 
results of this physics validation for both $D=20$ and $D=10$.

\subsubsection{Performance of the Network}
\label{sec:AE:results:perf}

\begin{figure}[!htb]
\begin{center}
\includegraphics[width=0.7\textwidth]{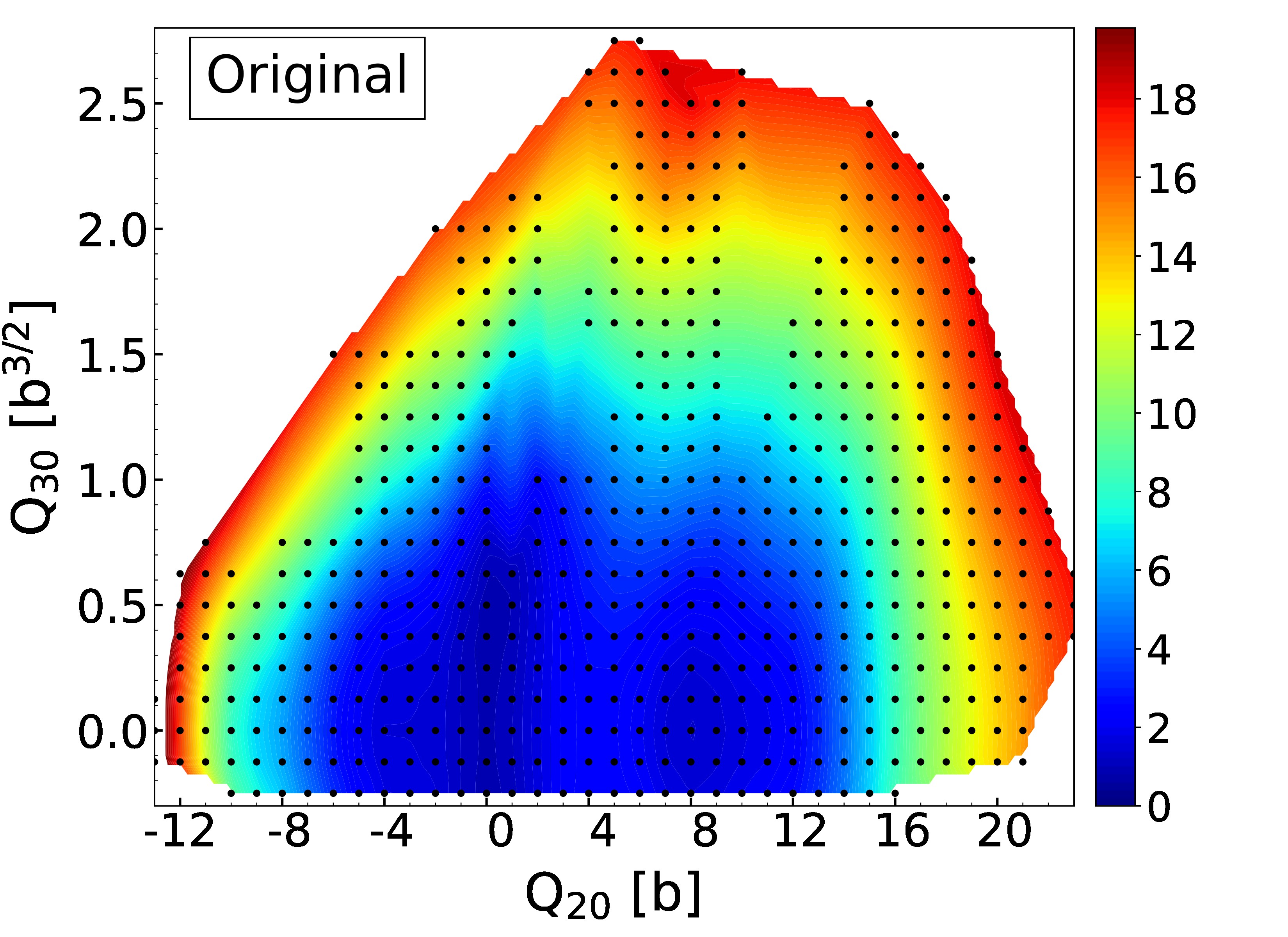}
\caption[]{Potential energy surface of $^{98}$Zr in the $(q_{20},q_{30})$ 
plane. Converged HFBTHO solutions are represented by black dots. Energies given 
by the color bar are in MeV relatively to the ground state.}
\label{fig:PES_orig}
\end{center}
\end{figure}

Figure \ref{fig:PES_orig} shows the initial potential energy surface in 
$^{98}$Zr used in this work. Using the HFBTHO solver, we performed a total of 
548 HFB calculations with constraints on the axial quadrupole, $q_{20 }= 
\langle\hat{Q}_{20}\rangle$, and axial octupole moment, $q_{30} = 
\langle\hat{Q}_{30}\rangle$. The mesh was: $-12.5\, \mathrm{b} \leq q_{20} 
\leq 25.0\, \mathrm{b}$ with steps $\delta q_{20} = 1\, \mathrm{b}$ and $0.0\, 
\mathrm{b}{3/2} \leq q_{30} \leq 3.0\, \mathrm{b}^{3/2}$ with step 
$\delta q_{30} = 0.125\,\mathrm{b}^{3/2}$. The black dots in 
Fig.~\ref{fig:PES_orig} indicate the location of the converged solutions. For 
each solution, the $\ncanP = \ncanPv$ highest-occupation proton and 
$\ncanN=\ncanNv$ highest-occupation neutron canonical wavefunctions were used 
as training data for the network\footnote{Since time-reversal symmetry is 
conserved, the Fermi energy is located around states with indices 
$\mu_p \approx 20$ and $\mu_n \approx 29$. Therefore, our choice implies that 
in our energy window, about 1/3 of all states are below the Fermi level and 
about 2/3 of them are above it.}.

For each of the losses discussed in Appendix~\ref{apx:metrics}, we trained the 
AE with the slightly modified Resnet 18 architecture described in 
\ssecref{sec:AE:archi:resnet}. It is important to keep in mind that the value 
of these losses should not be compared with one another. The only rigorous 
method to compare the performance of both networks would be to compute the 
many-body norm overlap across all the points in each case -- or to perform 
{\em a posteriori} physics validation with the reconstructed data, as will be 
shown in \sssecref{sec:AE:results:validation}.

\begin{figure}[!htb] 
\includegraphics[width=\textwidth]{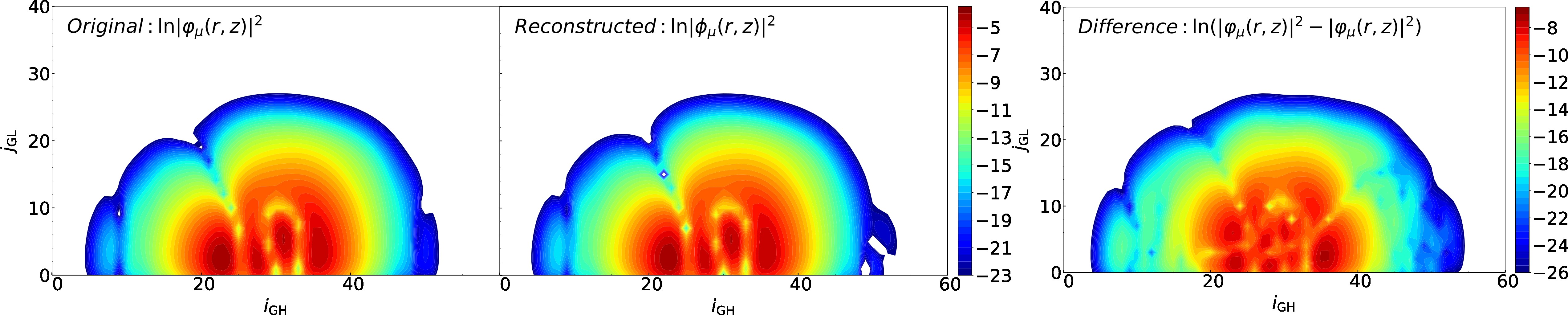}
\centering
\caption{Left: Contour plot of the logarithm of the squared norm of the neutron 
canonical wavefunction with occupation number $v_{\mu}^{2} = 0.945255$ (without 
the exponential factor). Middle: Same for the reconstructed wavefunction. 
Right: Logarithm of the difference between the squared norm of the original 
and reconstructed wavefunctions.}
\label{fig:wfs}
\end{figure}

To give an idea of the quality of the AE, we show in Fig.~\ref{fig:wfs} one 
example of the original and reconstructed canonical wavefunctions. 
Specifically, we consider the configuration $(q_{20},q_{30})=(-7.0\,\mathrm{b}, 
-0.25\,\mathrm{b}^{3/2})$ in the collective space and look at the neutron 
wavefunction with occupation number $v_{\mu}^{2} = 0.945255$, which is located 
near the Fermi surface. This example was obtained for an AE trained with the 
$d_{o}^{(0)}$ loss and compressed to $D=20$. The figure shows, in the left 
panel, the logarithm of the squared norm of the original wavefunction across 
the quadrature mesh, $\ln |\idcan_{\mu}|^2 \equiv 
\ln|T^{(\idcan_{\mu})}_{n_{\perp}n_{z}n_{\sigma}}|^2$, in the middle panel, the 
same quantity for the reconstructed wavefunction, and in the right panel the 
logarithm of the difference between the two. On this example, the AE can 
reconstruct the wavefunction with about 3\% error.

\subsubsection{Structure of the Latent Space}
\label{sec:AE:results:latent}

One of the advantages of AEs is the existence of a low-dimensional 
representation of the data. In principle, any visible structure in this latent 
space would be the signal that the network has properly learned, or encoded, 
some dominant features of the dataset. Here, our latent space has dimension 
$D=20$. This means that every canonical wavefunction, which is originally a 
matrix of size $n = N_{\perp}\times N_z$, is encoded into a single vector of 
size $D$. From a mathematical point of view, the encoder is thus a function 
\be
\begin{array}{rl}
\op{E}: \mathbb{R}^{n} & \longrightarrow \mathbb{R}^{D} \\[-2pt]
                \idcan & \longmapsto \fvec{v} = \op{E}(\idcan)
\end{array}
\label{eq:opE}
\ee

Let us consider some (scalar) quantity $P$ associated with the many-body state 
$\ket{\idNBodyOrig(\qvec)}$ at point $\qvec$. Such a quantity could be an 
actual observable such as the total energy but it could also be an auxiliary 
object such as the expectation value of the multipole moment operators. In 
fact, $P$ could also be a quantity associated with the individual degrees of 
freedom at point $\qvec$, for example the {\quasp} energies. In general terms, 
we can think of $P$ as the output value of the function
\be
\begin{array}{rl}
\op{P}: \mathbb{R}^{n} & \longrightarrow \mathbb{R} \\[-2pt]
                \idcan & \longmapsto P = \op{P}(\idcan)
\end{array}
\label{eq:opA}
\ee
For example, if $P$ represents the {\singp} canonical energies, then the 
function $\op{P}$ is the one that associates with each canonical wavefunction 
its {\singp} energy. Therefore, for every canonical wavefunction, there is a 
different value of $P$. Conversely, if $P = \braket{\hat{Q}_{20}}$, there is a 
single value for all the canonical wavefunctions at point $\qvec$. Since there 
is a vector in the latent space for each canonical wavefunction, and there is 
also a value for the quantity $P$ for each such function, we can then define 
the new function $\op{\mathcal{P}}$ acting on vectors of the {\em latent space} 
and defined as 
\be
\begin{array}{rl}
\op{\mathcal{P}}: \mathbb{R}^{D} & \longrightarrow \mathbb{R} \\[-2pt]
                        \fvec{v} & \longmapsto P = \op{\mathcal{P}}( \fvec{v})
\end{array}
\label{eq:opAlatent}
\ee
and it is straightforward to see that: $\op{P} = \op{\mathcal{P}}\circ\op{E}$. 
Our goal is now to try to analyze where various quantities $P$ are located in 
the latent space and whether one can identify some specific features of these 
locations.

\begin{figure}[!htb]
\includegraphics[width=\textwidth]{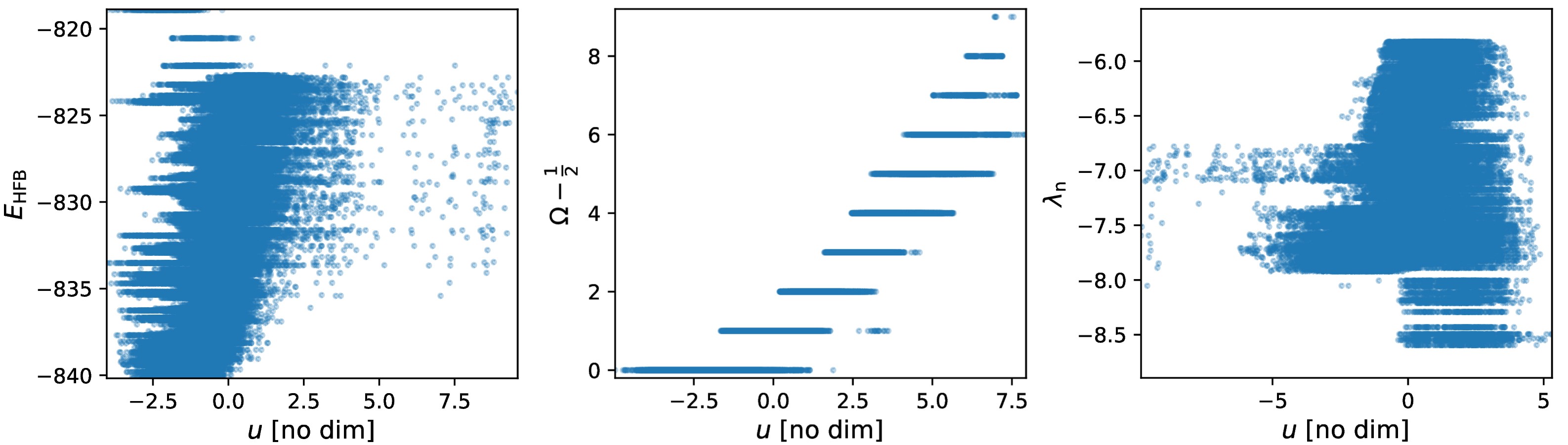}
\centering
\caption{One-dimensional projections of the $D$-dimensional linear fit for the 
total energy $E_{\rm HFB}$ (left panel), the projection $\Omega$ of the 
canonical state (middle panel) and the neutron Fermi energy $\lambda_n$ (right 
panel). Each point represents one of these quantities for a canonical 
wavefunction $\mu$ and a point $\qvec$ in the collective space.}
\label{fig:pca20}
\end{figure}

Since we have a total of $\ncanT = \ncanTv$ wavefunctions for each of the 
$N_p = 552$ points in the collective space, the encoder yields a set of 
$\ncanT \times N_p$ vectors of dimension $D$. This means that, in the latent 
space, every quantity $P$ above is also represented by a cloud of 
$\ncanT \times N_p$ such vectors. This is obviously impossible to visualize. 
For this reason, we introduce the following analysis. First, we perform a 
linear regression in the $D$-dimensional latent space of a few select 
quantities of interest $P$, that is, we write
\be
P = \gras{\alpha}\cdot\fvec{v} + b\,,
\ee
where $\gras{\alpha}$ is a $D$-dimensional vector, $\fvec{v}$ is the vector 
associated with the quantity $P$ in the latent space and $b\in\mathbb{R}$. The 
unit vector $\gras{u} = \gras{\alpha}/||\gras{\alpha}||$ can be interpreted as 
representing the leading direction in the latent space. The quantity 
$u = \gras{u}\cdot\fvec{v}$ is a scalar which we obtain easily from the result 
of the linear regression. We can thus plot the function $P: u \mapsto P(u)$. 
Examples of such functions are shown in Fig.~\ref{fig:pca20}. Each point in the 
figures represents the value $P = \op{\mathcal{P}}(\fvec{v})$ of some 
characteristic quantity at point $u = \gras{u}\cdot\fvec{v}$.

The three cases shown in Fig.~\ref{fig:pca20} illustrate that the network has 
not always identified relevant features. The case of $\Omega$, middle panel, is 
the cleanest: there is a clear slope as a function of $u$: if one sets $u = 1$, 
for example, then only values of $ 7/2 \leq \Omega  \leq 15/2$ are possible. 
Conversely, the AE has not really discovered any feature in the neutron Fermi 
energy (right panel): for any given value of $u$, there is a large range of 
possible values of Fermi energies. In the case of the total energy (left 
panel), the situation is somewhat intermediate: there is a faint slope 
suggesting a linear dependency of the energy as a function of $u$.

\subsubsection{Physics Validation}
\label{sec:AE:results:validation}

The results presented in the Section \ref{sec:AE:results:perf} suggest the AE 
has the ability to reproduce the canonical wavefunctions with good precision. 
To test this hypothesis, we recalculated the HFB solution at all the training, 
validation and testing points by substituting in the HFBTHO binary files the 
original canonical wavefunctions by the ones reconstructed by the AE. Recall 
that only the lowest $\ncanT$ wavefunctions with the largest occupation were 
encoded in the AE ($\ncanN = \ncanNv$ for neutrons and $\ncanP= \ncanPv$ for 
protons); the remaining ones were unchanged. In practice, their occupation is 
so small that their contribution to nuclear observables is very small ($<$ 10 
keV for the total energy, for example). 

\begin{figure}[!htb]
\begin{center}
\includegraphics[width=\textwidth]{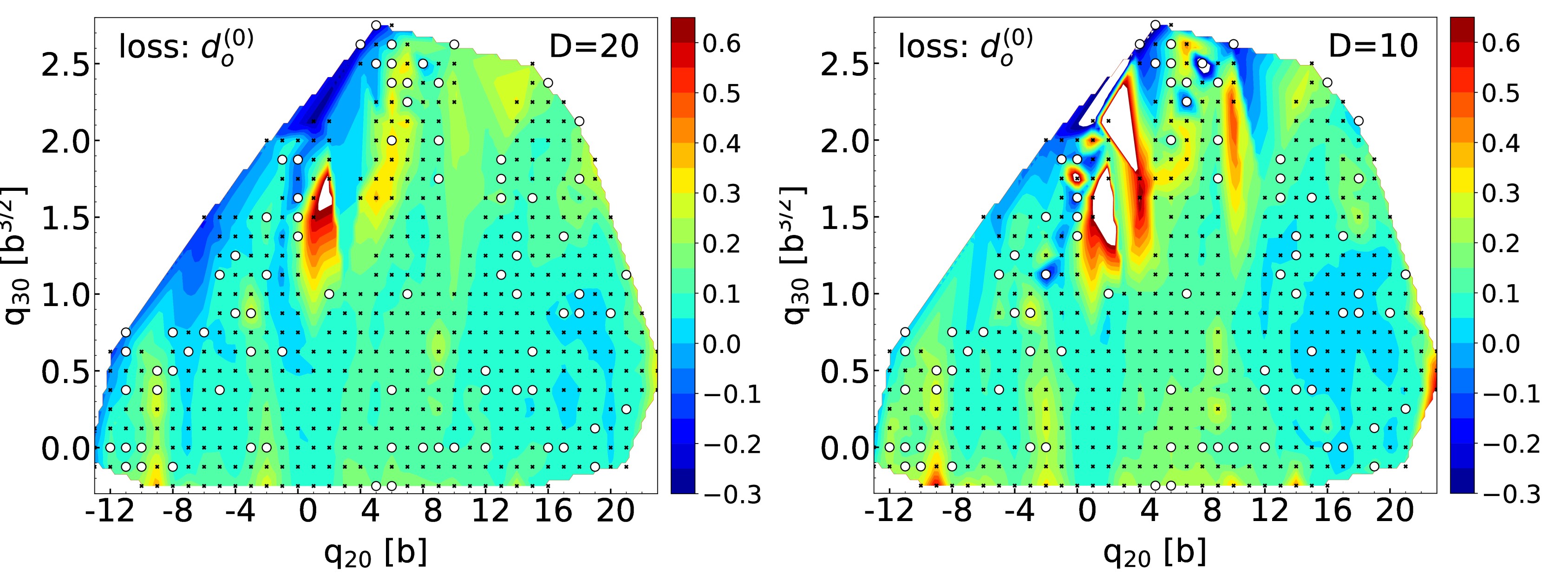}
\caption[]{Left: Potential energy surface in the $(q_{20},q_{30})$ plane for 
$^{98}$Zr obtained after replacing the first $\ncanN = \ncanNv$ and 
$\ncanP = \ncanPv$ highest-occupation canonical wavefunctions by their values 
reconstructed by the AE for a latent space of dimension $D=20$. The black dots 
show the location of the training points only, the white circles the location 
of the validation points. Right: same figure for a latent space of dimension 
$D=10$. For both figures, energies are given in MeV.}
\label{fig:PES}
\end{center}
\end{figure}

Figure~\ref{fig:PES} shows the error on the potential energy across the 
$(q_{20},q_{30})$ collective space obtained with the reconstructed canonical 
wavefunctions for latent spaces of dimension $D=20$ (left) and $D=10$ (right). 
In each case, we only show results obtained when using the $d_{o}^{(0)}$ loss, 
which gives the best results. The black crosses denote the location of all the 
original points; the white circles show the location of the validation points. 
Overall, the results are very encouraging. In both cases, most of the error is 
concentrated near regions of the PES where there are discontinuities (hence, 
the lack of converged solutions). Everywhere else, the error is small and 
mostly randomly distributed across the PES, that is, it is not systematically 
larger at the validation points. As expected, the quality of the reconstruction 
is a little worse when $D=10$: one can notice about a dozen of points for which 
the error is significantly larger, in absolute value. Examples include 
$(q_{20},q_{30}) = (-5.0\,\rm{b}, 1.125\,\rm{b}^{3/2})$ or $(q_{20},q_{30}) = 
(+8.0\,\rm{b}, 2.5\,\rm{b}^{3/2})$ in the validation set, and $(q_{20},q_{30}) 
= (0.0\,\rm{b}, 1.75\,\rm{b}^{3/2})$ or the region around $1\,\rm{b}\leq q_{20} 
\leq 4\,\rm{b}$ and $1.5\,\rm{b}^{3/2} \leq q_{30} \leq 2.25\,\rm{b}^{3/2}$ in 
the training set. These may suggest that for $D=10$, the loss may not have 
fully converged yet. Because of the existence of discontinuities near these 
points, this could also be the manifestation that our continuous AE cannot 
build a continuous representation of the data everywhere. However, the fact 
that an increase of the compression by a factor 2, from $D=20$ to $D=10$, does 
not substantially degrade the performance of the AE is very promising.

\begin{figure}[!htb]
\begin{center}
\includegraphics[width=\textwidth]{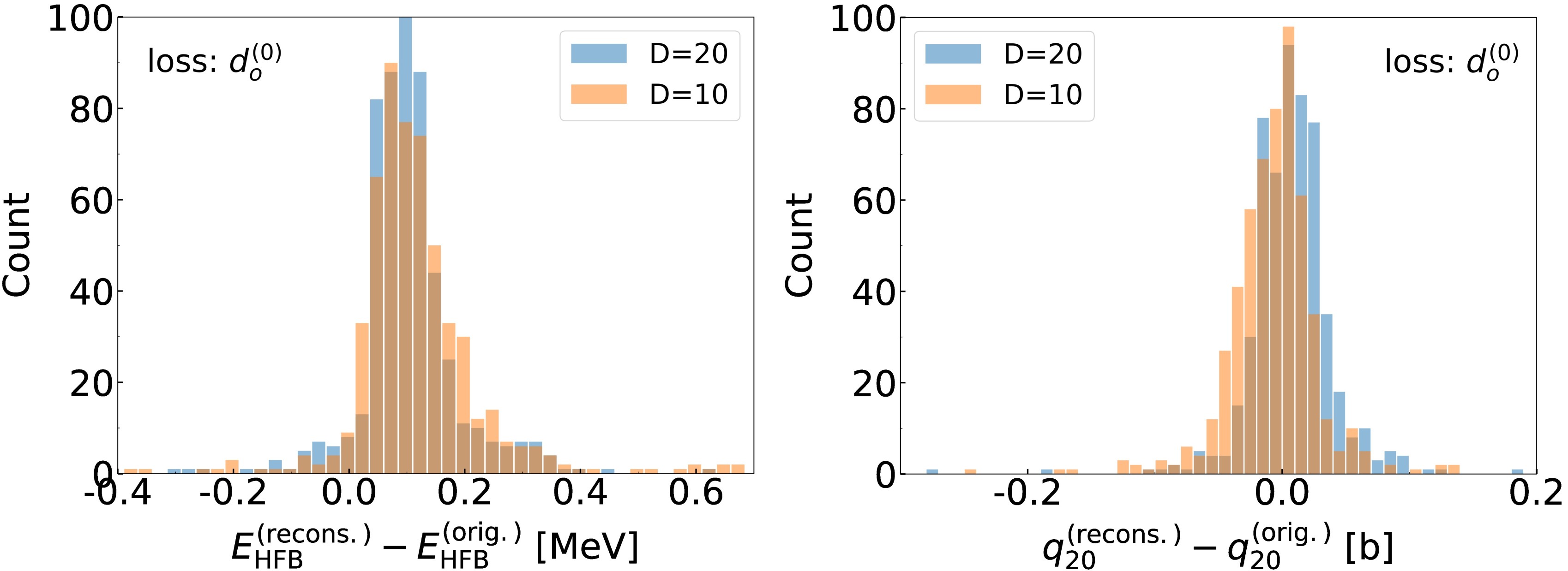}
\caption[]{Left: Histogram of the difference in total HFB energy between the 
original HFBTHO calculation and the result obtained by computing the energy in 
the canonical basis with the reconstructed wavefunctions (see text for 
details). Calculations were performed both for a $D=20$ and $D=10$ latent 
space. Right: Similar histogram for the expectation value of the axial 
quadrupole moment.}
\label{fig:multipoles}
\end{center}
\end{figure}

The two histograms in Fig.~\ref{fig:multipoles} give another measure of the 
quality of the AE. The histogram in the left shows the distribution of the 
error on the HFB energy for two sizes of the latent space, $D=20$ and $D=10$. 
In both cases, most of the error is less than $\pm 100$ keV, which is rather 
good. As mentioned before, the points with the higher error $\epsilon_E < -150$ 
keV or $\epsilon_E > 250$ keV do not correspond only to testing points. 
The histogram on the right shows the distribution of the error 
for $q_{20}$ and $q_{30}$ (in units of b and $b^{3/2}$, respectively).

\section{Conclusion}
\label{sec:conclu}

Extreme-scale calculations based on nuclear density functional methods relevant 
for, e.g., nuclear astrophysics simulations or uncertainty quantification 
remain computationally expensive and fraught with formal and practical issues 
associated with self-consistency or reduced collective spaces. In this article, 
we have analyzed two different techniques to build fast, efficient and accurate 
surrogate models, or emulators, or DFT objects. 

We first showed that Gaussian processes could reproduce reasonably well the 
values of the mean-field and pairing-field potentials of the HFB theory across 
a large two-dimensional potential energy surface. The absolute error on the 
total energy was within $\pm 100$ keV and the relative errors on the collective 
inertia tensor smaller than 5\%. However, GPs require the training data to be
``smoothly-varying'', i.e., they should not include phenomena such as nuclear 
scission or, more generally, discontinuities in the PES. It is well known that 
GPs are not reliable for extrapolation: such a technique can thus be very 
practical to densify (=interpolate) an existing potential energy surface but 
must not be applied outside its training range.

Although standard versions of GPs are fast and simple to use, incorporating 
more of the many of the existing correlations between the values of the HFB 
potentials may considerably increase the complexity of the emulator. In such a 
case, it is more natural to use directly deep-learning techniques. In this 
work, we reported the first application of autoencoders to emulate the 
canonical wavefunctions of the HFB theory. Autoencoders are a form of deep
neural network that compresses the input data, here the canonical 
wavefunctions, into a small-dimensional space called the latent representation. 
The encoder is trained simultaneously with a decoder by enforcing that the 
training data is left invariant after compression followed by decompression. In 
practice, the measure of such ``invariance'' is set by what is called the loss 
of the network. We discussed possible forms of the loss that are best adapted 
to learning quantum-mechanical wavefunctions of many-body systems such as 
nuclei. We showed that such an AE could successfully reduce the data into a 
space of dimension $D=10$ while keeping the total error on the energy lower 
than $\Delta E = 150$ keV (on average). The analysis of the latent space 
revealed well-identified structures in a few cases, which suggests the network 
can learn some of the physics underlying the data. This exploratory study 
suggests that AE could serve as reliable canonical wavefunctions generators. 
The next step will involve learning a full sequence of such wavefunctions, 
i.e., an ordered list, in order to emulate the full HFB many-body state.

\section*{Conflict of Interest Statement}
The authors declare that the research was conducted in the absence of any 
commercial or financial relationships that could be construed as a potential 
conflict of interest.

\section*{Author Contributions}
M.~Verriere led the study of autoencoders for canonical wavefunctions: design 
of the overall architecture of the autoencoders, development of the software 
stack and analysis of the results. N.~Schunck led the study of Gaussian process 
for mean-field potentials -- training and validation runs, code development and 
HFBTHO calculations -- and supervised the whole project. I.~Kim implemented, 
tested and trained different architectures of autoencoders and helped with the 
analysis of the results. P.~Marevi\'c developed an HFBTHO module to use canonical 
wavefunctions in HFB calculations. K.~Quinlan provided technical expertise 
about Gaussian processes and supervised M.~N'Go, who implemented and fitted 
Gaussian processes to mean-field potentials. D.~Regnier and R.D.~Lasseri 
provided technical consulting on deep neural network and the architecture of 
autoencoders. All authors contributed to the writing of the manuscript, and 
read and approved the submitted version.

\section*{Funding}
This work was performed under the auspices of the U.S.\ Department of Energy by 
Lawrence Livermore National Laboratory under Contract DE-AC52-07NA27344. 
Computing support came from the Lawrence Livermore National Laboratory (LLNL) 
Institutional Computing Grand Challenge program.

\appendix
\section{Overlap Between Non-Orthogonal Canonical Bases}
\label{apx:nonortho}

In \cite{haider1992microscopic} the norm overlap between two different HFB 
vacuua is expressed as a function of the single-particle overlap between the 
two respective sets of canonical wavefunctions and occupation numbers. This 
formula implicitly assumes that the canonical wavefunctions are orthornormal. 
When using canonical wavefunctions reconstructed by the AE, this property may 
not hold anymore and the Haider \& Gogny formula should not be used `as is' 
when evaluating the norm overlap. In this appendix, we show how to generalize 
it.

We recall that for a system with conserved time-reversal symmetry, the 
quasiparticle vacuum can be written \cite{ring2004nuclear,schunck2019energy}
\be
\label{eq:nonortho:vacuum}
\ket{\idNBodyOrig} = \prod_{\mu_>} \beta^{\dagger}_{\mu}\beta^{\dagger}_{\bar{\mu}}\ket{\idPV},
\ee
where the quasiparticle operators $(\hat{\beta}, \hat{\beta}^{\dagger})$ are 
obtained from a single-particle basis $(\aP{}, \cP{})$ by the Bogoliubov 
transformation $W$ of \eqref{eq:W} in \ssecref{subsec:canonical}. The 
Bloch-Messiah-Zumino decomposition of the matrix $W$ is a consequence of the 
fact that the quasiparticle operators should obey the same anticommutation 
relations as the particle operators. In the resulting canonical basis, the 
quasiparticle vacuum can be written in the BCS form,
\be
\ket{\idNBodyOrig} = \prod_{\mu_>} \left( u_\mu + v_\mu \cPc{\mu} \cPc{\bar{\mu}} \right) \ket{\idPV},
\label{eq:BCS}
\ee
where $\ket{\idPV}$ is the particle vacuum, $\bar{\mu}$ refers to the 
time-reversed partner of the state with index $\mu$, and the summation runs 
only over ``positive'' indices $\mu$. Implicit in this expression is the fact 
that the canonical wavefunctions associated with the operators $(\aPc{}, 
\cPc{})$ are orthonormal or, equivalently, that the operators $(\aPc{},\cPc{})$ 
anticommute. If these conditions are not verified, the form \eqref{eq:BCS} is
not valid and the formulas for the norm overlap given in 
\cite{haider1992microscopic} cannot apply. In our case, the fit of the AE gives 
a set of {\em reconstructed} canonical wavefunctions which we noted 
$\{ \idcanR_{\mu}(\rvec,\sigma) \}_{\mu}$ and are associated with a set of 
single-particle creation and annihilation operators $(\aPno{},\cPno{})$. 
Although we call these objects canonical orbitals, this is somewhat a misnomer 
since the wavefunctions are not necessarily orthonormal. As a consequence, one 
cannot define a BCS state \eqref{eq:BCS} with these operators. Our goal is to 
find a transformation of these single-particle operators that allows us to 
define a BCS state. 

Following the notations of \cite{haider1992microscopic}, we thus define the 
single-particle overlaps 
\be
\olapCanRec_{\mu\nu}^{(\idPno\idPno)} 
= \{\cPno{\mu}, \aPno{\nu}\}
= \sum_{\sigma}\int\diff\rvec\, \idcanR_{\mu}^{*}(\rvec,\sigma)\idcanR_{\nu}(\rvec,\sigma) .
\label{eq:sp_ovlp}
\ee
The set of all such overlaps define the overlap matrix 
$\olapCanRec^{(\idPno\idPno)}$. This overlap matrix is block diagonal as it 
satisfies the relations
\be
\olapCanRec^{(\idPno\idPno)}_{\mu\nu} = {\olapCanRec^{(\idPno\idPno)}_{\bar{\mu}\bar{\nu}}}^{*} ,
\qquad
\olapCanRec^{(\idPno\idPno)}_{\mu\bar{\nu}} = \olapCanRec^{(\idPno\idPno)}_{\bar{\mu}\nu} = 0.
\label{eq:sp_ovlp_mat}
\ee
From the {\singp} operators $(\aPno{},\cPno{})$, we can introduce a new set of 
{\quasp} operators $(\aQPno{},\cQPno{})$ through
\begin{subequations}
\label{eq:nonortho:BCS}
\begin{align}
\aQPno{\mu}       &= \orbCoefU_{\mu}\aPno{\mu}-\orbCoefV_{\mu}\cPno{\bar{\mu}}, 
\label{eq:chi}\\
\aQPno{\bar{\mu}} &= \orbCoefU_{\mu}\aPno{\bar{\mu}}+\orbCoefV_{\mu}\cPno{\mu}. 
\label{eq:chibar}
\end{align}
\end{subequations}
It is easy to see that these {\quasp} operators do not obey the Fermion 
anticommutation relation. In other words, the {\quasp} {\em spinors} associated 
with these operators are not orthogonal. We thus introduce the overlap matrix 
$\olapCanRec^{(\idQPno\idQPno)}_{\mu\nu}$ between any two such spinors 
$\mu,\nu>0$. Owing to \eqref{eq:chi}-\eqref{eq:chibar} and \eqref{eq:sp_ovlp} 
it is straightforward to show that it is given by 
\be
\olapCanRec^{(\idQPno\idQPno)}_{\mu\nu} = 
\orbCoefU_{\mu}\orbCoefU_{\nu}\olapCanRec^{(\idPno\idPno)}_{\mu\nu}
+
\orbCoefV_{\mu}\orbCoefV_{\nu}{\olapCanRec^{(\idPno\idPno)}_{\mu\nu}}^{*} ,
\label{eq:qp_ovlp}
\ee
and verify the same properties \eqref{eq:sp_ovlp_mat} as the 
single-particle overlap. We symmetrically orthogonalize the {\quasp} basis by 
eigendecomposing $\olapCanRec^{(\idQPno\idQPno)}_{\mu\nu}$~\footnote{In fact, 
we can limit ourself to compute the Cholesky decomposition of 
$\olapCanRec^{(\idQPno\idQPno)}_{\mu\nu}$, but we use the eigenvalues to check 
the rank and invert the matrix.}
\be
\olapCanRec^{(\idQPno\idQPno)}_{\mu\nu} = Q\Sigma^2Q^{\dagger},
\ee
where $Q$ is unitary and $\Sigma=\operatorname{diag}(\sigma_0,\sigma_1, \dots)$ 
with $\sigma_{\mu}>0$. We then construct a new orthogonal {\quasp} basis
\be
\label{eq:nonortho:orthodef}
\cQPyes{\mu} = \sum_{k} \cQPno{k} \big(Q\Sigma^{-1}\big)_{k\mu}
\ee
that satisfies the fermion commutation relations, 
$\{\cQPyes{\mu}, \aQPyes{\nu}\} = \delta_{\mu\nu}$. We can associate with these 
new {\quasp} operators $(\cQPyes{}, \aQPyes{})$ a quasiparticle vacuum of the 
type \eqref{eq:nonortho:vacuum}. We now need to find the Bogoliubov 
transformation $\tilde{W}$ (and its Bloch-Messiah decomposition) that relate 
the $(\cQPyes{}, \aQPyes{})$ to a properly orthonormal {\singp} basis. To this 
end, we first diagonalize the single-particle overlap matrix 
\be
\olapCanRec^{(\idPno\idPno)}_{\mu\nu} = R \tilde{\Sigma}^2 R^{\dagger},
\label{eq:sp_ovlp_diag}
\ee
which defines a new set of {\em particle} operators $(\aPyes{}, \cPyes{})$ 
through the relations
\begin{subequations}
\label{eq:nonortho:orbOrtho0}
\begin{align}
\cPyes{i}       &= \sum_{k} \cPno{k} \big( R\tilde{\Sigma}^{-1} \big)_{ki} 
\label{eq:f_to_a_dag}\\
\aPyes{\bar{i}} &= \sum_{k} \aPno{\bar{k}} \big( R\tilde{\Sigma}^{-1} \big)_{ki},
\label{eq:f_to_a}
\end{align}
\end{subequations}
By construction these new particle operators also satisfy the Fermion 
anti-commutation relations, $\{\cPyes{i}, \aPyes{j}\} = \delta_{ij}$. By 
inverting relations \eqref{eq:f_to_a_dag}-\eqref{eq:f_to_a}, using the 
expression \eqref{eq:chi}-\eqref{eq:chibar} relating the $(\cQPno{}, \aQPno{})$ 
to the $(\cPno{}, \aPno{})$ and using \eqref{eq:nonortho:orthodef}, these new 
particle operators can be related to the {\quasp} operators 
$(\cQPyes{}, \aQPyes{})$ through
\begin{subequations}
\begin{align}
\cQPyes{\mu}       &= \sum_{l} \cPyes{l}       \left[ \tilde{\Sigma}R^{\dagger} \orbCoefU Q\Sigma^{-1} \right]_{l\mu}
                             - \aPyes{\bar{l}} \left[ \tilde{\Sigma}R^{\dagger} \orbCoefV Q\Sigma^{-1} \right]_{l\mu} \\
\cQPyes{\bar{\mu}} &= \sum_{l} \cPyes{\bar{l}} \left[ \tilde{\Sigma}R^{\top}\orbCoefU Q^{*}\Sigma^{-1} \right]_{l\mu}
                             + \aPyes{l}       \left[ \tilde{\Sigma}R^{\top}\orbCoefV Q^{*}\Sigma^{-1} \right]_{l\mu}.
\end{align}
\end{subequations}
These two equations are the main result of this appendix. They show that we can 
extract from the non-orthogonal reconstructed, canonical wavefunctions a set of 
quasiparticle operators that obey the Fermion anticommutation relation, define 
a quasiparticle vacuum and are related to an orthonormal basis of the 
single-particle Hilbert space through the following Bogoliubov transformation
\be
\tilde{W} = \left(\begin{array}{cc}
\tilde{\Sigma}R^{\top}\orbCoefU{}Q^{*}\Sigma^{-1} & \tilde{\Sigma}R^{\dagger}\orbCoefV{}Q\Sigma^{-1} \\
\tilde{\Sigma}R^{\top}\orbCoefV{}Q^{*}\Sigma^{-1} & \tilde{\Sigma}R^{\dagger}\orbCoefU{}Q\Sigma^{-1}
\end{array}\right)
= \left(\begin{array}{cc}
\tilde{U} & \tilde{V}^{*} \\
\tilde{V} & \tilde{U}^{*}
\end{array}\right) \, .
\ee
This matrix only depends on the initial canonical occupations $\orbCoefU{}$ and 
$\orbCoefV{}$, as well as on the eigenvalues and eigenvectors of both the 
{\singp} overlap matrix \eqref{eq:sp_ovlp} and the {\quasp} overlap matrix 
\eqref{eq:qp_ovlp}.

From the new transformation $\tilde{W}$, we can define the one-body density 
matrix
\be
\rho 
= \tilde{V}^{*}\tilde{V}^{\top} 
= \tilde{\Sigma}R^{\dagger} \orbCoefV {\olapCanRec^{(\idQPno\idQPno)}}^{-1} \orbCoefV R\tilde{\Sigma}.
\ee
and put it into canonical form by diagonalizing it 
\be
\rho = \tilde{D}\occCanV^2\tilde{D}^{\dagger} .
\label{eq:rho}
\ee
The transformation $\tilde{D}$ defines the new canonical basis. By construction, 
these new canonical wavefunctions are expressed in the $(\aPyes{}, \cPyes{})$ 
basis, which is itself related to the original, non-orthogonal basis of the 
reconstructed wavefunctions $(\aPno{}, \cPno{})$ through 
\eqref{eq:f_to_a_dag}-\eqref{eq:f_to_a}. One can easily show that we have
\be
\tilde{D}_n(\rvec, \sigma) = \sum_k \idcanR_k(\rvec, \sigma) \big( R\tilde{\Sigma}^{-1}\tilde{D}\big)_{kn} \,.
\label{eq:phi_to_D}
\ee

At this point, we have obtained a set of genuine canonical wavefunctions 
$\tilde{D}_n(\rvec, \sigma)$ that are orthonormal and are associated with the 
new occupations $v_{n}$ defined by \eqref{eq:rho}. The relation between these 
canonical wavefunctions and the wavefunctions reconstructed by the AE is given 
by \eqref{eq:phi_to_D}. Thanks to this expression, we can now apply the 
Haider \& Gogny formulas for the norm overlap between two many-body states 
$\ket{\idNBodyOrig}$ and $\ket{\idNBodyRec}$. We find
\be
\braket{\idNBodyOrig | \idNBodyRec}
= 
\det\big( {\Sigma^{(\idNBodyOrig)}}^{-1} \big)
\det\big( {\Sigma^{(\idNBodyRec)}}^{-1} \big)
\det\big( \olapCan^{\idNBodyOrig\idNBodyRec} \big)
\det\big( \tilde{Z}^{(\idNBodyOrig\idNBodyRec)} \big) ,
\ee
where
\begin{subequations}
\begin{align}
\olapCan_{mn}^{(\idNBodyOrig\idNBodyRec)}
&= 
\sum_\sigma\int\diff{\rvec}\,
\tilde{D}_{m}^{(\idNBodyOrig)*}(\rvec,\sigma)
\tilde{D}_{n}^{(\idNBodyRec)}(\rvec, \sigma) ,
\\
Z^{(\idNBodyOrig\idNBodyRec)}
&= 
\occCanU^{(\idNBodyOrig)} \big( \olapCan^{(\idNBodyOrig\idNBodyRec)\dagger} \big)^{-1} \occCanU^{(\idNBodyRec)}
+ 
\occCanV^{(\idNBodyOrig)}\olapCan^{(\idNBodyOrig\idNBodyRec)} \occCanV^{(\idNBodyRec)} ,
\\
\occCanU^{(\idNBodyOrig/\idNBodyRec)}
&= 
\sqrt{1-\occCanV^{(\idNBodyOrig/\idNBodyRec)2}}.
\end{align}
\end{subequations}

\section{Metric Induced by an Inner Product}
\label{apx:metrics}

We present in this section the different notions of distance associated with an
inner product. We note $\braket{\avec | \bvec}$ the inner product between
two vectors $\avec$ and $\bvec$.
There are many examples of inner product
in nuclear physics, such as the overlap $\braket{\idNBodyOrig | \idNBodyRec}$ between two
many-body states $\ket{\idNBodyOrig}$ and $\ket{\idNBodyRec}$ or the overlap between
single-particle orbitals $\idcan(\rvec, \sigma)$ and $\idcanR(\rvec, \sigma)$
defined as
\begin{equation}
  \braket{\idcan | \idcanR}
  = \sum_\sigma\int\diff{\rvec}\,
      \idcan^*(\rvec, \sigma)
      \idcanR(\rvec, \sigma).
\end{equation}

Let us first recall some standard mathematics notations. The norm induced by
the inner product $\braket{\avec | \bvec}$ is defined in the usual way as
\begin{equation}
  \|\avec\| = \sqrt{\braket{\avec | \avec}}.
\end{equation}
We can then introduce the distance induced by the inner product as
\begin{equation}
  \label{eq:metrics:inducedMetric}
  \distInduced(\avec, \bvec) = \|\avec-\bvec\|.
\end{equation}
Note that this distance depends on the possible phase and norm of $\avec$
and $\bvec$. However, quantum-mechanical observables do not depend on either
of them. The norm-independent distance thus reads
\begin{equation}
  \label{eq:metrics:norminvMetric0}
  \distUnit(\avec, \bvec)
  = \left\|
      \frac{\avec}{\|\avec\|}
    - \frac{\bvec}{\|\bvec\|}
    \right\|,
\end{equation}
which can be rewritten as a function of the inner product between the two
normed vectors
\begin{equation}
  \label{eq:metrics:norminvMetric}
  \distUnit(\avec, \bvec)
  = \sqrt{2}
    \sqrt{1- \Re\left[ \Braket{ \frac{\avec}{\lVert\avec\rVert} | \frac{\bvec}{\lVert\bvec\rVert} } \right] }
  = \sqrt{2}
    \sqrt{1 - \left| \Braket{ \frac{\avec}{\lVert\avec\rVert} | \frac{\bvec}{\lVert\bvec\rVert} } \right|\cos\Theta},
\end{equation}
where $\Theta = \arg{\braket{\avec | \bvec}}$. Another choice for a
norm-invariant distance is the Great-Circle distance, also known as orthodromic
or spherical distance, that is defined as \cite{deza2009encyclopedia}
\begin{equation}
  \label{eq:metrics:orthoMetric}
  \distOrtho(\avec, \bvec)
  = \arccos\left(
      \Re\left[ \Braket{\frac{\avec}{\lVert\avec\rVert} | \frac{\bvec}{\lVert\bvec\rVert}} \right]%
    \right)
  = \arccos\left(
      \left| \Braket{\frac{\avec}{\lVert\avec\rVert} | \frac{\bvec}{\lVert\bvec\rVert}} \right|
      \cos\Theta%
    \right)
\end{equation}
The orthodromic distance is defined on the manifold of unit vectors. In the
case of real vector spaces, it can be interpreted as the angle between
$\avec$ and $\bvec$. Equations \eqref{eq:metrics:norminvMetric} and
\eqref{eq:metrics:orthoMetric} clearly show that both distances $\distUnit$ and
$\distOrtho$ still depend on the phase $\Theta$ between $\avec$ and
$\bvec$. To remove this dependency, we minimize each distance $\distUnit$
and $\distOrtho$ over $\Theta$. This gives the following two norm- and
phase-independent distances~\footnote{Note that they are distances over the
projective space $P(\mathcal{H})$, not over the 1-body Hilbert space
$\mathcal{H}$},
\begin{subequations}
  \label{eq:metrics:projMetric}
  \begin{align}
    \distProjUnit(\avec, \bvec)
    &= \sqrt{2}
       \sqrt{%
         1
       - \left|
           \Braket{\frac{\avec}{\lVert\avec\rVert} | \frac{\bvec}{\lVert\bvec\rVert}}%
         \right|%
       },
  \end{align}
\end{subequations}
and
\begin{subequations}
  \label{eq:metrics:projIntrMetric}
  \begin{align}
    \distProjOrtho(\avec, \bvec)
    &= \arccos%
         \left|
           \Braket{\frac{\avec}{\lVert\avec\rVert} | \frac{\bvec}{\lVert\bvec\rVert}}%
         \right|.%
  \end{align}
\end{subequations}
The distance $\distProjOrtho$ is an intrinsic metric and is named the 
Fubini–Study metric. It is a generalization of the Bloch sphere. 
Table~\ref{tab:metrics:distances} presents all the distances on the 1-body
Hilbert space between orbitals that we have considered in this work.


\begin{table}[!htb]
  \centering
  \begin{tabular}{c|l|c|c|c}
    \hline\hline
    & & \multicolumn{2}{c}{\textbf{Invariance}} & \\
    \textbf{Notation}
    & \textbf{Space}
    & \textbf{Norm}
    & \textbf{Phase}
    & \textbf{Definition} \\
    \hline
    $\distInduced$
    & $\mathcal{H}$
    & No
    & No
    & \eqref{eq:metrics:inducedMetric} \\
    $\distUnit$
    & Unit vectors of $\mathcal{H}$
    & Yes
    & No
    & \eqref{eq:metrics:norminvMetric0} \\
    $\distOrtho$
    & Unit vectors of $\mathcal{H}$
    & Yes
    & No
    & \eqref{eq:metrics:orthoMetric} \\
    $\distProjUnit$
    & Riemann sphere of $\mathcal{H}$
    & Yes
    & Yes
    & \eqref{eq:metrics:projMetric} \\
    $\distProjOrtho$
    & Riemann sphere of $\mathcal{H}$
    & Yes
    & Yes
    & \eqref{eq:metrics:projIntrMetric} \\
    \hline\hline
  \end{tabular}
  \caption{Different metrics can be defined on the set of orbitals.}
  \label{tab:metrics:distances}
\end{table}

All these distances are defined on the one-body Hilbert space of {\singp} 
wavefunctions. As a result, they do not depend on the occupation probability of 
canonical orbitals, in contrast to the many-body state which takes the BCS 
form. As already mentioned in the main text, determining such a dependency 
exactly from \eqref{eq:AE:archi:loss:exactdist} is not trivial and 
computationally demanding. Instead, we can adopt the approximation that the 
dependency should be proportional to some power $p$ of the occupation number 
$v_{\mu}^{2}$ associated with the current orbital,
\begin{equation}
\distGen^{(p)}(\idcan, \idcanR) = \big(v_{\mu}^{2}\big)^p \times \distGen(\idcan, \idcanR).
\end{equation}


\bibliographystyle{Frontiers-Vancouver}
\bibliography{biblio}

\providecommand{\noopsort}[1]{}
\begin{thebibliography}{116}
\expandafter\ifx\csname natexlab\endcsname\relax\def\natexlab#1{#1}\fi
\expandafter\ifx\csname urlstyle\endcsname\relax
  \expandafter\ifx\csname doi\endcsname\relax
  \def\doi#1{doi:\discretionary{}{}{}#1}\fi \else
  \expandafter\ifx\csname doi\endcsname\relax
  \def\doi{doi:\discretionary{}{}{}\begingroup \urlstyle{rm}\Url}\fi \fi
\expandafter\ifx\csname selectlanguage\endcsname\relax
  \def\selectlanguage#1{}\fi

\bibitem[{Eschrig(1996)}]{eschrig1996fundamentals}
Eschrig R.
\newblock {\em Fundamentals of Density Functional Theory\/} (Leipzig: Teubner)
  (1996).

\bibitem[{Schunck(2019)}]{schunck2019energy}
Schunck N.
\newblock {\em Energy Density Functional Methods for Atomic Nuclei.\/}.
\newblock {{IOP Expanding Physics}} ({Bristol, UK}: {IOP Publishing}) (2019).

\bibitem[{Schunck and Regnier(2022)}]{schunck2022theory}
Schunck N, Regnier D.
\newblock Theory of nuclear fission.
\newblock {\em Prog. Part. Nucl. Phys.\/} {\bf 125} (2022) 103963.
\newblock \doi{10.1016/j.ppnp.2022.103963}.

\bibitem[{Schunck and Robledo(2016)}]{schunck2016microscopic}
Schunck N, Robledo LM.
\newblock Microscopic theory of nuclear fission: A review.
\newblock {\em Rep. Prog. Phys.\/} {\bf 79} (2016) 116301.
\newblock \doi{10.1088/0034-4885/79/11/116301}.

\bibitem[{Kejzlar et~al.(2020)Kejzlar, Neufcourt, Nazarewicz, and
  Reinhard}]{kejzlar2020statistical}
Kejzlar V, Neufcourt L, Nazarewicz W, Reinhard PG.
\newblock Statistical aspects of nuclear mass models.
\newblock {\em J. Phys. G: Nucl. Part. Phys.\/} {\bf 47} (2020) 094001.
\newblock \doi{10.1088/1361-6471/ab907c}.

\bibitem[{Schunck et~al.(2020{\natexlab{a}})Schunck, O'Neal, Grosskopf,
  Lawrence, and Wild}]{schunck2020calibration}
Schunck N, O'Neal J, Grosskopf M, Lawrence E, Wild SM.
\newblock Calibration of energy density functionals with deformed nuclei.
\newblock {\em J. Phys. G: Nucl. Part. Phys.\/} {\bf 47} (2020{\natexlab{a}})
  074001.
\newblock \doi{10.1088/1361-6471/ab8745}.

\bibitem[{Erler et~al.(2012)Erler, Birge, Kortelainen, Nazarewicz, Olsen,
  Perhac et~al.}]{erler2012limits}
Erler J, Birge N, Kortelainen M, Nazarewicz W, Olsen E, Perhac AM, et~al.
\newblock The limits of the nuclear landscape.
\newblock {\em Nature\/} {\bf 486} (2012) 509.
\newblock \doi{10.1038/nature11188}.

\bibitem[{Ney et~al.(2020)Ney, Engel, Li, and Schunck}]{ney2020global}
Ney EM, Engel J, Li T, Schunck N.
\newblock Global description of $\beta^{-}$ decay with the axially deformed
  {{Skyrme}} finite-amplitude method: {{Extension}} to odd-mass and odd-odd
  nuclei.
\newblock {\em Phys. Rev. C\/} {\bf 102} (2020) 034326.
\newblock \doi{10.1103/PhysRevC.102.034326}.

\bibitem[{Mumpower et~al.(2016)Mumpower, Surman, McLaughlin, and
  Aprahamian}]{mumpower2016impact}
Mumpower MR, Surman R, McLaughlin GC, Aprahamian A.
\newblock The impact of individual nuclear properties on r-process
  nucleosynthesis.
\newblock {\em Prog. Part. Nucl. Phys.\/} {\bf 86} (2016) 86.
\newblock \doi{10.1016/j.ppnp.2015.09.001}.

\bibitem[{Perli{\'n}ska et~al.(2004)Perli{\'n}ska, Rohozi{\'n}ski, Dobaczewski,
  and Nazarewicz}]{perlinska2004local}
Perli{\'n}ska E, Rohozi{\'n}ski SG, Dobaczewski J, Nazarewicz W.
\newblock Local density approximation for proton-neutron pairing correlations:
  {{Formalism}}.
\newblock {\em Phys. Rev. C\/} {\bf 69} (2004) 014316.
\newblock \doi{10.1103/PhysRevC.69.014316}.

\bibitem[{Utama et~al.(2016)Utama, Piekarewicz, and Prosper}]{utama2016nuclear}
Utama R, Piekarewicz J, Prosper HB.
\newblock Nuclear mass predictions for the crustal composition of neutron
  stars: {{A Bayesian}} neural network approach.
\newblock {\em Phys. Rev. C\/} {\bf 93} (2016) 014311.
\newblock \doi{10.1103/PhysRevC.93.014311}.

\bibitem[{Utama and Piekarewicz(2017)}]{utama2017refining}
Utama R, Piekarewicz J.
\newblock Refining mass formulas for astrophysical applications: {{A Bayesian}}
  neural network approach.
\newblock {\em Phys. Rev. C\/} {\bf 96} (2017) 044308.
\newblock \doi{10.1103/PhysRevC.96.044308}.

\bibitem[{Utama and Piekarewicz(2018)}]{utama2018validating}
Utama R, Piekarewicz J.
\newblock Validating neural-network refinements of nuclear mass models.
\newblock {\em Phys. Rev. C\/} {\bf 97} (2018) 014306.
\newblock \doi{10.1103/PhysRevC.97.014306}.

\bibitem[{Niu and Liang(2018)}]{niu2018nuclear}
Niu ZM, Liang HZ.
\newblock Nuclear mass predictions based on {{Bayesian}} neural network
  approach with pairing and shell effects.
\newblock {\em Phys. Lett. B\/} {\bf 778} (2018) 48.
\newblock \doi{10.1016/j.physletb.2018.01.002}.

\bibitem[{Neufcourt et~al.(2019)Neufcourt, Cao, Nazarewicz, Olsen, and
  Viens}]{neufcourt2019neutron}
Neufcourt L, Cao Y, Nazarewicz W, Olsen E, Viens F.
\newblock Neutron {{Drip Line}} in the {{Ca Region}} from {{Bayesian Model
  Averaging}}.
\newblock {\em Phys. Rev. Lett.\/} {\bf 122} (2019) 062502.
\newblock \doi{10.1103/PhysRevLett.122.062502}.

\bibitem[{Lovell et~al.(2022)Lovell, Mohan, Sprouse, and
  Mumpower}]{lovell2022nuclear}
Lovell AE, Mohan AT, Sprouse TM, Mumpower MR.
\newblock Nuclear masses learned from a probabilistic neural network.
\newblock {\em Phys. Rev. C\/} {\bf 106} (2022) 014305.
\newblock \doi{10.1103/PhysRevC.106.014305}.

\bibitem[{Mumpower et~al.(2022)Mumpower, Sprouse, Lovell, and
  Mohan}]{mumpower2022physically}
Mumpower MR, Sprouse TM, Lovell AE, Mohan AT.
\newblock Physically interpretable machine learning for nuclear masses.
\newblock {\em Phys. Rev. C\/} {\bf 106} (2022) L021301.
\newblock \doi{10.1103/PhysRevC.106.L021301}.

\bibitem[{Niu et~al.(2019)Niu, Liang, Sun, Long, and Niu}]{niu2019predictions}
Niu ZM, Liang HZ, Sun BH, Long WH, Niu YF.
\newblock Predictions of nuclear $\beta$-decay half-lives with machine learning
  and their impact on r-process nucleosynthesis.
\newblock {\em Phys. Rev. C\/} {\bf 99} (2019) 064307.
\newblock \doi{10.1103/PhysRevC.99.064307}.

\bibitem[{Wang et~al.(2019)Wang, Pei, Liu, and Qiang}]{wang2019bayesian}
Wang ZA, Pei J, Liu Y, Qiang Y.
\newblock Bayesian {{Evaluation}} of {{Incomplete Fission Yields}}.
\newblock {\em Phys. Rev. Lett.\/} {\bf 123} (2019) 122501.
\newblock \doi{10.1103/PhysRevLett.123.122501}.

\bibitem[{Lovell et~al.(2020)Lovell, Mohan, and Talou}]{lovell2020quantifying}
Lovell AE, Mohan AT, Talou P.
\newblock Quantifying uncertainties on fission fragment mass yields with
  mixture density networks.
\newblock {\em J. Phys. G: Nucl. Part. Phys.\/} {\bf 47} (2020) 114001.
\newblock \doi{10.1088/1361-6471/ab9f58}.

\bibitem[{Lasseri et~al.(2020)Lasseri, Regnier, Ebran, and
  Penon}]{lasseri2020taming}
Lasseri RD, Regnier D, Ebran JP, Penon A.
\newblock Taming {{Nuclear Complexity}} with a {{Committee}} of {{Multilayer
  Neural Networks}}.
\newblock {\em Phys. Rev. Lett.\/} {\bf 124} (2020) 162502.
\newblock \doi{10.1103/PhysRevLett.124.162502}.

\bibitem[{Parr and Yang(1989)}]{parr1989density}
Parr R, Yang W.
\newblock {\em Density Functional Theory of Atoms and Molecules\/}.
\newblock International Series of Monographs on Chemistry (New York: Oxford
  University Press) (1989).

\bibitem[{Dreizler and Gross(1990)}]{dreizler1990density}
Dreizler R, Gross E.
\newblock {\em Density Functional Theory: An Approach to the Quantum Many-Body
  Problem\/} (Springer-Verlag) (1990).
\newblock \doi{10.1007/978-3-642-86105-5}.

\bibitem[{Engel(2007)}]{engel2007intrinsicdensity}
Engel J.
\newblock Intrinsic-density functionals.
\newblock {\em Phys. Rev. C\/} {\bf 75} (2007) 014306.
\newblock \doi{10.1103/PhysRevC.75.014306}.

\bibitem[{Barnea(2007)}]{barnea2007density}
Barnea N.
\newblock Density functional theory for self-bound systems.
\newblock {\em Phys. Rev. C\/} {\bf 76} (2007) 067302.
\newblock \doi{10.1103/PhysRevC.76.067302}.

\bibitem[{Engel et~al.(1975)Engel, Brink, Goeke, Krieger, and
  Vautherin}]{engel1975timedependent}
Engel YM, Brink DM, Goeke K, Krieger SJ, Vautherin D.
\newblock Time-dependent {{Hartree-Fock}} theory with {{Skyrme}}'s interaction.
\newblock {\em Nucl. Phys. A\/} {\bf 249} (1975) 215.
\newblock \doi{10.1016/0375-9474(75)90184-0}.

\bibitem[{Dobaczewski and Dudek(1996)}]{dobaczewski1996timeodd}
Dobaczewski J, Dudek J.
\newblock Time-{{Odd Components}} in the {{Rotating Mean Field}} and
  {{Identical Bands}}.
\newblock {\em Acta Phys. Pol. B\/} {\bf 27} (1996) 45.

\bibitem[{Bender et~al.(2003)Bender, Heenen, and
  Reinhard}]{bender2003selfconsistent}
Bender M, Heenen PH, Reinhard PG.
\newblock Self-consistent mean-field models for nuclear structure.
\newblock {\em Rev. Mod. Phys.\/} {\bf 75} (2003) 121.
\newblock \doi{10.1103/RevModPhys.75.121}.

\bibitem[{Lesinski et~al.(2007)Lesinski, Bender, Bennaceur, Duguet, and
  Meyer}]{lesinski2007tensor}
Lesinski T, Bender M, Bennaceur K, Duguet T, Meyer J.
\newblock Tensor part of the {{Skyrme}} energy density functional:
  {{Spherical}} nuclei.
\newblock {\em Phys. Rev. C\/} {\bf 76} (2007) 014312.
\newblock \doi{10.1103/PhysRevC.76.014312}.

\bibitem[{Schunck et~al.(2020{\natexlab{b}})Schunck, Quinlan, and
  Bernstein}]{schunck2020bayesian}
Schunck N, Quinlan KR, Bernstein J.
\newblock A {{Bayesian}} analysis of nuclear deformation properties with
  {{Skyrme}} energy functionals.
\newblock {\em J. Phys. G: Nucl. Part. Phys.\/} {\bf 47} (2020{\natexlab{b}})
  104002.
\newblock \doi{10.1088/1361-6471/aba4fa}.

\bibitem[{Dobaczewski et~al.(1984)Dobaczewski, Flocard, and
  Treiner}]{dobaczewski1984hartreefockbogolyubov}
Dobaczewski J, Flocard H, Treiner J.
\newblock Hartree-{{Fock-Bogolyubov}} description of nuclei near the
  neutron-drip line.
\newblock {\em Nucl. Phys. A\/} {\bf 422} (1984) 103.
\newblock \doi{10.1016/0375-9474(84)90433-0}.

\bibitem[{Dobaczewski et~al.(1996)Dobaczewski, Nazarewicz, Werner, Berger,
  Chinn, and Decharg{\'e}}]{dobaczewski1996meanfield}
Dobaczewski J, Nazarewicz W, Werner TR, Berger JF, Chinn CR, Decharg{\'e} J.
\newblock Mean-field description of ground-state properties of drip-line
  nuclei: {{Pairing}} and continuum effects.
\newblock {\em Phys. Rev. C\/} {\bf 53} (1996) 2809.
\newblock \doi{10.1103/PhysRevC.53.2809}.

\bibitem[{Vautherin and Brink(1972)}]{vautherin1972hartreefock}
Vautherin D, Brink DM.
\newblock Hartree-{{Fock Calculations}} with {{Skyrme}}'s {{Interaction}}.
  {{I}}. {{Spherical Nuclei}}.
\newblock {\em Phys. Rev. C\/} {\bf 5} (1972) 626.
\newblock \doi{10.1103/PhysRevC.5.626}.

\bibitem[{Dobaczewski and Dudek(1997)}]{dobaczewski1997solution}
Dobaczewski J, Dudek J.
\newblock Solution of the {{Skyrme-Hartree-Fock}} equations in the
  {{Cartesian}} deformed harmonic oscillator basis {{I}}. {{The}} method.
\newblock {\em Comput. Phys. Commun.\/} {\bf 102} (1997) 166.
\newblock \doi{10.1016/S0010-4655(97)00004-0}.

\bibitem[{Bender et~al.(2009)Bender, Bennaceur, Duguet, Heenen, Lesinski, and
  Meyer}]{bender2009tensor}
Bender M, Bennaceur K, Duguet T, Heenen PH, Lesinski T, Meyer J.
\newblock Tensor part of the {{Skyrme}} energy density functional. {{II}}.
  {{Deformation}} properties of magic and semi-magic nuclei.
\newblock {\em Phys. Rev. C\/} {\bf 80} (2009) 064302.
\newblock \doi{10.1103/PhysRevC.80.064302}.

\bibitem[{Hellemans et~al.(2012)Hellemans, Heenen, and
  Bender}]{hellemans2012tensor}
Hellemans V, Heenen PH, Bender M.
\newblock Tensor part of the {{Skyrme}} energy density functional. {{III}}.
  {{Time-odd}} terms at high spin.
\newblock {\em Phys. Rev. C\/} {\bf 85} (2012) 014326.
\newblock \doi{10.1103/PhysRevC.85.014326}.

\bibitem[{Ryssens et~al.(2015{\natexlab{a}})Ryssens, Hellemans, Bender, and
  Heenen}]{ryssens2015solution}
Ryssens W, Hellemans V, Bender M, Heenen PH.
\newblock Solution of the {{Skyrme-HF}}+{{BCS}} equation on a {{3D}} mesh,
  {{II}}: {{A}} new version of the {{Ev8}} code.
\newblock {\em Comput. Phys. Commun.\/} {\bf 187} (2015{\natexlab{a}}) 175.
\newblock \doi{10.1016/j.cpc.2014.10.001}.

\bibitem[{Valatin(1961)}]{valatin1961generalized}
Valatin JG.
\newblock Generalized {{Hartree-Fock Method}}.
\newblock {\em Phys. Rev.\/} {\bf 122} (1961) 1012.
\newblock \doi{10.1103/PhysRev.122.1012}.

\bibitem[{Mang(1975)}]{mang1975selfconsistent}
Mang HJ.
\newblock The self-consistent single-particle model in nuclear physics.
\newblock {\em Phys. Rep.\/} {\bf 18} (1975) 325.
\newblock \doi{10.1016/0370-1573(75)90012-5}.

\bibitem[{Blaizot and Ripka(1985)}]{blaizot1985quantum}
Blaizot JP, Ripka G.
\newblock {\em Quantum Theory of Finite Systems\/} (Cambridge: The MIT Press)
  (1985).

\bibitem[{Ring and Schuck(2004)}]{ring2004nuclear}
Ring P, Schuck P.
\newblock {\em The Nuclear Many-Body Problem\/}.
\newblock Texts and Monographs in Physics (Springer) (2004).

\bibitem[{Dobaczewski and Dudek(1995)}]{dobaczewski1995timeodd}
Dobaczewski J, Dudek J.
\newblock Time-odd components in the mean field of rotating superdeformed
  nuclei.
\newblock {\em Phys. Rev. C\/} {\bf 52} (1995) 1827.
\newblock \doi{10.1103/PhysRevC.52.1827}.

\bibitem[{Stoitsov et~al.(2005)Stoitsov, Dobaczewski, Nazarewicz, and
  Ring}]{stoitsov2005axially}
Stoitsov MV, Dobaczewski J, Nazarewicz W, Ring P.
\newblock Axially deformed solution of the {{Skyrme-Hartree-Fock-Bogolyubov}}
  equations using the transformed harmonic oscillator basis. {{The}} program
  {{HFBTHO}} (v1.66p).
\newblock {\em Comput. Phys. Commun.\/} {\bf 167} (2005) 43.
\newblock \doi{10.1016/j.cpc.2005.01.001}.

\bibitem[{Heyde and Wood(2011)}]{heyde2011shape}
Heyde K, Wood JL.
\newblock Shape coexistence in atomic nuclei.
\newblock {\em Rev. Mod. Phys.\/} {\bf 83} (2011) 1467.
\newblock \doi{10.1103/RevModPhys.83.1467}.

\bibitem[{Nakatsukasa et~al.(2016)Nakatsukasa, Matsuyanagi, Matsuo, and
  Yabana}]{nakatsukasa2016timedependent}
Nakatsukasa T, Matsuyanagi K, Matsuo M, Yabana K.
\newblock Time-dependent density-functional description of nuclear dynamics.
\newblock {\em Rev. Mod. Phys.\/} {\bf 88} (2016) 045004.
\newblock \doi{10.1103/RevModPhys.88.045004}.

\bibitem[{Griffin and Wheeler(1957)}]{griffin1957collective}
Griffin JJ, Wheeler JA.
\newblock Collective {{Motions}} in {{Nuclei}} by the {{Method}} of {{Generator
  Coordinates}}.
\newblock {\em Phys. Rev.\/} {\bf 108} (1957) 311.
\newblock \doi{10.1103/PhysRev.108.311}.

\bibitem[{Wa~Wong(1975)}]{wawong1975generatorcoordinate}
Wa~Wong C.
\newblock Generator-coordinate methods in nuclear physics.
\newblock {\em Phys. Rep.\/} {\bf 15} (1975) 283.
\newblock \doi{10.1016/0370-1573(75)90036-8}.

\bibitem[{Reinhard and Goeke(1987)}]{reinhard1987generator}
Reinhard PG, Goeke K.
\newblock The generator coordinate method and quantised collective motion in
  nuclear systems.
\newblock {\em Rep. Prog. Phys.\/} {\bf 50} (1987) 1.
\newblock \doi{10.1088/0034-4885/50/1/001}.

\bibitem[{Verriere and Regnier(2020)}]{verriere2020timedependent}
Verriere M, Regnier D.
\newblock The {{Time-Dependent Generator Coordinate Method}} in {{Nuclear
  Physics}}.
\newblock {\em Front. Phys.\/} {\bf 8} (2020) 1.
\newblock \doi{10.3389/fphy.2020.00233}.

\bibitem[{Brink and Weiguny(1968)}]{brink1968generator}
Brink DM, Weiguny A.
\newblock The generator coordinate theory of collective motion.
\newblock {\em Nucl. Phys. A\/} {\bf 120} (1968) 59.
\newblock \doi{10.1016/0375-9474(68)90059-6}.

\bibitem[{Onishi and Une(1975)}]{onishi1975local}
Onishi N, Une T.
\newblock Local {{Gaussian Approximation}} in the {{Generator Coordinate
  Method}}.
\newblock {\em Prog. Theor. Phys.\/} {\bf 53} (1975) 504.
\newblock \doi{10.1143/PTP.53.504}.

\bibitem[{Une et~al.(1976)Une, Ikeda, and Onishi}]{une1976collective}
Une T, Ikeda A, Onishi N.
\newblock Collective {{Hamiltonian}} in the {{Generator Coordinate Method}}
  with {{Local Gaussian Approximation}}.
\newblock {\em Prog. Theor. Phys.\/} {\bf 55} (1976) 498.
\newblock \doi{10.1143/PTP.55.498}.

\bibitem[{Bloch and Messiah(1962)}]{bloch1962canonical}
Bloch C, Messiah A.
\newblock The canonical form of an antisymmetric tensor and its application to
  the theory of superconductivity.
\newblock {\em Nucl. Phys.\/} {\bf 39} (1962) 95.
\newblock \doi{10.1016/0029-5582(62)90377-2}.

\bibitem[{Zumino(1962)}]{zumino1962normal}
Zumino B.
\newblock Normal {{Forms}} of {{Complex Matrices}}.
\newblock {\em J. Math. Phys.\/} {\bf 3} (1962) 1055--1057.
\newblock \doi{10.1063/1.1724294}.

\bibitem[{Marevi{\'c} et~al.(2022)Marevi{\'c}, Schunck, Ney, Navarro~P{\'e}rez,
  Verriere, and O'Neal}]{marevic2022axiallydeformed}
Marevi{\'c} P, Schunck N, Ney EM, Navarro~P{\'e}rez R, Verriere M, O'Neal J.
\newblock Axially-deformed solution of the {Skyrme-Hartree-Fock-Bogoliubov}
  equations using the transformed harmonic oscillator basis (iv) {HFBTHO}
  (v4.0): A new version of the program.
\newblock {\em Comput. Phys. Commun.\/} {\bf 276} (2022) 108367.
\newblock \doi{10.1016/j.cpc.2022.108367}.

\bibitem[{Drischler et~al.(2020)Drischler, Furnstahl, Melendez, and
  Phillips}]{drischler2020how}
Drischler C, Furnstahl RJ, Melendez JA, Phillips DR.
\newblock How {{Well Do We Know}} the {{Neutron-Matter Equation}} of {{State}}
  at the {{Densities Inside Neutron Stars}}? {{A Bayesian Approach}} with
  {{Correlated Uncertainties}}.
\newblock {\em Phys. Rev. Lett.\/} {\bf 125} (2020) 202702.
\newblock \doi{10.1103/PhysRevLett.125.202702}.

\bibitem[{Kravvaris et~al.(2020)Kravvaris, Quinlan, Quaglioni, Wendt, and
  Navr{\'a}til}]{kravvaris2020quantifying}
Kravvaris K, Quinlan KR, Quaglioni S, Wendt KA, Navr{\'a}til P.
\newblock Quantifying uncertainties in neutron-$\alpha$ scattering with chiral
  nucleon-nucleon and three-nucleon forces.
\newblock {\em Phys. Rev. C\/} {\bf 102} (2020) 024616.
\newblock \doi{10.1103/PhysRevC.102.024616}.

\bibitem[{Acharya and Bacca(2022)}]{acharya2022gaussian}
Acharya B, Bacca S.
\newblock Gaussian process error modeling for chiral effective-field-theory
  calculations of $np\leftrightarrow d\gamma$ at low energies.
\newblock {\em Physics Letters B\/} {\bf 827} (2022) 137011.
\newblock \doi{10.1016/j.physletb.2022.137011}.

\bibitem[{Pastore et~al.(2017)Pastore, Shelley, Baroni, and
  Diget}]{pastore2017new}
Pastore A, Shelley M, Baroni S, Diget CA.
\newblock A new statistical method for the structure of the inner crust of
  neutron stars.
\newblock {\em J. Phys. G: Nucl. Part. Phys.\/} {\bf 44} (2017) 094003.
\newblock \doi{10.1088/1361-6471/aa8207}.

\bibitem[{Kortelainen et~al.(2010)Kortelainen, Lesinski, Mor{\'e}, Nazarewicz,
  Sarich, Schunck et~al.}]{kortelainen2010nuclear}
Kortelainen M, Lesinski T, Mor{\'e} J, Nazarewicz W, Sarich J, Schunck N,
  et~al.
\newblock Nuclear energy density optimization.
\newblock {\em Phys. Rev. C\/} {\bf 82} (2010) 024313.
\newblock \doi{10.1103/PhysRevC.82.024313}.

\bibitem[{Kortelainen et~al.(2012)Kortelainen, McDonnell, Nazarewicz, Reinhard,
  Sarich, Schunck et~al.}]{kortelainen2012nuclear}
Kortelainen M, McDonnell J, Nazarewicz W, Reinhard PG, Sarich J, Schunck N,
  et~al.
\newblock Nuclear energy density optimization: {{Large}} deformations.
\newblock {\em Phys. Rev. C\/} {\bf 85} (2012) 024304.
\newblock \doi{10.1103/PhysRevC.85.024304}.

\bibitem[{Kortelainen et~al.(2014)Kortelainen, McDonnell, Nazarewicz, Olsen,
  Reinhard, Sarich et~al.}]{kortelainen2014nuclear}
Kortelainen M, McDonnell J, Nazarewicz W, Olsen E, Reinhard PG, Sarich J,
  et~al.
\newblock Nuclear energy density optimization: {{Shell}} structure.
\newblock {\em Phys. Rev. C\/} {\bf 89} (2014) 054314.
\newblock \doi{10.1103/PhysRevC.89.054314}.

\bibitem[{Higdon et~al.(2015)Higdon, McDonnell, Schunck, Sarich, and
  Wild}]{higdon2015bayesian}
Higdon D, McDonnell JD, Schunck N, Sarich J, Wild SM.
\newblock A {{Bayesian}} approach for parameter estimation and prediction using
  a computationally intensive model.
\newblock {\em J. Phys. G: Nucl. Part. Phys.\/} {\bf 42} (2015) 034009.
\newblock \doi{10.1088/0954-3899/42/3/034009}.

\bibitem[{McDonnell et~al.(2015)McDonnell, Schunck, Higdon, Sarich, Wild, and
  Nazarewicz}]{mcdonnell2015uncertainty}
McDonnell JD, Schunck N, Higdon D, Sarich J, Wild SM, Nazarewicz W.
\newblock Uncertainty {{Quantification}} for {{Nuclear Density Functional
  Theory}} and {{Information Content}} of {{New Measurements}}.
\newblock {\em Phys. Rev. Lett.\/} {\bf 114} (2015) 122501.
\newblock \doi{10.1103/PhysRevLett.114.122501}.

\bibitem[{Neufcourt et~al.(2018)Neufcourt, Cao, Nazarewicz, and
  Viens}]{neufcourt2018bayesian}
Neufcourt L, Cao Y, Nazarewicz W, Viens F.
\newblock Bayesian approach to model-based extrapolation of nuclear
  observables.
\newblock {\em Phys. Rev. C\/} {\bf 98} (2018) 034318.
\newblock \doi{10.1103/PhysRevC.98.034318}.

\bibitem[{Neufcourt et~al.(2020{\natexlab{a}})Neufcourt, Cao, Giuliani,
  Nazarewicz, Olsen, and Tarasov}]{neufcourt2020proton}
Neufcourt L, Cao Y, Giuliani S, Nazarewicz W, Olsen E, Tarasov OB.
\newblock Beyond the proton drip line: {{Bayesian}} analysis of proton-emitting
  nuclei.
\newblock {\em Phys. Rev. C\/} {\bf 101} (2020{\natexlab{a}}) 014319.
\newblock \doi{10.1103/PhysRevC.101.014319}.

\bibitem[{Neufcourt et~al.(2020{\natexlab{b}})Neufcourt, Cao, Giuliani,
  Nazarewicz, Olsen, and Tarasov}]{neufcourt2020quantified}
Neufcourt L, Cao Y, Giuliani SA, Nazarewicz W, Olsen E, Tarasov OB.
\newblock Quantified limits of the nuclear landscape.
\newblock {\em Phys. Rev. C\/} {\bf 101} (2020{\natexlab{b}}) 044307.
\newblock \doi{10.1103/PhysRevC.101.044307}.

\bibitem[{Rasmussen and Williams(2006)}]{rasmussen2006gaussian}
Rasmussen CE, Williams CKI.
\newblock {\em Gaussian Processes for Machine Learning\/}.
\newblock Adaptive Computation and Machine Learning ({Cambridge, Mass}: {MIT
  Press}) (2006).

\bibitem[{Bartel et~al.(1982)Bartel, Quentin, Brack, Guet, and
  H{\aa}kansson}]{bartel1982better}
Bartel J, Quentin P, Brack M, Guet C, H{\aa}kansson HB.
\newblock Towards a better parametrisation of {{Skyrme-like}} effective forces:
  {{A}} critical study of the {{SkM}} force.
\newblock {\em Nucl. Phys. A\/} {\bf 386} (1982) 79.
\newblock \doi{10.1016/0375-9474(82)90403-1}.

\bibitem[{Schunck et~al.(2014)Schunck, Duke, Carr, and
  Knoll}]{schunck2014description}
Schunck N, Duke D, Carr H, Knoll A.
\newblock Description of induced nuclear fission with {{Skyrme}} energy
  functionals: {{Static}} potential energy surfaces and fission fragment
  properties.
\newblock {\em Phys. Rev. C\/} {\bf 90} (2014) 054305.
\newblock \doi{10.1103/PhysRevC.90.054305}.

\bibitem[{Schunck(2013{\natexlab{a}})}]{schunck2013density}
Schunck N.
\newblock Density {{Functional Theory Approach}} to {{Nuclear Fission}}.
\newblock {\em Acta Phys. Pol. B\/} {\bf 44} (2013{\natexlab{a}}) 263.
\newblock \doi{10.5506/APhysPolB.44.263}.

\bibitem[{Warda and Robledo(2011)}]{warda2011microscopic}
Warda M, Robledo LM.
\newblock Microscopic description of cluster radioactivity in actinide nuclei.
\newblock {\em Phys. Rev. C\/} {\bf 84} (2011) 044608.
\newblock \doi{10.1103/PhysRevC.84.044608}.

\bibitem[{Warda et~al.(2018)Warda, Zdeb, and Robledo}]{warda2018cluster}
Warda M, Zdeb A, Robledo LM.
\newblock Cluster radioactivity in superheavy nuclei.
\newblock {\em Phys. Rev. C\/} {\bf 98} (2018) 041602(R).
\newblock \doi{10.1103/PhysRevC.98.041602}.

\bibitem[{Matheson et~al.(2019)Matheson, Giuliani, Nazarewicz, Sadhukhan, and
  Schunck}]{matheson2019cluster}
Matheson Z, Giuliani SA, Nazarewicz W, Sadhukhan J, Schunck N.
\newblock Cluster radioactivity of $_{118}^{294}\mathrm{Og}_{176}$.
\newblock {\em Phys. Rev. C\/} {\bf 99} (2019) 041304.
\newblock \doi{10.1103/PhysRevC.99.041304}.

\bibitem[{Dubray and Regnier(2012)}]{dubray2012numerical}
Dubray N, Regnier D.
\newblock Numerical search of discontinuities in self-consistent potential
  energy surfaces.
\newblock {\em Comput. Phys. Commun.\/} {\bf 183} (2012) 2035.
\newblock \doi{10.1016/j.cpc.2012.05.001}.

\bibitem[{Sadhukhan(2020)}]{sadhukhan2020microscopic}
Sadhukhan J.
\newblock Microscopic {{Theory}} for {{Spontaneous Fission}}.
\newblock {\em Front. Phys.\/} {\bf 8} (2020) 567171.
\newblock \doi{10.3389/fphy.2020.567171}.

\bibitem[{Bruinsma et~al.(2020)Bruinsma, Perim, Tebbutt, Hosking, Solin, and
  Turner}]{bruinsma2020scalable}
Bruinsma W, Perim E, Tebbutt W, Hosking S, Solin A, Turner R.
\newblock Scalable {{Exact Inference}} in {{Multi-Output Gaussian Processes}}.
\newblock {\em Proceedings of the 37th {{International Conference}} on
  {{Machine Learning}}\/} ({PMLR}) (2020), {\em Proceedings of {{Machine
  Learning Research}}\/}, vol. 119, 1190.

\bibitem[{{\'A}lvarez et~al.(2012){\'A}lvarez, Rosasco, and
  Lawrence}]{alvarez2012kernels}
{\'A}lvarez MA, Rosasco L, Lawrence ND.
\newblock Kernels for {{Vector-Valued Functions}}: {{A Review}}.
\newblock {\em MAL\/} {\bf 4} (2012) 195.
\newblock \doi{10.1561/2200000036}.

\bibitem[{Bender et~al.(2020)Bender, Bernard, Bertsch, Chiba, Dobaczewski,
  Dubray et~al.}]{bender2020future}
Bender M, Bernard R, Bertsch G, Chiba S, Dobaczewski J, Dubray N, et~al.
\newblock Future of nuclear fission theory.
\newblock {\em J. Phys. G: Nucl. Part. Phys.\/} {\bf 47} (2020) 113002.
\newblock \doi{10.1088/1361-6471/abab4f}.

\bibitem[{Schunck(2013{\natexlab{b}})}]{schunck2013microscopic}
Schunck N.
\newblock Microscopic description of induced fission.
\newblock {\em J. Phys.: Conf. Ser.\/} {\bf 436} (2013{\natexlab{b}}) 012058.
\newblock \doi{10.1088/1742-6596/436/1/012058}.

\bibitem[{Ryssens et~al.(2015{\natexlab{b}})Ryssens, Heenen, and
  Bender}]{ryssens2015numerical}
Ryssens W, Heenen PH, Bender M.
\newblock Numerical accuracy of mean-field calculations in coordinate space.
\newblock {\em Phys. Rev. C\/} {\bf 92} (2015{\natexlab{b}}) 064318.
\newblock \doi{10.1103/PhysRevC.92.064318}.

\bibitem[{Jin et~al.(2017)Jin, Bulgac, Roche, and
  Wlaz{\l}owski}]{jin2017coordinatespace}
Jin S, Bulgac A, Roche K, Wlaz{\l}owski G.
\newblock Coordinate-space solver for superfluid many-fermion systems with the
  shifted conjugate-orthogonal conjugate-gradient method.
\newblock {\em Phys. Rev. C\/} {\bf 95} (2017) 044302.
\newblock \doi{10.1103/PhysRevC.95.044302}.

\bibitem[{Regnier et~al.(2017)Regnier, Dubray, Schunck, and
  Verri{\`e}re}]{regnier2017microscopic}
Regnier D, Dubray N, Schunck N, Verri{\`e}re M.
\newblock Microscopic description of fission dynamics: {{Toward}} a {{3D}}
  computation of the time dependent {{GCM}} equation.
\newblock {\em EPJ Web Conf.\/} {\bf 146} (2017) 04043.
\newblock \doi{10.1051/epjconf/201714604043}.

\bibitem[{Zhao et~al.(2021)Zhao, Nik{\v s}i{\'c}, and
  Vretenar}]{zhao2021microscopic}
Zhao J, Nik{\v s}i{\'c} T, Vretenar D.
\newblock Microscopic self-consistent description of induced fission:
  {{Dynamical}} pairing degree of freedom.
\newblock {\em Phys. Rev. C\/} {\bf 104} (2021) 044612.
\newblock \doi{10.1103/PhysRevC.104.044612}.

\bibitem[{Lau et~al.(2022)Lau, Bernard, and Simenel}]{lau2022smoothing}
Lau NWT, Bernard RN, Simenel C.
\newblock Smoothing of one- and two-dimensional discontinuities in potential
  energy surfaces.
\newblock {\em Phys. Rev. C\/} {\bf 105} (2022) 034617.
\newblock \doi{10.1103/PhysRevC.105.034617}.

\bibitem[{Baldi(2012)}]{baldi2012autoencoders}
Baldi P.
\newblock Autoencoders, unsupervised learning, and deep architectures.
\newblock {\em Proceedings of ICML workshop on unsupervised and transfer
  learning\/} (JMLR Workshop and Conference Proceedings) (2012), 37--49.

\bibitem[{Burda et~al.(2015)Burda, Grosse, and
  Salakhutdinov}]{burda2015importance}
Burda Y, Grosse R, Salakhutdinov R.
\newblock Importance weighted autoencoders.
\newblock {\em arXiv preprint arXiv:1509.00519\/}  (2015).

\bibitem[{Chen et~al.(2012)Chen, Xu, Weinberger, and
  Sha}]{chen2012marginalized}
Chen M, Xu Z, Weinberger K, Sha F.
\newblock Marginalized denoising autoencoders for domain adaptation.
\newblock {\em arXiv preprint arXiv:1206.4683\/}  (2012).

\bibitem[{Gong et~al.(2019)Gong, Liu, Le, Saha, Mansour, Venkatesh
  et~al.}]{gong2019memorizing}
Gong D, Liu L, Le V, Saha B, Mansour MR, Venkatesh S, et~al.
\newblock Memorizing normality to detect anomaly: Memory-augmented deep
  autoencoder for unsupervised anomaly detection.
\newblock {\em Proceedings of the IEEE/CVF International Conference on Computer
  Vision\/} (2019), 1705--1714.

\bibitem[{Bengio et~al.(2013)Bengio, Courville, and
  Vincent}]{bengio2013representation}
Bengio Y, Courville A, Vincent P.
\newblock Representation learning: A review and new perspectives.
\newblock {\em IEEE transactions on pattern analysis and machine
  intelligence\/} {\bf 35} (2013) 1798--1828.

\bibitem[{Zhang et~al.(2014)Zhang, Du, and Zhang}]{zhang2014saliency}
Zhang F, Du B, Zhang L.
\newblock Saliency-guided unsupervised feature learning for scene
  classification.
\newblock {\em IEEE Transactions on Geoscience and Remote Sensing\/} {\bf 53}
  (2014) 2175--2184.

\bibitem[{Yu et~al.(2017)Yu, Hong, Rui, and Tao}]{yu2017multitask}
Yu J, Hong C, Rui Y, Tao D.
\newblock Multitask autoencoder model for recovering human poses.
\newblock {\em IEEE Transactions on Industrial Electronics\/} {\bf 65} (2017)
  5060--5068.

\bibitem[{{von Neuman} and {Wigner}(1929)}]{vonneuman1929uber}
{von Neuman} J, {Wigner} E.
\newblock {Uber merkw{\"u}rdige diskrete Eigenwerte. Uber das Verhalten von
  Eigenwerten bei adiabatischen Prozessen}.
\newblock {\em Physikalische Zeitschrift\/} {\bf 30} (1929) 467--470.

\bibitem[{Teller(1937)}]{teller1937crossing}
Teller E.
\newblock The {{Crossing}} of {{Potential Surfaces}}.
\newblock {\em J. Phys. Chem.\/} {\bf 41} (1937) 109.
\newblock \doi{10.1021/j150379a010}.

\bibitem[{{Longuet-Higgins} et~al.(1958){Longuet-Higgins}, {\"O}pik, Pryce, and
  Sack}]{longuet-higgins1958studies}
{Longuet-Higgins} HC, {\"O}pik U, Pryce MHL, Sack RA.
\newblock Studies of the {{Jahn-Teller Effect}}. {{II}}. {{The Dynamical
  Problem}}.
\newblock {\em Proc. R. Soc. Lond. A\/} {\bf 244} (1958) 1.
\newblock \doi{10.1098/rspa.1958.0022}.

\bibitem[{{Longuet-Higgins}(1975)}]{longuet-higgins1975intersection}
{Longuet-Higgins} H.
\newblock The intersection of potential energy surfaces in polyatomic
  molecules.
\newblock {\em Proc. R. Soc. Lond. A\/} {\bf 344} (1975) 147.
\newblock \doi{10.1098/rspa.1975.0095}.

\bibitem[{Domcke et~al.(2011)Domcke, Yarkony, and K\"oppel}]{domcke2011conical}
Domcke W, Yarkony D, K\"oppel H, editors.
\newblock {\em Conical Intersections: Theory, Computation and Experiment\/}.
\newblock Advanced Series in Physical Chemistry: Volume 17 (World Scientific)
  (2011).
\newblock \doi{10.1142/7803}.

\bibitem[{Larson et~al.(2020)Larson, Sj\"oqvist, and
  \"Ohberg}]{larson2020intersections}
Larson J, Sj\"oqvist E, \"Ohberg P.
\newblock {\em Intersections in Physics. An Introduction to Synthetic Gauge
  Theories\/}.
\newblock Lecture Notes in Physics (Springer) (2020).
\newblock \doi{10.1007/978-3-030-34882-3}.

\bibitem[{Haider and Gogny(1992)}]{haider1992microscopic}
Haider Q, Gogny D.
\newblock Microscopic approach to the generator coordinate method with pairing
  correlations and density-dependent forces.
\newblock {\em J. Phys. G: Nucl. Part. Phys.\/} {\bf 18} (1992) 993.
\newblock \doi{10.1088/0954-3899/18/6/003}.

\bibitem[{Ioffe and Szegedy(2015)}]{ioffe2015batch}
Ioffe S, Szegedy C.
\newblock Batch normalization: Accelerating deep network training by reducing
  internal covariate shift.
\newblock {\em International conference on machine learning\/} (PMLR) (2015),
  448--456.

\bibitem[{Srivastava et~al.(2014)Srivastava, Hinton, Krizhevsky, Sutskever, and
  Salakhutdinov}]{srivastava2014dropout}
Srivastava N, Hinton G, Krizhevsky A, Sutskever I, Salakhutdinov R.
\newblock Dropout: a simple way to prevent neural networks from overfitting.
\newblock {\em The journal of machine learning research\/} {\bf 15} (2014)
  1929--1958.

\bibitem[{He et~al.(2016)He, Zhang, Ren, and Sun}]{he2016deep}
He K, Zhang X, Ren S, Sun J.
\newblock Deep residual learning for image recognition.
\newblock {\em Proceedings of the IEEE conference on computer vision and
  pattern recognition\/} (2016), 770--778.

\bibitem[{Krizhevsky et~al.(2012)Krizhevsky, Sutskever, and
  Hinton}]{krizhevsky2012imagenet}
Krizhevsky A, Sutskever I, Hinton GE.
\newblock Imagenet classification with deep convolutional neural networks.
\newblock {\em Advances in neural information processing systems\/} {\bf 25}
  (2012).

\bibitem[{Zeiler and Fergus(2014)}]{zeiler2014visualizing}
Zeiler MD, Fergus R.
\newblock Visualizing and understanding convolutional networks.
\newblock {\em European conference on computer vision\/} (Springer) (2014),
  818--833.

\bibitem[{Sermanet et~al.(2013)Sermanet, Eigen, Zhang, Mathieu, Fergus, and
  LeCun}]{sermanet2013overfeat}
Sermanet P, Eigen D, Zhang X, Mathieu M, Fergus R, LeCun Y.
\newblock Overfeat: Integrated recognition, localization and detection using
  convolutional networks.
\newblock {\em arXiv preprint arXiv:1312.6229\/}  (2013).

\bibitem[{Szegedy et~al.(2017)Szegedy, Ioffe, Vanhoucke, and
  Alemi}]{szegedy2017inception}
Szegedy C, Ioffe S, Vanhoucke V, Alemi AA.
\newblock Inception-v4, inception-resnet and the impact of residual connections
  on learning.
\newblock {\em Thirty-first AAAI conference on artificial intelligence\/}
  (2017).

\bibitem[{Li et~al.(2018)Li, Xu, Taylor, Studer, and
  Goldstein}]{li2018visualizing}
Li H, Xu Z, Taylor G, Studer C, Goldstein T.
\newblock Visualizing the loss landscape of neural nets.
\newblock {\em Advances in neural information processing systems\/} {\bf 31}
  (2018).

\bibitem[{Zhang et~al.(2022)Zhang, Wu, Zhang, Zhu, Lin, Zhang
  et~al.}]{zhang2022resnet}
Zhang H, Wu C, Zhang Z, Zhu Y, Lin H, Zhang Z, et~al.
\newblock Resnet: Split-attention networks.
\newblock {\em Proceedings of the IEEE/CVF Conference on Computer Vision and
  Pattern Recognition\/} (2022), 2736--2746.

\bibitem[{Radosavovic et~al.(2020)Radosavovic, Kosaraju, Girshick, He, and
  Doll{\'a}r}]{radosavovic2020designing}
Radosavovic I, Kosaraju RP, Girshick R, He K, Doll{\'a}r P.
\newblock Designing network design spaces.
\newblock {\em Proceedings of the IEEE/CVF conference on computer vision and
  pattern recognition\/} (2020), 10428--10436.

\bibitem[{Cubuk et~al.(2020)Cubuk, Zoph, Shlens, and Le}]{cubuk2020randaugment}
Cubuk ED, Zoph B, Shlens J, Le QV.
\newblock Randaugment: Practical automated data augmentation with a reduced
  search space.
\newblock {\em Proceedings of the IEEE/CVF conference on computer vision and
  pattern recognition workshops\/} (2020), 702--703.

\bibitem[{Yun et~al.(2019)Yun, Han, Oh, Chun, Choe, and Yoo}]{yun2019cutmix}
Yun S, Han D, Oh SJ, Chun S, Choe J, Yoo Y.
\newblock Cutmix: Regularization strategy to train strong classifiers with
  localizable features.
\newblock {\em Proceedings of the IEEE/CVF international conference on computer
  vision\/} (2019), 6023--6032.

\bibitem[{Zhang et~al.(2017)Zhang, Cisse, Dauphin, and
  Lopez-Paz}]{zhang2017mixup}
Zhang H, Cisse M, Dauphin YN, Lopez-Paz D.
\newblock mixup: Beyond empirical risk minimization.
\newblock {\em arXiv preprint arXiv:1710.09412\/}  (2017).

\bibitem[{Wickramasinghe et~al.(2021)Wickramasinghe, Marino, and
  Manic}]{wickramasinghe2021resnet}
Wickramasinghe CS, Marino DL, Manic M.
\newblock Resnet autoencoders for unsupervised feature learning from
  high-dimensional data: Deep models resistant to performance degradation.
\newblock {\em IEEE Access\/} {\bf 9} (2021) 40511.
\newblock \doi{10.1109/ACCESS.2021.3064819}.

\bibitem[{He et~al.(2015)He, Zhang, Ren, and Sun}]{he2015delving}
He K, Zhang X, Ren S, Sun J.
\newblock Delving deep into rectifiers: Surpassing human-level performance on
  imagenet classification.
\newblock {\em Proceedings of the IEEE international conference on computer
  vision\/} (2015), 1026--1034.

\bibitem[{Kingma and Ba(2014)}]{kingma2014adam}
Kingma DP, Ba J.
\newblock Adam: A method for stochastic optimization.
\newblock {\em arXiv preprint arXiv:1412.6980\/}  (2014).

\bibitem[{Deza and Deza(2009)}]{deza2009encyclopedia}
Deza M, Deza E.
\newblock {\em Encyclopedia of Distances\/} (Springer) (2009).
\newblock \doi{10.1007/978-3-642-00234-2}.

\end{thebibliography}

\end{document}